%% file: it_journal.tex
\documentclass[journal]{IEEEtran}
\usepackage{bbm,dsfont,mathrsfs,fixmath}
\usepackage{hyperref}
\usepackage{amsmath,amssymb,graphicx}
\usepackage{cite}
\usepackage{amstext,amsfonts,amssymb}
\usepackage{epsfig}
\usepackage{exscale}
\usepackage{mathtools}
\usepackage{enumerate}
\usepackage{cite}
\usepackage{graphicx}
\usepackage{amssymb}
\usepackage{array}
\usepackage{stfloats}
\usepackage{balance}
\usepackage{dsfont}
\usepackage{bbm}
\usepackage{ifpdf}
\usepackage{stfloats,color}
\usepackage{dsfont}

\include{packages}

\include{commandes}

\usepackage[T1]{fontenc}

\title{Collaborative Information Bottleneck}

\author{Matias Vera,~\IEEEmembership{Student~Member,~IEEE},
 Leonardo Rey Vega,\\~\IEEEmembership{Member,~IEEE},
  and~ Pablo~Piantanida,~\IEEEmembership{Senior~Member,~IEEE}
  \thanks{The material in this paper was in part presented at the 2015 and at the 2017 IEEE International Symposium on Information Theory.}
  \thanks{Matias Vera is with Facultad de Ingenier\'ia, Universidad de Buenos Aires, Buenos Aires, Argentina. Email: mvera@fi.uba.ar.}
    \thanks{Leonardo Rey Vega is with Universidad de Buenos Aires and CSC-CONICET, Buenos Aires, Argentina. Email:  lrey@fi.uba.ar.}
     \thanks{P. Piantanida is with CentraleSup\'elec -  French National Center for Scientific Research (CNRS) - Universit\'{e} Paris-Sud,  3 Rue Joliot-Curie, F-91190 Gif-sur-Yvette, France and with Montreal Institute for Learning Algorithms (MILA) at Universit\'e de Montr\'eal, 2920 Chemin de la Tour, Montr\'{e}al, QC H3T 1N8, Canada. Email: pablo.piantanida@centralesupelec.fr.} 
  \thanks{The work of Leonardo Rey Vega work was supported by grants PIP11220150100578CO from CONICET, UBACyT 20020170100470BA from University of Buenos Aires and PICS MoSiME from CNRS. The work of Matias Vera was supported by Peruilh PhD Scholarship from Universidad de Buenos Aires. This project has received funding from the European Union Horizon 2020 research and innovation programme under the Marie Sklodowska-Curie grant agreement No 792464.}
  \thanks{Copyright (c) 2017 IEEE. Personal use of this material is permitted.  However, permission to use this material for any other purposes must be obtained from the IEEE by sending a request to pubs-permissions@ieee.org.}
}

\allowdisplaybreaks

\begin{document}
        
\maketitle

\begin{abstract}
This paper investigates a multi-terminal source coding problem under a logarithmic loss fidelity which does not necessarily lead to an additive distortion measure. The problem is motivated by an extension of the Information Bottleneck method to a multi-source scenario where several encoders have to  build cooperatively rate-limited descriptions of their sources in order to maximize information with respect to other unobserved (hidden) sources. More precisely, we study fundamental information-theoretic limits of the so-called: (i) \emph{Two-way Collaborative Information Bottleneck} (TW-CIB) and (ii) the \emph{Collaborative Distributed Information Bottleneck} (CDIB) problems. The TW-CIB problem consists of two distant encoders that separately observe marginal (dependent) components $X_1$ and $X_2$ and can  cooperate through multiple exchanges of limited information with the aim of extracting information about \emph{hidden} variables $(Y_1,Y_2)$, which can be  arbitrarily dependent on $(X_1,X_2)$. On the other hand, in CDIB  there are two cooperating encoders which separately observe $X_1$ and $X_2$ and a third node which can listen to the exchanges between the two encoders in order to obtain information about a hidden variable $Y$. The  \emph{relevance} (figure-of-merit) is measured in terms of a normalized (per-sample) multi-letter mutual information metric (\emph{log-loss} fidelity) and an interesting tradeoff arises by constraining the \emph{complexity} of descriptions, measured in terms of the rates needed for the exchanges between the encoders and decoders involved. Inner and outer bounds to the complexity-relevance region of these problems are derived from which optimality is characterized for several cases of interest. Our resulting theoretical complexity-relevance regions are finally evaluated for binary symmetric and Gaussian statistical models, showing theoretical tradeoffs between the complexity-constrained descriptions and their relevance with respect to the hidden variables.

\end{abstract}

\begin{IEEEkeywords}
Multi-terminal source coding; Logarithmic loss; Distributed source coding; Noisy rate-distortion; Side information; Interactive lossy source coding; Information Bottleneck;  Shannon theory.
\end{IEEEkeywords}

\IEEEpeerreviewmaketitle

\section{Introduction}
In the last years we have witnessed a monumental proliferation of digital data, leading to new efforts in the understanding of the fundamental principles behind the discovery of relevant information from massive data sets.  A good data representation is paramount for performing large-scale data processing and analysis in a computationally efficient (e.g. minimizing communication resources and time of computation) and statistically meaningful manner~\cite{Chandrasekaran2013Computational}. In addition to reducing computation time, proper data representations can decrease storage requirements, which translates into reduced inter-node communication allowing to take advantage of different information sources (multi-view analysis) to improve prediction  performance. 

The challenge of identifying \emph{relevant rate-limited} information from observed samples, that is the statistical useful information that those observations provide about other \emph{hidden} variables of interest, is to obtain compressed descriptions that are good enough statistics for inference of these hidden variables. This raises fundamental questions about the information-theoretic principles underlying the process of discovering valuable and relevant knowledge in the form of structured information. In that sense, the standard rate-distortion function of lossy source coding~\cite{Shannon1993CodingTheoremsForADiscreteSourceWithAFidelityCriterion} provides an interesting starting point as a means to understand fundamental information-theoretic tradeoffs between \emph{relevance} (quality of data descriptions) and \emph{complexity} (size in terms of bits of the descriptions). Relevance can be linked to an appropriate (non-additive) fidelity measure that captures the meaningful characteristics of unobserved data while complexity can be associated to the size of the data descriptions generated from the observed samples.

In this paper, we investigate the fundamental information-theoretic limits of a collaborative and distributed source coding problem with a (not necessarily additive) log-loss fidelity, which is motivated by the Information Bottleneck problem~\cite{tishby99}. As opposed to a centralized setting, in our present framework each source observes only a fragment of the total data set to process, where subsets of data tuples (possibly overlapping) are available at different sites. This distributed setup typically imposes a set of constraints on the decoders which are absent in the centralized setup and that could prohibit the transfer of raw data from each of the sites to a central location. We approach this challenging problem from an information-theoretic perspective, studying the exchanges of data descriptions between sites or agents subject to communication (information rates) constraints.

\subsection{Related Work}


The idea of obtaining good descriptions of a hidden variable through the compression of an observed depedent one can be formalized through the \emph{noisy source-coding} problem introduced in~\cite{1057738}, where the functions that generate the appropriate descriptions corresponds to the class of rate-limited encoders that compress the observation $X$ with the goal of minimizing a fidelity (distortion) measure with respect to an unobserved variable $Y$. The optimal rate-distortion tradeoff region follows from the function~\cite{Shannon1993CodingTheoremsForADiscreteSourceWithAFidelityCriterion}:
\begin{equation*}
R(D)= \inf\limits_{p_{U|X}: \  \mathbb{E}[d(U,Y)]\leq D } I(X;U), 
\end{equation*} 
where $d:\mathcal{U}\times \mathcal{Y}  \rightarrow \mathbb{R}_+$ is a per-letter distortion (or \emph{loss}) measure and $p_{U|X}:\mathcal{X}\rightarrow \mathcal{P}(\mathcal{U})$ is conditional distribution that satisfies the Markov chain $U \mkv X \mkv Y$. Several distortion functions could be of interest in practice such as the Hamming or quadratic loss. In particular, taking the loss $d(u,y)=-\log p_{Y|U}(y|u)$ with $D=H(Y)-\mu$ yields an interesting case of an additive (over the source samples)  mutual information as the (single-letter) distortion measure. This measure of relevance was first proposed in~\cite{tishby99} giving birth to the \emph{Information Bottleneck} method. The main idea behind it is finding a compressed description $f(X^n)$ of the data $X^n=(X_1,\dots,X_n)$ with coding rate $\log | f| \leq nR$ subject to a constraint on the mutual information $I\big(f(X^n);Y_i\big)\geq \mu$, where $Y_i$ depends on $X_i$, and $\mu$ is the minimal level of \emph{relevance} required and $R$ is the coding rate. As pointed out in~\cite{Courtade_2014}, this notion of relevance boils down to noisy lossy source coding with \emph{logarithmic loss} distortion, from which the optimal tradeoff region (rates of complexity $R$ and relevance $\mu$) follows from the rate-relevance function:
\begin{equation*}
R(\mu)= \inf\limits_{p_{U|X}: \  I(U;Y)\geq \mu} I(X;U),  
\end{equation*} 
where $p_{U|X}:\mathcal{X}\rightarrow \mathcal{P}(\mathcal{U})$ forms a Markov chain $U \mkv X \mkv Y$. The function $\mu\mapsto R(\mu)$ (or its dual $R\mapsto \mu(R)$) provides a curve similar to the rate-distortion curve, that provides all tradeoffs between coding rates and levels of information w.r.t. hidden variable $Y$. Interestingly, the same single-letter characterization is also the optimal characterization when the relevance is measured by a multi-letter mutual information $I\big(f(X^n);Y^n\big)\geq n\mu$ with $Y^n=(Y_1,\dots,Y_n)$ which is, in general, a non-additive distortion~\cite{witsenhausen75}. This was also observed in \cite{Courtade_2014}. The rate-relevance function given by  for the classical information bottleneck given by $R(\mu)$ can then be though either, as point-to-point noisy source coding problem with additive single letter distorion given by $d(u,y)=-\log p_{Y|U}(y|u)$ or with multi-letter fidelity criterion given by $I\big(f(X^n);Y^n\big)$ as discussed above.

In line with the above mentioned works and modeling the structure of data and its hidden variables by independent and identically distributed samples draw from a known distribution, this paper aims at understanding how proper distributed data descriptions translates into reduced inter-encoder communication when there are several parties involved which observe dependent sources and are interested in extracting useful information about other hidden variables. This clearly should be done by taking advantage of the dependence between the different information sources to recover a good enough statistic that summarizes relevant information about some unobserved hidden variables using cooperation and interaction among all parties involved.


It is worth to further emphasize our motivation behind the use of a multi-letter (non-additive) mutual information as a measure of relevance. Although in principle more difficult to analyze, it appears to be  more natural and appealing from a practical perspective, as it allows the possibility of better exploring temporal dependences in the metric of relevance induced by the encoding mapping with respect to the case where an additive metric is considered as in~\cite{tishby99}. Despite the fact both additive and non-additive relevances lead asymptotically to the same mathematical problem (the reader may be refer to~\cite{GeorgISIT2015B} for further details), the multi-letter form of the relevance is connected to a variety of interesting problems in information theory. More precisely, the multi-letter (non-additive) relevance becomes: the asymptotic exponent corresponding to the second type error probability of distributed testing against independence~\cite{1057194,Gil2015}; the  asymptotic characterization of images of sets via noisy channels~\cite{1055800} and is also related to the Hypercontraction of the Markov operator~\cite{ahlswede1976} and gambling problems~\cite{Erkip1998efficiency}.

The distributed (non-cooperative) setting of the source coding problem with logarithmic loss distortion, was first investigated in~\cite{Courtade_2014}, where a complete characterization of the complexity-relevance region was derived, solving completely the Berger-Tung problem~\cite{PhD-Tung} under this specific distortion metric. Moreover, the well-known longstanding open CEO problem~\cite{berger_ceo_1996} was also completely solved under this distortion metric. The CEO problem is in fact a well-studied problem which has received a lot of interest in the last years because of its relevance to distributed sensing schemes, specially for the quadratic Gaussian case~\cite{prabhakaran_rate_2004,oohama_rate-distortion_2005}. A multi-terminal source coding problem --fundamentally different from previous distributed source coding problems-- termed \emph{information-theoretic biclustering} was also investigated in~\cite{DBLP:journals/corr/PichlerPM16}. In this setting, several distributed  (non-cooperative) encoders are interested in maximizing, as much as possible, redundant information among their observations. Equally important is the impact that cooperation and interaction can have in distributed source coding scenarios. In this sense, the seminal work by Kaspi \cite{kaspi_two-way_1985} has sparked some interest in the recents years, where several papers in the fields of distributed function computation and rate-distortion theory were published~\cite{ma_ishwar2011,Permuter_2010,Chia_2012,ours_2015}.



\subsection{Contributions}

In summary, we will consider a multi-point source coding problem where the dependence between the observed and hidden sources can be exploited through cooperation and interaction. As we will be interested in a multi-letter fidelity criterion given by the mutual information between the generated descriptions and the hidden variables we can see that our general setting can be interpreted as a multi-point information bottleneck problem generalizing the above discussed classical point-to-point information bottleneck to a distributed setting.

\begin{figure}[t]
	\centering{\includegraphics[width=0.48\textwidth]{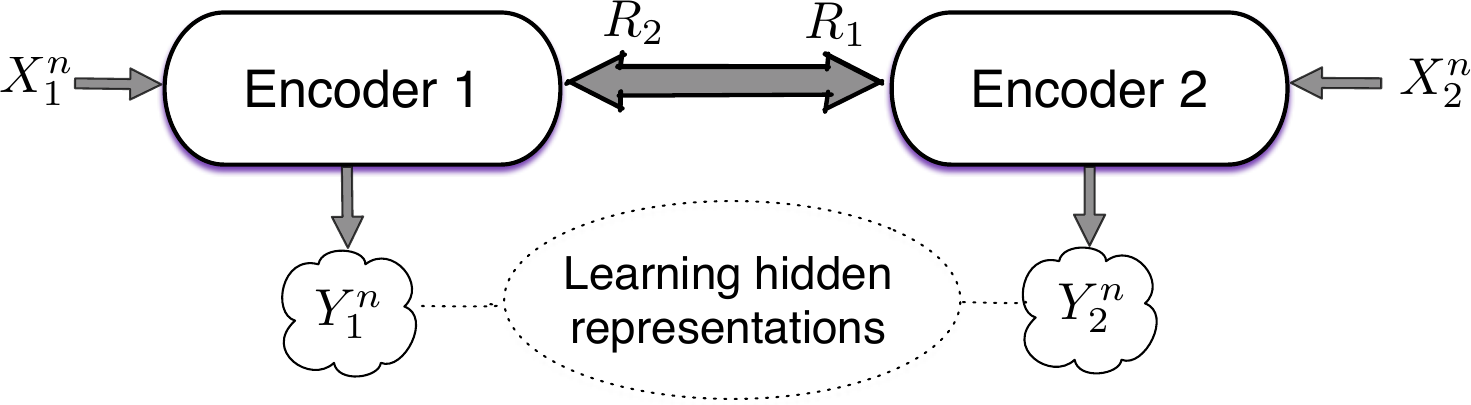}}
	\caption{Two-way Collaborative Information Bottleneck (TW-CIB).}
\label{fig:db1}
\end{figure}

In more precise terms, in this paper, we first study the so-called \emph{Two-way Collaborative Information Bottleneck} (TW-CIB) problem, as described in Fig.~\ref{fig:db1}.  This scenario consists of two distant encoders that separately observe marginal components $X_1^n$ and $X_2^n$ of a joint memoryless process and wish to cooperate through multiple exchanges of limited (complexity) rate with the goal of extracting relevant information about some hidden variables $(Y_1^n,Y_2^n)$, which can be arbitrarily dependent on $(X_1^n,X_2^n)$. The relevance of the information extracted is measured in terms of the normalized multi-letter mutual information between the generated descriptions and the corresponding \emph{hidden} variables. We characterize the set of all feasible rates of complexity and relevance, for an arbitrary number of exchange rounds. This result is particularized to some binary symmetric and all possible Gaussian statistical models. In particular, the analysis of the binary symmetric case (even for the simpler \emph{half-round} case) appears to be rather involved.


\begin{figure}[t]
	\centering{\includegraphics[width=0.48\textwidth]{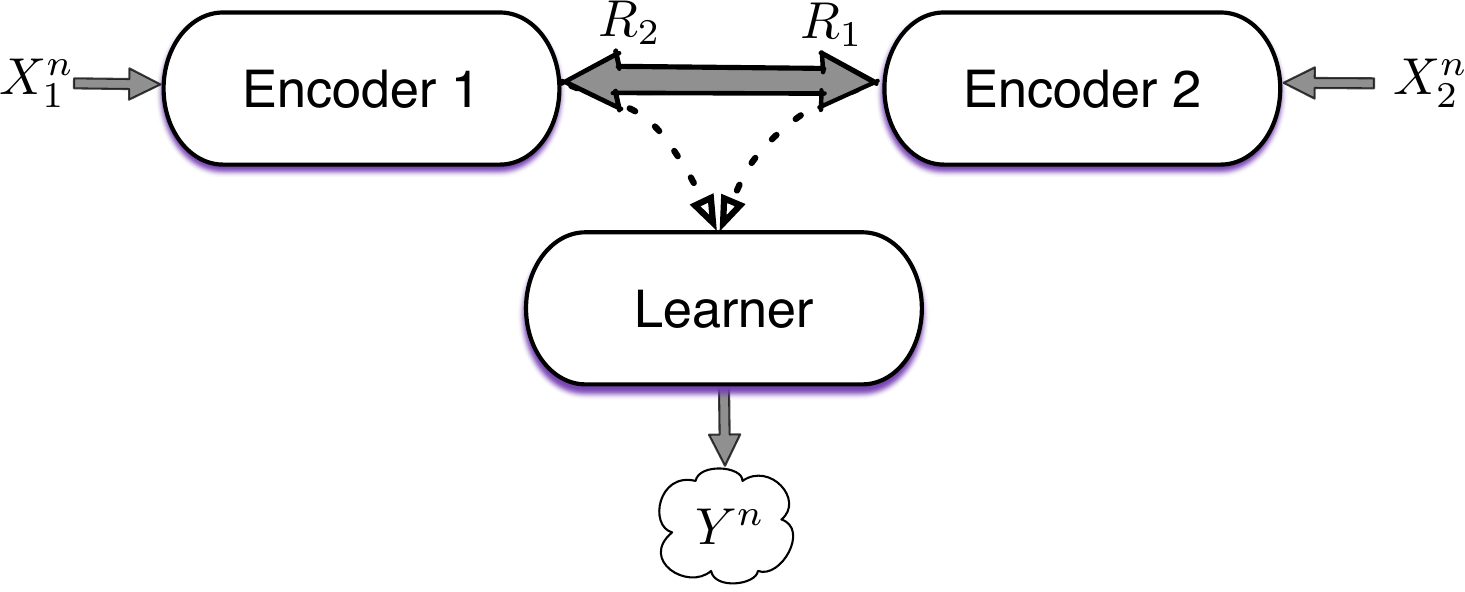}}
	\caption{Collaborative Distributed Information Bottleneck (CDIB).}
\label{fig:db2}
\end{figure}

Then, we investigate the so-called \emph{Collaborative Distributed Information Bottleneck} (CDIB) problem, as described in Fig.~\ref{fig:db2}. This differs from the above scenario in that only a single decoder which is not part of the encoders is considered. Still the decoder wishes to use  descriptions from sources  $X_1^n$ and $X_2^n$ to maximize the  multi-letter mutual information with respect to the \emph{hidden} (relevant) variable $Y^n$. This scenario can be identified as being the natural extension of the previous works~\cite{Courtade_2014, our_isit15}. However, in the present setting, encoders $1$ and $2$ can interactively cooperate by exchanging pieces of information that should be informative enough about $Y$ but without becoming too complex in order to be transmitted and recovered at the decoder. The central  difficult arises in finding the way to  explicitly exploit the correlation present between the variables $(X_1,X_2,Y)$ to reduce the cost of communication. We begin by deriving an inner bound to the complexity-relevance region of this problem  that is valid for any number of exchanges between the encoders. To this end, we use a cooperative binning procedure to allow explicit cooperation between encoders while guaranteeing successful decoding at the decoder. This can be achieved despite of the fact that the decoder has not side information. Then, we provide an outer bound which proves to be tight if either $X_1\mkv Y\mkv X_2$ or $X_2\mkv X_1 \mkv Y$ when only one round of exchange is allowed. Our results are finally applied to the Gaussian case. 

The rest of the paper is organized as follows. In Section~\ref{subsec:CRL}, we introduce the Two-way Collaborative Information Bottleneck (TW-CIB) problem and provide the optimal characterization of the set of achievable complexity-relevance tradeoffs. In Section~\ref{subsec:CDIB}, we introduce the Collaborative Distributed Information Bottleneck (CDIB) problem and provide inner and outer bounds to the corresponding set of achievable complexity-relevance tradeoffs. Optimal characterizations are provided in the two specific cases mentioned above. Proofs of the several outer bounds presented in the paper are relegated to Section~\ref{sec:conv} while the inner bounds are developed in the appendices. Gaussian models are investigated in Section~\ref{sec:gaussian} while the binary symmetric model for the TW-CIB problem is studied in Section~\ref{sec:binary}. Finally, in Section~\ref{sec:summary} the conclusions are presented.
 		
\subsection*{Conventions and Notations }
We use upper-case letters to denote random variables and lower-case letters to denote realizations of random variables. With ${x^n}$ and $X^n$ we denote vectors and random vectors of $n$ components, respectively. The $i$-th component of vector $x^n$ is denoted interchangeably as $x_i$ or $x_{[i]}$ and with $x_{[s:t]}$ we denote the components with indices ranging from $s$ to $t$ with $s\leq t$.  All alphabets  are assumed to be finite, except for the Gaussian models discussed in Section~\ref{sec:gaussian}. Entropy is denoted by $H(\cdot)$, differential entropy by $h(\cdot)$, binary entropy by $h_2(\cdot)$ and mutual information by $I(\cdot;\cdot)$. If $X$, $Y$ and $V$ are three random variables on some alphabets their probability distribution is denoted by $p_{XYV}$. When clear from context we will simple denote $p_{X}(x)$ with $p(x)$. With $\mathcal{P}\left(\mathcal{X}\right)$ we denote the set of probability distributions over alphabet $\mathcal{X}$.  If the probability distribution of random variables $X,Y,V$ satisfies $p(x|yv)=p(x|y)$ for each $x,y,v$, then they form a Markov chain, which is denoted by $X\mkv Y\mkv V$. When $Z_1$ and $Z_2$ are independent random variables we will denote it as $Z_1\perp Z_2$. Conditional variance of $Z_1$ given $Z_2$ is denoted by $\textrm{Var}[Z_1|Z_2]$. The set of strong typical sequences associated with random variable $X$ is denoted by  $\mathcal{T}^n_{[X]\epsilon}$, where $\epsilon>0$. Given $x^n$, the conditional strong typical set given $x^n$ is denoted as $\mathcal{T}^n_{[Y|X]\epsilon}(x^n)$. Typical and conditional typical sets are denoted as $\mathcal{T}_{\epsilon}^n$ when clear from the context. The cardinality of set $\mathcal{A}$ is denoted by $|\mathcal{A}|$ and with $2^{\mathcal{A}}$ we denote its power set. The complement of $\mathcal{A}$ is denoted by $\mathcal{A}^{c}$. With $\mathbb{R}_{\geq 0}$ and $\mathbb{Z}_{\geq 0}$ we denote the real and integer numbers greater than $0$, respectively. If $a$ and $b$ are real numbers, with $a\ast b$ we denote $a(1-b)+b(1-a)$. We denote $[a]^{+}=\max{\left\{a,0\right\}}$ when $a\in\mathbb{R}$. All logarithms are taken in base 2. 

We finally introduce some convenient notation that will be used through the paper. Let $V_{1,l}$ and $V_{2,l}$ be a sequence of  random variables and let:  
\begin{align*}
W_{1,l}& \triangleq \big\{V_{1,k},V_{2,k}\big\}_{k=1}^{l-1}\;\text{ for } l\in[1:K+1],\\ W_{2,l}&\triangleq \big\{W_{1,l},V_{1,l}\big\}\;\text{ for } l\in[1:K].
\end{align*}
This definition will help to simplify the expressions of the inner and outer bounds of this paper. It will be clear from the following sections that while each $V_{1,l}$, $V_{2,l}$ will be used in the generation of the descriptions in encoders 1 and 2 at time $l$, $W_{1,l}$ and $W_{2,l}$ will represent the set of descriptions generated and recovered at both encoders 1 and 2 up to time $l$.

\section{Two-way Collaborative Information Bottleneck} \label{subsec:CRL}

We begin by introducing  the so-called {\it Two-way Collaborative Information Bottleneck} (TW-CIB) problem and then state the optimal characterization of the corresponding complexity-relevance region. 

\subsection{Problem statement}
Consider $(X_1^n,X_2^n,Y_1^n,Y_2^n)$ to be sequences of $n$ i.i.d. copies of random variables $(X_{1},X_{2},Y_{1},Y_{2})$ distributed according to $p(x_{1},x_{2},y_{1},y_{2})$ taking values on $\mathcal{X}_1\times\mathcal{X}_2\times\mathcal{Y}_1\times\mathcal{Y}_2$, where $\mathcal{X}_i,\mathcal{Y}_i$ with $i\in\{1,2\}$ are finite alphabets. First, encoder $1$ generates a (representation) description, based on its observed input sequence $X_1^n=(X_{11},\dots,X_{1n})$ and transmits it to encoder 2. After correctly recovering this description, encoder $2$ generates a description  based on its observed input sequence $X_2^n$ and the recovered message from encoder 1 and transmits a description to encoder 2. This process is repeated at both encoders, where each new description is generated based on the observed source realization and the previous description  recovered up to that time. The generation of the description at encoder 1 (based on the observed source and previous history) and the recovering at encoder 2 is referred to as a \emph{half-round}. The addition of the generation of the description at encoder 2 and its recovering at encoder 1 constitutes what we shall call simply a \emph{round}. After $K$ rounds have been completed, the information exchange between both encoders concludes. It is expected that the level of relevant information that decoder 1 has gathered about the \emph{hidden representation} variable $Y_1^n$ is above a required value $\mu_1\geq 0$. Similarly, decoder 2 requires a minimum value of relevant information about sequence $Y_2^n$ of $\mu_2\geq 0$. This problem can be graphically represented as in Fig.~\ref{fig:db1}. A mathematical formulation of the described process is given below.

 \begin{definition}[$K$-step code and complexity-relevance region of the TW-CIB problem]
 A $K$-step $n$-length TW-CIB code, for the network model in Fig.~\ref{fig:db1}, is defined by a sequence of encoder mappings: 
 \begin{IEEEeqnarray*}{lcl}
 f_l\, &: &\, \mathcal{X}_1^n\times \mathcal{J}_1\times\cdots\times\mathcal{J}_{l-1} \longrightarrow \mathcal{I}^l \ , \\
 g_l \, &: &\,\mathcal{X}_2^n\times \mathcal{I}_1\times\cdots\times\mathcal{I}_{l} \longrightarrow \mathcal{J}_l  
 \end{IEEEeqnarray*}
 with $l\in[1:K]$ and message sets: $\mathcal{I}_l \triangleq \left\{1,2,\dots,|\mathcal{I}_l|\right\}$ and $\mathcal{J}_l \triangleq \left\{1,2,\dots,|\mathcal{J}_l|\right\}$. In compact form we denote a $K$-step interactive source coding by $(n,\mathcal{F})$ where $\mathcal{F}$ denote the set of encoders mappings.

An $4$-tuple $\left(R_1,R_2,\mu_1,\mu_2\right)\in\mathbb{R}^4_{\geq 0}$ is said to be $K$-achievable if $\forall\varepsilon>0$ exists $n_0(\varepsilon)$, such that $\forall\,n>n_0(\varepsilon)$ exists a $K$-step TW-CIB code $(n,\mathcal{F})$ with complexity rates satisfying: 
  \begin{equation}
  \label{eq:rate_ml_1}
 \frac{1}{n}\sum_{l=1}^K\log{|\mathcal{I}_l|}\leq R_1+\varepsilon\ ,\ \  \frac{1}{n}\sum_{l=1}^K\log{|\mathcal{J}_l|}\leq R_2+\varepsilon,
  \end{equation} 
 and normalized multi-letter relevance conditions:
 \begin{equation}
  \label{eq:relevance_ml_1}
 \mu_i-\epsilon\leq \frac{1}{n}I\left(Y_i^n;I^KJ^KX_i^n\right)\ ,\ i\in\{1,2\}.
 \end{equation} 
 The $K$-step \emph{complexity-relevance region} $ \mathcal{R}_{\mbox{\tiny TW-CIB}}(K)$ for the TW-CIB problem is defined as:
  \begin{align*}
 \mathcal{R}_{\mbox{\tiny TW-CIB}}(K)&\triangleq\big\{(R_1,R_2,\mu_1,\mu_2): (R_1,R_2,\mu_1,\mu_2)\  \mbox{is}\nonumber\\
 &\qquad K\mbox{-achievable} \big\}.
 \end{align*} 
 \end{definition}
 \begin{remark}
By the memoryless property of $Y_i^n$, the relevance condition can be equivalently written as:
\begin{equation*}
\frac{1}{n}H(Y_i^n|I^KJ^KX_i^n)\leq \mu_i^\prime +\epsilon,\ i\in\{1,2\}
\end{equation*}
 where $ \mu_i^\prime\triangleq H(Y_i)-\mu_i$. In this way, the TW-CIB problem can be recast in the conventional interactive rate-distortion  problem~\cite{kaspi_two-way_1985} using \emph{logarithmic-loss distortion}~\cite{Courtade_2014}, where  at encoder 1 we put a ``soft'' decoder whose outputs are probability distributions on $\mathcal{Y}_i^n$ (refer to~\cite[Lemmas 18, 19]{DBLP:journals/corr/PichlerPM16} for further details). The descriptions $(I^K,J^K)$ can be considered as the indices of the family of probability distributions that the decoder can output. It is also easily shown that, restricting the output probability distributions to products ones should not reduce the optimal complexity-relevance region. 
 \end{remark}
 \begin{remark}
 $ \mathcal{R}_{\mbox{\tiny TW-CIB}}(K)$ depends on the ordering in the encoding procedure. Above we have defined the encoding functions $\left\{f_l,g_l\right\}_{l=1}^K$ assuming encoder 1 acts first, followed by encoder  2, and the process beginning again at encoder 1. We could consider all possible orderings and take  $ \mathcal{R}_{\mbox{\tiny TW-CIB}}(K)$ to be the union of the achievable complexity-relevance pairs over all possible encoding orderings. For sake of clarity and simplicity, we shall not pursue this further.
 \end{remark}
\begin{remark}
 It is straightforward to check that $ \mathcal{R}_{\mbox{\tiny TW-CIB}}(K)$ is convex and closed.
\end{remark}
\begin{remark}
We could consider the case in which the number of rounds is arbitrary. In that case we can define the ultimate complexity-relevance region as:
\begin{IEEEeqnarray*}{lCl}
\mathcal{R}_{\mbox{\tiny TW-CIB}}& \triangleq  &\bigcup_{K\in\mathbb{Z}> 0}\mathcal{R}_{\mbox{\tiny TW-CIB}}(K)\\
&=&\left\{(R_1,R_2,\mu_1,\mu_2): (R_1,R_2,\mu_1,\mu_2)\  \mbox{is}\right.\nonumber\\
&&\qquad\left. K\mbox{-achievable for some}\ K\in\mathbb{Z}> 0
\right\}.\nonumber
\end{IEEEeqnarray*}
The set limiting operation in the above equation can be easily seen to be well-defined.
\end{remark}

\subsection{Characterization of the complexity-relevance region}

The next theorem provides the characterization of $\mathcal{R}_{\mbox{\tiny TW-CIB}}(K)$ in terms of single-letters expressions:

\begin{theorem}[Characterization of the complexity-relevance region for TW-CIB]\label{theo:p1}
Consider an arbitrary pmf $p(x_1,x_2,y_1,y_2)$. The corresponding region  $\mathcal{R}_{\mbox{\tiny TW-CIB}}(K)$  is the set of  tuples $\left(R_1,R_2,\mu_1,\mu_2\right)\in\mathbb{R}^4_{\geq 0}$ such that there exists auxiliary random variables $\left\{V_{1,l},V_{2,l}\right\}_{l=1}^K$  satisfying:
\begin{IEEEeqnarray*}{rCl}
R_1&\geq& I(X_1;W_{1,K+1}|X_2),\\
R_2&\geq& I(X_2;W_{1,K+1}|X_1),\\
\mu_1&\leq& I(Y_1;W_{1,K+1}X_1),\\
\mu_2&\leq& I(Y_2;W_{1,K+1}X_2),
\end{IEEEeqnarray*}
taking values in finite discrete alphabets $\mathcal{V}_{1,l}$ and $\mathcal{V}_{2,l}$ and satisfying  Markov chains:
\begin{align}
V_{1,l}&\mkv (X_1,W_{1,l})\mkv (X_2,Y_1,Y_2),\label{eq:markv1-p1}\\
V_{2,l}&\mkv (X_2,W_{2,l})\mkv (X_1,Y_1,Y_2)\label{eq:markv2-p1}
\end{align}
for $l\in[1:K]$. The auxiliary random variables can be restricted to take values in finite alphabets with cardinalities bounds given by: 
\begin{align*}
|\mathcal{V}_{1,l}|&\leq|\mathcal{X}_1||\mathcal{W}_{1,l}|+3\ ,\textrm{ for  $l=[1:K]$}\\
|\mathcal{V}_{2,l}|&\leq|\mathcal{X}_2||\mathcal{W}_{2,l}|+3\ ,\textrm{ for  $l=[1:K-1]$}\\
|\mathcal{V}_{2,K}|&\leq|\mathcal{X}_2||\mathcal{W}_{2,K}|+1.
\end{align*}
where $|\mathcal{W}_{1,l}|=\prod_{i=1}^{l-1}|\mathcal{V}_{1,i}||\mathcal{V}_{2,i}|$ and $|\mathcal{W}_{2,l}|=|\mathcal{W}_{1,l}||\mathcal{V}_{2,l}|$ for  $l=[1:K]$.
\end{theorem}
\begin{IEEEproof}
The proof of the achievability is given in Appendix \ref{sec:achiev} while the converse part is relegated to the next section.
\end{IEEEproof}

\begin{remark}
It is inmediate to see that the point-to-point classical information bottleneck problemm, where observing $X_1$ we are interested in extracting information about $Y_2$ can be seen as an special case of Theorem \ref{theo:p1} when $X_2,Y_1=\varnothing$.
\end{remark}

\section{Collaborative Distributed Information Bottleneck}\label{subsec:CDIB}

We begin by introducing the so-called {\it Collaborative Distributed Information Bottleneck} (CDIB) problem and then provide bounds to the optimal complexity-relevance region. Special cases for which these bounds are tight are also discussed.

\subsection{Problem statement}

Consider $(X_1^n,X_2^n,Y^n)$ be sequences of $n$ i.i.d. copies of random variables $(X_{1},X_{2},Y)$ distributed according to $p(x_{1},x_{2},y)$ taking values on $\mathcal{X}_1\times\mathcal{X}_2\times\mathcal{Y}$. We will consider a cooperative setup in which $X_1^n$ and $X_2^n$ are observed at encoders $1$ and $2$, respectively, and a third party referred as {\it the decoder} wishes to ``learn'' the \emph{hidden representation} variable $Y^n$. Encoders $1$ and $2$ cooperatively and interactively generate representations that are perfectly heard by the decoder, through a noiseless but rate-limited broadcast link, as shown in Fig.~\ref{fig:db2}. The cooperation between encoders $1$ and $2$ permits to save rate during the exchanges and at the same time maintaining an appropriate level of relevance between the generated descriptions and the hidden variable $Y^n$. Encoders 1 and 2 interact as in the TW-CIB problem. After they ceased to exchange their descriptions, the decoder attempts to recover the descriptions generated at encoders 1 and 2, which should have some predefined level of information with respect to $Y^n$.

\begin{definition}[$K$-step code  and complexity-relevance region of the CDIB problem]
A $K$-step $n$-length CDIB code, for the network model in Fig.~\ref{fig:db2}, is defined by a sequence of encoder mappings: 
\begin{IEEEeqnarray*}{lcl}
f_l \,&: &\, \mathcal{X}_1^n\times \mathcal{J}_1\times\cdots\times\mathcal{J}_{l-1}  \longrightarrow \mathcal{I}_l \ , \\
g_l \, &: &\,\mathcal{X}_2^n\times \mathcal{I}_1\times\cdots\times\mathcal{I}_{l}  \longrightarrow \mathcal{J}_l  \ , 
\end{IEEEeqnarray*}
with $l\in[1:K]$ and message sets: $\mathcal{I}_l \triangleq  \left\{1,2,\dots,|\mathcal{I}_l|\right\}$ and $\mathcal{J}_l \triangleq  \left\{1,2,\dots,|\mathcal{J}_l|\right\}$. In compact form we denote a $K$-step CDIB code by $(n,\mathcal{F})$ where $\mathcal{F}$ denote the set of encoders mappings.

A $3$-tuple $\left(R_1,R_2,\mu\right)\in\mathbb{R}^3_{\geq 0}$ is said to be $K$-achievable if $\forall\varepsilon>0$ exists $n_0(\varepsilon)$, such that $\forall\,n>n_0(\varepsilon)$ exists a $K$-step source code $(n,\mathcal{F})$ with rates satisfying: 
 \begin{equation}
 \label{eq:rate_ml-p2}
\frac{1}{n}\sum_{l=1}^K\log{|\mathcal{I}_l|}\leq R_1+\varepsilon\ , \ \ \ \frac{1}{n}\sum_{l=1}^K\log{|\mathcal{J}_l|}\leq R_2+\varepsilon
 \end{equation} 
and normalized multi-letter relevance at the decoder:
\begin{equation} \label{eq:relevance_ml-p2}
\mu-\epsilon\leq \frac{1}{n}I\left(Y^n;I^KJ^K\right).
\end{equation} 
The $K$-step \emph{complexity-relevance region} $\mathcal{R}_{\mbox{\tiny CDIB}}(K)$ is defined as:
 \begin{equation*}
\mathcal{R}_{\mbox{\tiny CDIB}}(K) \triangleq  \big\{(R_1,R_2,\mu)\  \mbox{is}\ K\mbox{-achievable} \big\}.
\end{equation*} 
\end{definition}
It is clearly seen that Remarks 2, 3 and 4 also apply to this problem. In fact, when $X_1\mkv Y\mkv X_2$, this problem can be seen as a cooperative and interactive CEO problem~\cite{berger_ceo_1996} with logarithmic loss~\cite{Courtade_2014}. The main difference with respect to these previous works is that the present setting allows cooperation between encoders 1 and 2. This could lead to savings in rate and/or gains in the achievable relevance levels through an adequate use of the structure of the statistical dependence between sources $X_1,X_2$ and $Y$.

\subsection{Bounds to the complexity-relevance region}

We now state the following inner bound to the complexity-relevance region $\mathcal{R}_{\mbox{\tiny CDIB}}(K)$.

\begin{theorem}[Inner bound to $\mathcal{R}_{\mbox{\tiny CDIB}}(K)$]
\label{theo:inner_bound}
Consider $\mathcal{R}_{\mbox{\tiny CDIB}}^{\mbox{\tiny inner}}(K)$ to be the region of tuples $(R_1,R_2,\mu)\in\mathbb{R}^3_{\geq 0}$ such that there exists auxiliary random variables $\left\{V_{1,l},V_{2,l}\right\}_{l=1}^K$ satisfying:
\begin{IEEEeqnarray*}{rCl}
R_1&\geq& I(X_1;W_{1,K+1}|X_2),\\
R_2&\geq& I(X_2;V_{2,K}|W_{2,K})+I(X_2;W_{2,K}|X_1),\\
R_1+R_2&\geq& I(X_1X_2;W_{1,K+1}),\\
\mu&\leq& I(Y;W_{1,K+1}),\label{eq-p1-sl-rel-in}
\end{IEEEeqnarray*} 
taking values in finite discrete alphabets $\mathcal{V}_{1,l}$ and $\mathcal{V}_{2,l}$ and satisfying  Markov chains:
\begin{align}
V_{1,l}&\mkv (X_1,W_{1,l})\mkv (X_2,Y),\label{eq:markv1-p2}\\
V_{2,l}&\mkv (X_2,W_{2,l})\mkv (X_1,Y)\label{eq:markv2-p2}
\end{align}
for $l\in[1:K]$. The auxiliary random variables can be restricted to take values in alphabets verifying: 
\begin{align*}
|\mathcal{V}_{1,l}|&\leq|\mathcal{X}_1||\mathcal{W}_{1,l}|+4\ ,\textrm{ for  $l=[1:K-1]$}\\
|\mathcal{V}_{2,l}|&\leq|\mathcal{X}_2||\mathcal{W}_{2,l}|+4\ ,\textrm{ for  $l=[1:K-1]$}\\
|\mathcal{V}_{1,K}|&\leq|\mathcal{X}_1||\mathcal{W}_{1,K}|+3,\\
|\mathcal{V}_{2,K}|&\leq|\mathcal{X}_2||\mathcal{W}_{2,K}|+1.
\end{align*}
where $|\mathcal{W}_{1,l}|=\prod_{i=1}^{l-1}|\mathcal{V}_{1,i}||\mathcal{V}_{2,i}|$ and $|\mathcal{W}_{2,l}|=|\mathcal{W}_{1,l}||\mathcal{V}_{2,l}|$ for  $l=[1:K]$. Then, $\mathcal{R}_{\mbox{\tiny CDIB}}^{\mbox{\tiny inner}}(K)\subseteq\mathcal{R}_{\mbox{\tiny CDIB}}(K)$. 
\end{theorem}
\begin{IEEEproof}
See Appendix \ref{sec:achiev}.
\end{IEEEproof}
\begin{remark}
As shown in the Appendix this region is achievable using a special cooperative binning between encoders 1 and 2 which was inspired by previous work in~\cite{ours_2015}. After the information exchange is accomplished, the decoder needs to recover the descriptions generated at encoders 1 and 2. At each round, for example encoder 2, generates its own description after having recovered the ones generated at encoder 1 at the present and previous rounds. So, instead of binning only its last generated description, it can also consider in its binning what he already knows from its past descriptions and the ones from encoder 1 (see Appendix \ref{sec:achiev}). This allows for an explicit cooperation between encoders 1 and 2 in order to help the decoder to recover both descriptions despite of the fact that it does not have side information and without penalizing the rate $R_1$ (e.g. observe the rate constraint on $R_1$ is conditioned on $X_2$, which corresponds to the minimum rate needed by encoder 2 to recover the descriptions generated at encoder 1). Note also that the rate expressions corresponding  to $R_1$ and $R_2$ can be written as:
\begin{align*}
R_1&\geq\sum_{l=1}^K I\left(X_1;V_{1,l}|W_{1,l}X_2\right),\\
R_2&\geq I\left(X_2;V_{2,K}|W_{2,K}\right)-I\left(X_2;V_{2,K}|W_{2,K}X_1\right)\nonumber\\
&\qquad+\sum_{l=1}^{K} I\left(X_2;V_{2,l}|W_{2,l}X_1\right),
\end{align*}
where the sequential nature of the coding is revealed. We see that for every round $l$ both rates equations present terms $ I\left(X_1;V_{1,l}|W_{1,l}X_2\right)$ and $I\left(X_2;V_{2,l}|W_{2,l}X_1\right)$ which correspond to the minimum rates that encoder 2 (encoder 1) needs in order to recover the last description generated by encoder 1 (encoder 2). However, the rate equation $R_2$ presents a penalizing term that involves the description generated at encoder 2 in round $K$. This term appears because the last description generated at encoder 2 will not get benefit from further cooperative binning given that there are not any more rounds. As the decoder has not side information, the encoder 2 has to send an excess rate to compensate for that and help him to recover all generated descriptions. It is clear that for the TW-CIB problem, this cooperative binning is not needed because an external decoder (i.e. different from the encoders) is not present and both encoders --before generating a new description-- know (with probability close to 1) the descriptions generated in previous rounds.
\end{remark}

The following result gives us an outer bound to  the complexity-relevance region  $\mathcal{R}_{\mbox{\tiny CDIB}}(K)$ in the special case that $X_1\mkv Y\mkv X_2$.

\begin{theorem}[Outer bound to $\mathcal{R}_{\mbox{\tiny CDIB}}(K)$]
\label{theo:outer_bound}
Assume that we have the Markov chain $X_1\mkv Y\mkv X_2$. Let  $\mathcal{R}_{\mbox{\tiny CDIB}}^{\mbox{\tiny outer}}(K)$ to be the region of tuples $(R_1,R_2,\mu)\in\mathbb{R}^3_{\geq 0}$ such that there exists auxiliary random variables $\left\{V_{1,l},V_{2,l}\right\}_{l=1}^K$ simultaneously satisfying:
\begin{IEEEeqnarray}{rCl}
R_1&\geq& I(X_1;W_{1,K+1}|X_2),\label{eq:convR1}\\
R_2&\geq& \left[I(X_2;W_{1,K+1}|Y)-I(Y;W_{2,K})+\mu\right]^{+},\label{eq:convR2}\,\,\,\,\,\,\,\,\\
R_1+R_2&\geq& I(X_1X_2;W_{1,K+1}|Y)+\mu,\nonumber\\
\mu&\leq& I(Y;W_{1,K+1}),\nonumber
\end{IEEEeqnarray}
satisfying the Markov chains \eqref{eq:markv1-p2} and \eqref{eq:markv2-p2} for $l\in[1:K]$ and taking values in finite discrete alphabets $\mathcal{V}_{1,l}$ and $\mathcal{V}_{2,l}$ with cardinalities bounded by: 
\begin{align*}
|\mathcal{V}_{1,l}|&\leq|\mathcal{X}_1||\mathcal{W}_{1,l}|+4\ ,\textrm{ for  $l=[1:K]$}\\
|\mathcal{V}_{2,l}|&\leq|\mathcal{X}_2||\mathcal{W}_{2,l}|+4\ ,\textrm{ for  $l=[1:K-1]$}\\
|\mathcal{V}_{2,K}|&\leq|\mathcal{X}_2||\mathcal{W}_{2,K}|+1.
\end{align*}
where $|\mathcal{W}_{1,l}|=\prod_{i=1}^{l-1}|\mathcal{V}_{1,i}||\mathcal{V}_{2,i}|$ and $|\mathcal{W}_{2,l}|=|\mathcal{W}_{1,l}||\mathcal{V}_{2,l}|$ for  $l=[1:K]$. Then, $\mathcal{R}_{\mbox{\tiny CDIB}}^{\mbox{\tiny outer}}(K)\supseteq\mathcal{R}_{\mbox{\tiny CDIB}}(K)$. 
\end{theorem}
\begin{IEEEproof}
The proof is relegated to Section~\ref{sec:conv}.
\end{IEEEproof}

In general, it appears not possible to show that $\mathcal{R}_{\mbox{\tiny CDIB}}^{\mbox{\tiny outer}}(K)=\mathcal{R}_{\mbox{\tiny CDIB}}^{\mbox{\tiny inner}}(K)$ for every $K\in\mathbb{Z}_{\geq 0}$ when $X_1\mkv Y\mkv X_2$. However, this is indeed the case when $K=1$ that is, the interaction between encoders 1 and 2 is restricted to only one round.

\begin{remark}
The Markov chain $X_1\mkv Y\mkv X_2$ turns our problem into the interactive-cooperative CEO problem. This approach has a well-known converse for the sum-rate~\cite[Theorem 3.1]{prabhakaran04} which has been proved for an additive distortion but can be easily re-adapted. However, the sum-rate constraint provided in this paper is tighter.  To check this,  we can ignore conditions \eqref{eq:convR1} and \eqref{eq:convR2}. Then the corner points of $\mathcal{R}_{\mbox{\tiny CDIB}}^{\mbox{\tiny outer}}(K)$ are: 
\begin{align*}
Q_A&=[I(X_1X_2;W_{1,K+1}),I(Y;W_{1,K+1})],\\
Q_B&=[I(X_1X_2;W_{1,K+1}|Y),0],
\end{align*}
where these components correspond to the sum-rate and relevance, respectively.
The resulting corner points meet simultaneously: $\mu\leq I(Y;U)$ and 
\begin{equation*}
R_1+R_2\geq I(Y;U)+I(X_1;U|YZ)+I(X_2;U|YZ),
\end{equation*}
where $Z$ is a random variable independent of $(X_1,X_2,Y)$, and $U$ satisfying $Y\mkv(X_1,X_2,Z)\mkv U$ and $X_1\mkv(Y,U,Z)\mkv X_2$. To show this, let us assume that $Z\triangleq z$ almost surely, i.e., $Z$ is a degenerated random variable, and set $U\triangleq W_{1,K+1}$ and $Z\triangleq z$ or $U\triangleq u$ for the corner points $Q_A$ and $Q_B$, respectively.
\end{remark}

\subsection{Characterization of the complexity-relevance region when $X_1\mkv Y\mkv X_2$ with $K=1$}

\begin{theorem}[Complexity-relevance region when $X_1\mkv Y\mkv X_2$ with $K=1$]
\label{thm:k=1}
Assume $K=1$ and $X_1\mkv Y\mkv X_2$, then $\mathcal{R}_{\mbox{\tiny CDIB}}^{\mbox{\tiny outer}}(1)=\mathcal{R}_{\mbox{\tiny CDIB}}^{\mbox{\tiny inner}}(1)=\mathcal{R}_{\mbox{\tiny CDIB}}(1)$.
\end{theorem}

\begin{IEEEproof}
The proof of the equality between the regions provided in Theorems~\ref{theo:inner_bound} and~\ref{theo:outer_bound} is postponed to the next section.
\end{IEEEproof}
\begin{remark}[The role of cooperation]
\label{rem:cooperacion}
The region $\mathcal{R}_{\mbox{\tiny CDIB}}(1)$ can be written as:
\begin{align*}
R_1\;&\geq\;I(X_1;V_1|X_2),\\
R_2\;&\geq\;I(X_2;V_2|V_1),\\ 
R_1+R_2\;&\geq\;I(X_1X_2;V_1V_2),\\
\mu\;&\leq\;I(Y;V_1V_2),
\end{align*}
with $V_1$ and $V_2$ taking values in finite alphabets $\mathcal{V}_1$ and $\mathcal{V}_2$ and satisfying $V_1\mkv X_1\mkv (X_2,Y)$, $V_2\mkv (V_1,X_2) \mkv (X_1,Y)$. It is worth to compare this with the non-cooperative CEO rate-distortion region under  logarithmic loss~\cite[Theorem 3]{Courtade_2014}. As it is well known, that region can be expressed in terms of rates $R_1$, $R_2$ and relevance $\mu$, instead of logarithmic loss distortion level $\mu^\prime$. In this manner, we can write the following (non-cooperative) complexity-relevance region $\mathcal{R}_{\mbox{\tiny DNCRL}}$ as:
\begin{align*}
R_1\;&\geq\;I(X_1;V_1|V_2),\\
R_2\;&\geq\;I(X_2;V_2|V_1),\\ 
R_1+R_2\;&\geq\;I(X_1X_2;V_1V_2),\\
\mu\;&\leq\;I(Y;V_1V_2),
\end{align*}
where  $V_1$ and $V_2$ take values in finite alphabets $\mathcal{V}_1$ and $\mathcal{V}_2$ satisfying: $V_1\mkv X_1\mkv (X_2,Y)$ and $V_2\mkv X_2\mkv (X_1,Y)$ form Markov chains. It is clearly seen that $\mathcal{R}_{\mbox{\tiny CDIB}}(1)\supseteq\mathcal{R}_{\mbox{\tiny DNCRL}}$. First, note that:
\begin{align*}
I(X_1;V_1|V_2)&=I(V_1;X_2|V_2)+I(X_1;V_1|X_2)\\
&\geq I(X_1;V_1|X_2).
\end{align*}
Secondly, the set of probability distributions over which $\mathcal{R}_{\mbox{\tiny CDIB}}(1)$ is constructed is greater than the one corresponding to $\mathcal{R}_{\mbox{\tiny DNCRL}}$. This is seen in the requirement of the auxiliary random variable $V_2$, which in the cooperative case can depend on $V_1$, reflecting the possibility of cooperation between the encoders. 
\end{remark}

\subsection{Characterization of the complexity-relevance region when $X_1\mkv X_2\mkv Y$ with  $K=1$} \label{sec:alternate}

\begin{definition} Let $\hat{\mathcal{R}}_{\mbox{\tiny CDIB}}(1)$ be the set of tuples $(R_1,R_2,\mu)\in\mathbb{R}^3_{\geq 0}$ such that there exists a joint pmf $p(x_1,x_2,y,v_1,v_2)$ that preserves the joint distribution of the sources $(X_1,X_2,Y)$ and
\begin{IEEEeqnarray}{rCl}
R_1&\geq& I(X_1;V_1),\label{eq-missing-1}\\
R_2&\geq& I(X_2;V_2|V_1),\\
\mu&\leq& I(Y;V_1V_2),\label{eq-missing-3}
\end{IEEEeqnarray}
with auxiliary random variables $V_1,V_2$ satisfying:
\begin{equation}
V_1\mkv X_1\mkv (X_2,Y)\ ,\ \ V_2\mkv (V_1, X_1, X_2) \mkv Y.
\label{eq:second_mkv_cond}
\end{equation}
Similarly, let $\tilde{\mathcal{R}}_{\mbox{\tiny CDIB}}(1)$ be the set of tuples $(R_1,R_2,\mu)\in\mathbb{R}^3_{\geq 0}$ verifying~\eqref{eq-missing-1}-\eqref{eq-missing-3} such that there exists a joint pmf $p(x_1,x_2,y,v_1,v_2)$ that preserves the joint pmf of the sources $(X_1,X_2,Y)$ while satisfying:
\begin{equation}
V_1\mkv X_1\mkv (X_2,Y)\ ,\ \ V_2\mkv (V_1,X_2)\mkv (X_1,Y).
\label{eq:third_mkv_cond}
\end{equation}
\end{definition}

Theorems \ref{theo:alternative_inner} and \ref{theo:optimal_X1X2Y} will imply the characterization of the corresponding complexity-relevance region. We present first Theorem \ref{theo:alternative_inner} which gives us inner and outer bounds for $\mathcal{R}_{\mbox{\tiny CDIB}}(1)$ for arbitrary random sources $X_1,X_2,Y$.
\begin{theorem}
\label{theo:alternative_inner}
Assume $K=1$ and arbitrary random variables $X_1,X_2,Y$. Then, we have 
\begin{equation*}
\tilde{\mathcal{R}}_{\mbox{\tiny CDIB}}(1)\subseteq\mathcal{R}_{\mbox{\tiny CDIB}}(1)\subseteq\hat{\mathcal{R}}_{\mbox{\tiny CDIB}}(1).
\end{equation*}
\end{theorem}
\begin{IEEEproof}
The proof is relegated to the next section.
\end{IEEEproof}
The following result implies  that $\tilde{\mathcal{R}}_{\mbox{\tiny CDIB}}(1)=\hat{\mathcal{R}}_{\mbox{\tiny CDIB}}(1)$ when $X_1\mkv X_2\mkv Y$.

\begin{theorem}
\label{theo:optimal_X1X2Y}
Assume $K=1$ and $X_1\mkv X_2\mkv Y$. Then $\hat{\mathcal{R}}_{\mbox{\tiny CDIB}}(1)\subseteq\tilde{\mathcal{R}}_{\mbox{\tiny CDIB}}(1)$.
\end{theorem}

\begin{IEEEproof}
Assume that $(R_1,R_2,\mu)\in\hat{\mathcal{R}}_{\mbox{\tiny CDIB}}(1)$. Then, there exists a pmf 
\begin{equation*}
p(x_1,x_2,y ,v_1,v_2 )=p(x_1,x_2,y)p(v_1|x_1)p(v_2|x_1,x_2,v_1)
\end{equation*}
such that: $R_1\geq I(X_1;V_1)$ $R_2\geq I(X_2;V_2|V_1)$ and $\mu\leq I(Y;V_1V_2)$. Consider the pmf
\begin{equation*}
\tilde{p}(x_1,x_2,y,v_1,v_2)=p(x_1,x_2,x_3)p(v_1|x_1)\tilde{p}(v_2|x_2,v_1),
\end{equation*} 
where
\begin{align*}
\tilde{p}(v_2|x_2,v_1)&\triangleq  \frac{p(x_2,v_1,v_2)}{p(x_2,v_1)}\\
&=\frac{\sum_{x_1'}p(x_1',x_2)p(v_1|x_1')p(v_2|x_1'x_2v_1)}{\sum_{x_1'}p(x_1',x_2)p(v_1|x_1')}.
\end{align*}
By assumption this pmf preserves the sources $(X_1,X_2,Y)$ while satisfying  (\ref{eq:third_mkv_cond}). Moreover, it can be shown without difficulty that $\tilde{p}(x_1,v_1)=p(x_1,v_1)$ and $\tilde{p}(x_2,v_1,v_2)=p(x_2,v_1,v_2)$. This implies that $I(X_1;V_1)$ and $I(X_2;V_2|V_1)$ are preserved. If we further assume that $X_1\mkv X_2\mkv Y$, we can write:
\begin{align*}
\tilde{p}(y,v_1,v_2)&= \sum_{x_1,x_2}p(x_1,x_2,y)p(v_1|x_1)\times\nonumber\\
&\quad\quad\frac{\sum_{x_1'}p(x_1',x_2)p(v_1|x_1')p(v_2|x_1'x_2v_1)}{\sum_{x_1'}p(x_1',x_2)p(v_1|x_1')},\\
&=\sum_{x_2}p(y|x_2)\sum_{x_1}p(x_1,x_2)p(v_1|x_1)\times\nonumber\\
&\quad\quad\frac{\sum_{x_1'}p(x_1',x_2)p(v_1|x_1')p(v_2|x_1'x_2v_1)}{\sum_{x_1'}p(x_1',x_2)p(v_1|x_1')}\\
&=\sum_{x_1',x_2}p(x_1',x_2,y)p(v_1|x_1')p(v_2|x_1'x_2v_1)\\ &=p(y,v_1,v_2).
\end{align*}
As a consequence,  the term $I(Y;V_1V_2)$ is also preserved and thus $(R_1,R_2,\mu)\in\tilde{\mathcal{R}}_{\mbox{\tiny CDIB}}(1)$.
\end{IEEEproof}

The next corollary immediately follows.
\begin{corollary}
\label{coro:1}
Provided that $X_1\mkv X_2\mkv Y$, we have $\tilde{\mathcal{R}}_{\mbox{\tiny CDIB}}(1)=\hat{\mathcal{R}}_{\mbox{\tiny CDIB}}(1)=\mathcal{R}_{\mbox{\tiny CDIB}}(1)$.
\end{corollary}
It is easily seen that for achieving any $(R_1,R_2,\mu)\in\tilde{\mathcal{R}}_{\mbox{\tiny CDIB}}(1)$ it is not necessary to use binning. First encoder 1 sends its description which can recovered at encoder 2 and the decoder. Then, encoder 2 uses this description --as a coded side information which is also available at the decoder-- to generate and sends its own one to the decoder. The previous claim shows this coding scheme is optimal when $X_1\mkv X_2\mkv Y$. As $\mathcal{R}_{\mbox{\tiny CDIB}}^{\mbox{\tiny inner}}(1)$ is also achievable and $\tilde{\mathcal{R}}_{\mbox{\tiny CDIB}}(1)\subseteq\mathcal{R}_{\mbox{\tiny CDIB}}^{\mbox{\tiny inner}}(1)$ (which is trivial to show), we can state an alternative characterization of the complexity-relevance region. 

\begin{corollary}[Alternative characterization of  $\mathcal{R}_{\mbox{\tiny CDIB}}(1)$ when $X_1\mkv X_2\mkv Y$]
\label{coro:2}
Assume $K = 1$ and that $X_1\mkv X_2\mkv Y$ form a Markov chain, then $\tilde{\mathcal{R}}_{\mbox{\tiny CDIB}}(1)=\mathcal{R}_{\mbox{\tiny CDIB}}^{\mbox{\tiny inner}}(1)=\mathcal{R}_{\mbox{\tiny CDIB}}(1)=\hat{\mathcal{R}}_{\mbox{\tiny CDIB}}(1)$.
\end{corollary}
\begin{IEEEproof}
Follows easily from the above discussion. An alternative proof of this Corollary is presented in Appendix~\ref{sec:equivalence}.
\end{IEEEproof}
\begin{remark}
From the previous results it should be clear that the coding procedure presented in Theorem  \ref{theo:inner_bound} is clearly optimal for both cases $X_1\mkv Y\mkv X_2$ and $X_1\mkv X_2\mkv Y$. The first Markov chain corresponds to the typical one considered in the CEO problem~\cite{ELGamal-Kim-book}. This would be the case where, for example, the hidden variable $Y$ is related with the observed variables $X_1$ and $X_2$ through and additive model: $X_1=Y+Z_1$, $X_2=Y+Z_2$ where $Z_1$ and $Z_2$ are independent random variables. For example this situation could appear in a sensor network setting where $X_1$ and $X_2$ are observed in two geographically separated nodes and in which the fusion center (node 3) desires to obtain a good representation of the hidden variable $Y$. The case in which  $X_1\mkv X_2\mkv Y$ can represent also the case of the distributed sensor network setting, in which the measurements in one of sensors ($X_2$) is most informative with respect to the hidden variable $Y$ that the ones in the other ($X_1$). This could represent a situation in which the hidden variable $Y$ models a physical phenomenon which originates in given point of space and in which the statistical dependence with variables $X_1$ and $X_2$ at the points of measurements (the sites where nodes 1 and 2 are positioned) depends strongly of their distance to the point of origin. If node 2 is closer than node 1 to the point of origin of $Y$, $X_2$ would have a stronger statistical dependence with $Y$ and the given Markov chain can be a useful approximate model of this situation. 
\end{remark}

\begin{remark}
It is worth to mention that the cardinality of the auxiliary variables in this case can be bounded in two different ways. The auxiliary random variables involved in the representation of $\hat{\mathcal{R}}_{\mbox{\tiny CDIB}}(1)$ can be restricted to take values in alphabets satisfying: 
\begin{equation*}
|\mathcal{V}_1|\leq|\mathcal{X}_1|+3\ ,\qquad |\mathcal{V}_2|\leq|\mathcal{X}_1||\mathcal{X}_2||\mathcal{V}_1|+1\ .
\end{equation*}
While the auxiliary random variables involved in the representation of  $\tilde{\mathcal{R}}_{\mbox{\tiny CDIB}}(1)$ can be restricted to take values in alphabets verifying: 
\begin{equation*}
|\mathcal{V}_1|\leq|\mathcal{X}_1|+3\ ,\qquad |\mathcal{V}_2|\leq|\mathcal{X}_2||\mathcal{V}_1|+1\ .
\end{equation*}
\end{remark}

\section{Converses in Theorems~\ref{theo:p1}, \ref{theo:outer_bound}, ~\ref{thm:k=1} and~\ref{theo:alternative_inner}}
\label{sec:conv}
In this section, we provide the proofs to the converses of Theorems~\ref{theo:p1}, \ref{theo:outer_bound} and~\ref{theo:alternative_inner}. Together with the inner bounds obtained in Appendix~\ref{sec:achiev} these results imply the characterization of the corresponding complexity-relevance regions in Theorems~\ref{theo:p1},~\ref{thm:k=1} and Corollary~\ref{coro:2}. 

\subsection{Converse result for Theorem~\ref{theo:p1}}
If a tuple $(R_1,R_2,\mu_1,\mu_2)$ is achievable, then  for all $\varepsilon >0$ there exists $n_0(\varepsilon)$, such that $\forall\,n>n_0(\varepsilon)$  there exists a code $(n,\mathcal{F})$ with rates and relevance satisfying \eqref{eq:rate_ml_1} and \eqref{eq:relevance_ml_1}. For $t=[1:n]$, define variables:
\begin{align*}
V_{1,1,t}&\triangleq  \left(I_{1},X_{1[1:t-1]},X_{2[t+1:n]}\right)\\
V_{1,l,t}&\triangleq  I_{l},\quad \forall\; l\in[2:K]\\
V_{2,l,t}&\triangleq  J_{l},\quad \forall\; l\in[1:K].
\end{align*}
These auxiliary random variables satisfy, for $t=[1:n]$ the Markov conditions (\ref{eq:markv1-p1}) and (\ref{eq:markv2-p1}) and are similar to the choices made in \cite{kaspi_two-way_1985}. In that sense, the converse proof follows along similar lines as in~\cite{kaspi_two-way_1985}. However, for sake of completeness we provide the proof. 
 
\subsubsection{Constraint on rate $R_1$}
For the first rate, we have
\begin{align*}
n(R_1+&\varepsilon)
	\geq H\left(I^K\right)\\ 
	&\stackrel{(a)}{\geq}	 I\left(I^KJ^K;X_1^n|X_2^n\right)\\
		&=\sum_{t=1}^n I\left(I^KJ^K;X_{1t}|X_2^n X_{1[1:t-1]}\right)\\
			&=\sum_{t=1}^n I\left(I^KJ^KX_{1[1:t-1]}X_{2[1:t-1]}X_{2[t+1:n]};X_{1t}|X_{2t}\right)\\
			&=\sum_{t=1}^n I\left(W_{1,K+1[t]}X_{2[1:t-1]};X_{1t}|X_{2t}\right)\\
	&\stackrel{(b)}{\geq}\sum_{t=1}^n I\left(W_{1,K+1[Q]};X_{1[Q]}|X_{2[Q]},Q=t\right)\\
	&\stackrel{(c)}{=} nI\left(\widetilde{W}_{1,K+1};X_{1}|X_{2}\right),
\end{align*}
where
\begin{itemize}
\item step~$(a)$ follows from the fact that $J^K=(J_1,\dots,J_K)$ is function of $I^K=(I_1,\dots,I_K)$ and $X_2^n$;
\item step~$(b)$ follows from the use of a time sharing random variable $Q$ uniformly distributed over the set $[1:n]$ and independent of the other variables and from the non-negativity of mutual information;
\item step~$(c)$ follows by defining a new random variable $\widetilde{W}_{1,K+1}\triangleq  (W_{1,K+1[Q]} ,Q)$.
\end{itemize}

\subsubsection{Constraint on rate $R_2$}
The analysis is similar to the case for $R_1$ and for that reason is omitted. The final result is:
\begin{equation*}
n(R_2+\varepsilon)\geq nI\left(\widetilde{W}_{1,K+1};X_{2}|X_{1}\right).
\end{equation*}

\subsubsection{Constraint on relevance $\mu_1$}
For the first relevance, we have
\begin{align*}
n(\mu_1-\varepsilon)&	\leq  
 \sum_{t=1}^nI\left(Y_{1t};I^KJ^KX_1^n|Y_{1[t+1:n]}\right)\\
	&= \sum_{t=1}^nI\left(Y_{1t};I^KJ^KX_{1[1:t-1]}X_{1t}X_{1[t+1:n]}Y_{1[t+1:n]}\right)\\
	&\stackrel{(a)}{\leq}\sum_{t=1}^nI\left(Y_{1t};W_{1,K+1[t]}X_{1[t+1:n]}Y_{1[t+1:n]}X_{1t}\right)\\
	&\stackrel{(b)}{=}	\sum_{t=1}^nI\left(Y_{1[Q]};W_{1,K+1[Q]}X_{1[Q]}|Q=t\right)\\
	&\stackrel{(c)}{=}	nI\left(Y_1;\widetilde{W}_{1,K+1}X_1\right),
\end{align*}
where
\begin{itemize}
\item step $(a)$ follows from the definition of $W_{1,K+1[t]}$ and non-negativity of mutual information;
\item step~$(b)$ follows from the Markov chain $Y_{1t}\mkv\left(W_{1,K+1[t]},X_{1t}\right)\mkv\left(X_{1[t+1:n]},Y_{1[t+1:n]}\right)$ and the use of a time sharing random variable $Q$ uniformly distributed over the set $[1:n]$ and independent of the other variables;
\item step~$(c)$ follows by letting a new random variables $\widetilde{W}_{1,K+1}\triangleq  (W_{1,K+1[Q]} ,Q)$.
\end{itemize}

\subsubsection{Relevance $\mu_2$}
Again, the analysis is similar to the one for $\mu_1$. Following similar steps, we obtain:
\begin{equation*}
n(\mu_2-\varepsilon)\leq nI\left(Y_2;\widetilde{W}_{1,K+1}X_2\right).
\end{equation*}

\subsection{Converse result for Theorem~\ref{theo:outer_bound}}

If a tuple $(R_1,R_2,\mu)$ is achievable, then  for all $\varepsilon >0$ there exists $n_0(\varepsilon)$, such that $\forall\,n>n_0(\varepsilon)$  there exists a code $(n,\mathcal{F})$ with rates and relevance satisfying \eqref{eq:rate_ml-p2} and \eqref{eq:relevance_ml-p2}. For $t=[1:n]$, define variables:
\begin{align*}
V_{1,1,t}&\triangleq  \left(I_{1},Y_{[1:t-1]},X_{2[t+1:n]}\right)\\
V_{1,l,t}&\triangleq  I_{l},\quad \forall\; l\in[2:K]\\
V_{2,l,t}&\triangleq  J_{l},\quad \forall\; l\in[1:K].
\end{align*}
These auxiliary random variables satisfy, for $t=[1:n]$, the Markov conditions (\ref{eq:markv1-p2}) and (\ref{eq:markv2-p2}).
\subsubsection{Constraint on rate $R_1$}
For the first rate, we have
\begin{align*}
n&(R_1+\varepsilon)
	\geq H\left(I^K\right)\\
&\stackrel{(a)}{\geq}	 I\left(I^KJ^K;X_1^n|X_2^n\right)\\
	&=\sum_{t=1}^n I\left(I^KJ^K;X_{1t}|X_2^n X_{1[1:t-1]}\right)\\
	&=\sum_{t=1}^n I\left(I^KJ^KX_{1[1:t-1]}X_{2[1:t-1]}X_{2[t+1:n]};X_{1t}|X_{2t}\right)\\
	&\stackrel{(b)}{=}\sum_{t=1}^n I\left(W_{1,K+1[t]}X_{1[1:t-1]}X_{2[t+1:n]};X_{1t}|X_{2t}\right)\\
	&\stackrel{(c)}{\geq}\sum_{t=1}^n I\left(W_{1,K+1[Q]};X_{1[Q]}|X_{2[Q]},Q=t\right)\\
	&\stackrel{(d)}{=} nI\left(\widetilde{W}_{1,K+1};X_{1}|X_{2}\right),
\end{align*}
where
\begin{itemize}
\item step~$(a)$ follows from the fact that $J^K$ is function of $I^K$ and $X_2^n$;
\item step~$(b)$ use the Markov chain 
$Y_{[1:t-1]}\mkv$\\ $\left(I^K,J^K,X_{2}^nX_{1[1:t-1]}\right)\mkv X_{1t}$;
\item step~$(c)$ follows from the use of a time sharing random variable $Q$ uniformly distributed over the set $[1:n]$ independent of the other variables;
\item step~$(d)$ follows by letting a new random variable $\widetilde{W}_{1,K+1}\triangleq  (W_{1,K+1[Q]} ,Q)$.
\end{itemize}

\subsubsection{Constraint on rate  $R_2$}
For the second rate, we have
\begin{align*}
&n(R_2+\varepsilon) \geq  H\left(J^K\right)\\
	&\geq H\left(J^{K-1} |X_1^nY^n\right)  +H\left(J_K|I^KJ^{K-1} \right)\\
		&\stackrel{(a)}{=} I\left(J^{K-1};X_2^n|X_1^nY^n\right)+I\left(J_K;X_2^nY^n|I^KJ^{K-1} \right)\\
			&= I\left(J^{K-1};X_2^n|X_1^nY^n\right)+I\left(J_K;X_2^n|I^KJ^{K-1}Y^n\right)\nonumber\\
&\qquad{-I\left(Y^n;I^KJ^{K-1}\right)+I\left(Y^n;I^KJ^K\right)}\\
&\stackrel{(b)}{\geq} \sum_{t=1}^n\Big[ I\left( J^{K-1};X_{2t}|X_{2[t+1:n]}Y^nX_1^n\right)\nonumber\\
&\qquad+I\left( J_K;X_{2t}|I^KJ^{K-1}X_{2[t+1:n]}Y^n\right)\nonumber\\
&\qquad- I\left(Y_{t};I^KJ^{K-1}|Y_{[1:t-1]} \right)\Big]+ n(\mu-\varepsilon)\\
&\stackrel{(c)}{=}\sum_{t=1}^n \Big[ I\left( J^{K-1}I^K;X_{2t}|X_{2[t+1:n]}Y^nX_1^n\right)\nonumber\\
&\qquad+ I\left( J_K;X_{2t}|I^KJ^{K-1}X_{2[t+1:n]}Y_{[1:t]}\right)\nonumber\\
&\qquad-I\left(Y_{t};I^KJ^{K-1}Y_{[1:t-1]} \right)\Big]+ n(\mu-\varepsilon)\\
&\stackrel{(d)}{\geq }\sum_{t=1}^n \Big[ I\left( J^{K-1}I^KX_{2[t+1:n]}Y_{[1:t-1]};X_{2t}|Y_t\right)\nonumber\\
&\qquad+ I\left( J_K;X_{2t}|I^KJ^{K-1}X_{2[t+1:n]}Y_{[1:t]}\right) \nonumber\\
&\qquad{  - I\left(Y_{t};I^KJ^{K-1}Y_{[1:t-1]}X_{2[t+1:n]} \right) \Big] + n(\mu-\varepsilon)}\\
&\stackrel{(e)}{=}\sum_{t=1}^n \Big[I\left(W_{2,K[Q]};X_{2[Q]}|Y_{[Q]},Q=t\right)\nonumber\\
&\qquad+  I\left( V_{2K[Q]};X_{2[Q]}|W_{2,K[Q]}Y_{[Q]},Q=t\right) \nonumber\\
&\qquad{-I\left(Y_{[Q]};W_{2,K[Q]}|Q=t \right)\Big]+ n(\mu-\varepsilon)}\\
&\stackrel{(f)}{=}  nI\left(\widetilde{W}_{1,K+1};X_{2}|Y\right)-nI\left(Y;\widetilde{W}_{2,K}\right)+ n(\mu-\varepsilon),
\end{align*}
where
\begin{itemize}
\item step~$(a)$ follows from the fact that by definition of the code $I_{l}$ is function of $J^{l-1}$ and $X_1^n$, and $J_{l}$ is function of $I^{l}$ and $X_2^n$;
\item step~$(b)$ follows from \eqref{eq:relevance_ml-p2} and the chain rule for mutual information and entropy;
\item step~$(c)$ follows from the fact that by definition of the code $I^K$ is function of $J^{K-1}$ and $X_1^n$, and the Markov chains $Y_{[t+1:n]}\mkv\left(I^K,J^{K-1},X_{2[t+1:n]},Y_{[1:t]}\right)\mkv X_{2t}$ and $Y_{[t+1:n]}\mkv\left(I^K,J^K,X_{2[t+1:n]},Y_{[1:t]}\right)\mkv X_{2t}$;
\item step~$(d)$ follows from the fact that $X_1\mkv Y\mkv X_2$ and non-negativity of mutual information;
\item step~$(e)$ follows from the use of a time sharing random variable $Q$ uniformly distributed over the set $[1:n]$ and independent of the other variables;
\item step~$(f)$ follows by letting a new random variables $\widetilde{W}_{1,K+1}\triangleq  (W_{1,K+1[Q]} ,Q)$, $\widetilde{W}_{2,K}\triangleq  (W_{2,K[Q]},Q)$.
\end{itemize}

\subsubsection{Constraint on sum-Rate $R_1+R_2$}
For the sum-rate, we have
\begin{align*}
n&(R_1+R_2+2\varepsilon)\nonumber\\
	&\stackrel{(a)}{\geq}	I\left(I^KJ^K;Y^n\right)+I\left(I^KJ^K;X_1^nX_2^n|Y^n\right)  \\
	&\stackrel{(b)}{\geq } n(\mu-\varepsilon)+\sum_{t=1}^n I\left(I^KJ^KX_{2[t+1:n]}Y_{[1:t-1]};X_{1t}X_{2t}|Y_t \right)\\
	&\stackrel{(c)}{= } n(\mu-\varepsilon) + \sum_{t=1}^n I\left(W_{1,K+1[Q]};X_{1[Q]}X_{2[Q]}|Y_{[Q]},Q=t\right)\\
	&\stackrel{(d)}{= }  n(\mu-\varepsilon)+ n I(\widetilde{W}_{1,K+1};X_{1}X_{2}|Y),
\end{align*}
where
\begin{itemize}
\item step~$(a)$ follows from definition of the code  $I^K$ and $J^K$ are functions of $X_1^n$ and $X_2^n$;
\item step~$(b)$ follows from \eqref{eq:relevance_ml-p2};
\item step~$(c)$ follows from the use of a time sharing random variable $Q$ uniformly distributed over the set $[1:n]$ independent of the other variables;
\item step~$(d)$ follows by letting a new random variables: $\widetilde{W}_{1,K+1}\triangleq  (W_{1,K+1[Q]} ,Q)$.
\end{itemize}

\subsubsection{Constraint on the relevance $\mu$}
Finally, for the relevance, we have
\begin{IEEEeqnarray*}{rCl}
n(\mu-\epsilon)
	&\leq& \;I\left(Y^n;I^KJ^K\right)\\
	&=& \sum_{t=1}^nI\left(Y_t;I^KJ^K|Y_{[1:t-1]}\right)\\
	&\leq&\sum_{t=1}^nI\left(Y_t;I^KJ^KY_{[1:t-1]}X_{2[t+1:n]}\right)\\
	&\stackrel{(a)}{=}&	\sum_{t=1}^nI\left(Y_{[Q]};W_{1,K+1}|Q=t\right)\\
	&\stackrel{(b)}{=}&	nI\left(Y;\widetilde{W}_{1,K+1}\right),
\end{IEEEeqnarray*}
where
\begin{itemize}
\item step~$(a)$ follows from the use of a time sharing random variable $Q$ uniformly distributed over the set $[1:n]$ independent of the other variables;
\item step~$(b)$ follows by letting a new random variables $\widetilde{W}_{1,K+1}\triangleq  (W_{1,K+1[Q]} ,Q)$.
\end{itemize}

In this way we conclude the proof that  $\mathcal{R}_{\mbox{\tiny CDIB}}^{\mbox{\tiny outer}}(K) \supseteq\mathcal{R}_{\mbox{\tiny CDIB}}(K)$. The fact that $\mathcal{R}_{\mbox{\tiny CDIB}}^{\mbox{\tiny inner}}(K) \subseteq\mathcal{R}_{\mbox{\tiny CDIB}}(K)$ is given in Appendix \ref{sec:achiev}.

\subsection{Proof of Theorem~\ref{thm:k=1}} \label{sec:case_l=1}

We now show that  $\mathcal{R}_{\mbox{\tiny CDIB}}^{\mbox{\tiny outer}}(1)=\mathcal{R}_{\mbox{\tiny CDIB}}^{\mbox{\tiny inner}}(1)=\mathcal{R}_{\mbox{\tiny CDIB}}(1)$ which implies Theorem~\ref{thm:k=1}. When $K=1$ we have that $\mathcal{R}_{\mbox{\tiny CDIB}}^{\mbox{\tiny inner}}(1)$ reads as:

\begin{IEEEeqnarray*}{rCl}
R_1 &\geq & I(X_1;V_1|X_2),\\
R_2 &\geq& I(X_2;V_2|V_1),\\ 
R_1+R_2 &\geq& I(X_1X_2;V_1V_2),\\
\mu &\leq & I(Y;V_1V_2),
\end{IEEEeqnarray*}
with $V_1$ and $V_2$ taking values in finite alphabets $\mathcal{V}_1$ and $\mathcal{V}_2$ and satisfying $V_1\mkv X_1\mkv (X_2,Y)$, $V_2\mkv (V_1,X_2) \mkv (X_1,Y)$. Similarly $\mathcal{R}_{\mbox{\tiny CDIB}}^{\mbox{\tiny outer}}(1)$ can be written as:
\begin{IEEEeqnarray*}{rCl}
R_1&\geq&I(X_1;U_1|X_2), \\
R_2&\geq&[I(X_2;U_2|YU_1)-I(Y;U_1)+\mu]^+,\\
R_1+R_2&\geq&I(X_1X_2;U_1U_2|Y)+\mu, \\
\mu&\leq&I(Y;U_1U_2),
\end{IEEEeqnarray*}
with the auxiliary variables $U_1$ and $U_2$ taking values in finite alphabets $\mathcal{U}_1$ and $\mathcal{U}_2$ and satisfying: $U_1\mkv X_1\mkv (X_2,Y)$ and $U_2\mkv (U_1,X_2) \mkv (X_1,Y)$ (note that $U_1\mkv Y\mkv X_2$ because of $X_1\mkv Y\mkv X_2$). From the previous results it is clear that $\mathcal{R}_{\mbox{\tiny CDIB}}^{\mbox{\tiny inner}}(1)\subseteq\mathcal{R}_{\mbox{\tiny CDIB}}^{\mbox{\tiny outer}}(1)$. Similarly to~\cite{Courtade_2014}, it can be shown that $\mathcal{R}_{\mbox{\tiny CDIB}}^{\mbox{\tiny inner}}(1)\supseteq\mathcal{R}_{\mbox{\tiny CDIB}}^{\mbox{\tiny outer}}(1)$. This is accomplished by showing that, when we fix a distribution on $(U_1,U_2)$ for every point $(R_1,R_2,\mu)\in\mathcal{R}_{\mbox{\tiny CDIB}}^{\mbox{\tiny outer}}(1)$, we can find an appropriate distribution $(V_1,V_2)$ such that $(R_1,R_2,\mu)\in\mathcal{R}_{\mbox{\tiny CDIB}}^{\mbox{\tiny inner}}(1)$. To this end, we study the extreme points (see Appendix \ref{app:mu_D} for a definition) and \emph{directions}~\cite{rockafellar_convex_1970} of  the restriction of $\mathcal{R}_{\mbox{\tiny CDIB}}^{\mbox{\tiny outer}}(1)$ over the assumed distribution of $(U_1,U_2)$. The details are given in Appendix \ref{sec:extreme}.

\subsection{Converse result for Theorem~\ref{theo:alternative_inner}}

The proof that  $\tilde{\mathcal{R}}_{\mbox{\tiny CDIB}}(1)\subseteq\mathcal{R}_{\mbox{\tiny CDIB}}(1)$ follows from simple multiterminal coding arguments and for that reason is omitted. The relevance level can be obtained using the same ideas that those in Appendix \ref{sec:achiev}. For $\mathcal{R}_{\mbox{\tiny CDIB}}(1)\subseteq\hat{\mathcal{R}}_{\mbox{\tiny CDIB}}(1)$ assume that  $(R_1,R_2,\mu)\in\mathcal{R}_{\mbox{\tiny CDIB}}(1)$ , then  for all $\varepsilon >0$ there exists $n_0(\varepsilon)$, such that $\forall\,n>n_0(\varepsilon)$  there exists a code $(n,\mathcal{F})$ with rates and relevance satisfying \eqref{eq:rate_ml-p2} and \eqref{eq:relevance_ml-p2}. For each $t=[1:n]$, we define random variables:
\begin{equation}
\label{eq:def_aux_1}
V_{1,t}\triangleq  \left(I_1,Y_{[1:t-1]},X_{2[1:t-1]}\right),\qquad
V_{2,t}\triangleq  J_1.
\end{equation}
It is easy to show that these choices verify (\ref{eq:second_mkv_cond}). Using similar steps as in the previous converse proofs, we can easily  obtain the following bounds:
\begin{IEEEeqnarray*}{rCl}
R_1+\epsilon &\geq& \frac{1}{n}\sum_{t=1}^n I\left(I_1 X_{2[1:t-1]}Y_{[1:t-1]};X_{1t}\right),\\
R_2+\epsilon &\geq&\frac{1}{n}\sum_{t=1}^n I\left(X_{2t};J_1\big| I_1X_{2[1:t-1]}Y_{[1:t-1]}\right),\\
\mu-\epsilon &\leq& \frac{1}{n}\sum_{t=1}^n  I\left(Y_{t};I_1J_1X_{2[1:t-1]}Y_{[1:t-1]}\right).
\end{IEEEeqnarray*}
From a time-sharing argument and using (\ref{eq:def_aux_1}) we get the rate conditions corresponding to $\hat{\mathcal{R}}_{\mbox{\tiny CDIB}}(1)$.

\section{Gaussian Source Models}
\label{sec:gaussian}

In this section, we study Gaussian models between source samples and hidden representations. Although the above achievability results are strictly valid for random variables taking values on finite alphabets, the results can be applied to continuous random variables with sufficiently well behaved probability density function (e.g. Gaussian random variables).  A simple sequence of coding schemes consisting of a quantization procedure over the sources and appropriate test channels (with diminishing quantization steps) followed by  coding schemes as the ones presented in this paper will suffice (e.g. see~\cite{ELGamal-Kim-book}). 

\subsection{Gaussian TW-CIB model}
Let $(X_1,X_2,Y_1,Y_2)$ be Gaussian random variables with zero-mean. We will assume without loss of generality that we can write:
\begin{equation}
\label{eq:model_gaussian_TWI}
\left[\begin{array}{c}Y_1\\Y_2\end{array}\right]=\mathbf{A}\cdot\left[\begin{array}{c}X_1\\X_2\end{array}\right]+\left[\begin{array}{c}Z_1\\Z_2\end{array}\right],\ \ \mathbf{A}=\left[\begin{array}{cc} a_{11}& a_{12}\\ a_{21}& a_{22}\end{array}\right],
\end{equation}
where $Z_1\perp (X_1,X_2)$ and $Z_2\perp(X_1,X_2)$ and matrix $\mathbf{A}$ can be obtained from the correlation structure of the random variables. To this end, define:
\begin{align*}
a_{12}&\triangleq  \frac{\sigma_{y_1}}{\sigma_{x_2}}\frac{\rho_{x_2 y_1}-\rho_{x_1 y_1}\rho_{x_1 x_2}}{1-\rho_{x_1 x_2}^2}\\
a_{21}&\triangleq  \frac{\sigma_{y_2}}{\sigma_{x_1}}\frac{\rho_{x_1y_2}-\rho_{x_2 y_2}\rho_{x_1 x_2}}{1-\rho_{x_1 x_2}^2}\ ,
\end{align*}
where $\sigma^2_b$ denotes the variance of a random variable $B$, and $\rho_{b_1b_2}$ represents the \emph{Pearson product-moment correlation coefficient} between random variables $B_1$ and $B_2$. As $(X_1,X_2,Y_1,Y_2)$ are jointly Gaussian, then $Z_1$ and $Z_2$ are Gaussian as well. It is easy to check that:
\begin{equation*}
\sigma_{z_1}^2=\frac{\sigma_{y_1}^2\beta}{1-\rho_{x_1 x_2}^2}\ ,\qquad\sigma_{z_2}^2=\frac{\sigma_{y_2}^2\delta}{1-\rho_{x_1 x_2}^2}\ ,
\end{equation*}
where 
\begin{align*}
\beta\triangleq  &1-\rho_{x_1 x_2}^2-\rho_{x_1 y_1}^2-\rho_{x_2 y_1}^2+2\rho_{x_1 x_2}\rho_{x_1 y_1}\rho_{x_2 y_1}\ ,\\
\delta\triangleq  &1-\rho_{x_1 x_2}^2-\rho_{x_2 y_2}^2-\rho_{x_1 y_2}^2+2\rho_{x_1 x_2}\rho_{x_2 y_2}\rho_{x_1 y_2}\ .
\end{align*}
We are ready to present our first result.
\begin{theorem}[Complexity-relevance region for the Gaussian TW-CIB model] \label{gauss_posta}
When $(X_1,X_2,Y_1,Y_2)$ are jointly Gaussian, for any $K$, $\mathcal{R}_{\mbox{\tiny TW-CIB}}(K)$ is given by:
\begin{align}
R_1&\geq\frac{1}{2}\log\left(\frac{(1-\rho_{x_1x_2}^2)(1-\rho_{x_2y_2}^2)-\delta}{2^{-2\mu_2}(1-\rho_{x_1x_2}^2)-\delta}\right)\nonumber\\
0\leq\mu_2&<\frac{1}{2}\log\left(\frac{1-\rho_{x_1x_2}^2}{\delta}\right),\label{eq:R_1_mu_2}\\
R_2&\geq\frac{1}{2}\log\left(\frac{(1-\rho_{x_1x_2}^2)(1-\rho_{x_1y_1}^2)-\beta}{2^{-2\mu_1}(1-\rho_{x_1x_2}^2)-\beta}\right)\nonumber\\
0\leq\mu_1&<\frac{1}{2}\log\left(\frac{1-\rho_{x_1x_2}^2}{\beta}\right).
\label{eq:R_2_mu_1}
\end{align}
\end{theorem}

\begin{IEEEproof}
We first consider the converse.

\textit{Converse:} Assume $(R_1,R_2,\mu_1,\mu_2)\in\mathcal{R}_{\mbox{\tiny TW-CIB}}(K)$. Consider the relevance level $\mu_1$. Using (\ref{eq:relevance_ml_1}):
\begin{align}
&\mu_1-\epsilon\leq \frac{1}{n}I\left(Y_1^n;I^KJ^KX_1^n\right),\nonumber\\
&= h(Y_1)-\frac{1}{n}h\left(Y_1^n\big|I^KJ^KX_1^n\right),\nonumber\\
&= \frac{1}{2}\log{\left(2\pi e\sigma_{y_1}^2\right)}-\underbrace{\frac{1}{n}h\left(a_{12}X_2^n+Z_1^n\big|I^KJ^KX_1^n\right)}_{(a)}.
\label{eq:cond_ mu}
\end{align}
From the equation for $R_2$ and using the fact that $J^K$ is function of $X_2^n$ and $I^K$ it is not difficult to obtain:
\begin{align}
&R_2+\epsilon\geq \frac{1}{n}I\left(X_2^n;I^KJ^K\big|X_1^n\right),\nonumber\\
&= h(X_2|X_1)-\frac{1}{n}h\left(X_2^n\big|I^KJ^KX_1^n\right),\nonumber \\
&= \frac{1}{2}\log{\left(2\pi e \textrm{Var}[X_2|X_1]\right)}-\underbrace{\frac{1}{n}h\left(X_2^n\big|I^KJ^KX_1^n\right)}_{(b)}.
\label{eq:cond_R_2}
\end{align}
As $Z_1^n\perp \left(I^KJ^K\right)$ we can link $(a)$ and $(b)$ using the conditional EPI \cite{rioul_information_2011} to write:
\begin{align*}
&2^{\frac{2}{n}h \left(a_{12}X_2^n+Z_1^n\big|I^KJ^KX_1^n\right)}\nonumber\\
 &\qquad\geq
 a_{12}^2 2^{\frac{2}{n}h\left(X_2^n\big|I^KJ^KX_1^n\right)}+2\pi e\sigma_{z_1}^2.
\end{align*}
From (\ref{eq:cond_ mu}) and (\ref{eq:cond_R_2}) we can write:
\begin{equation*}
R_2+\epsilon\geq\frac{1}{2}\log{\left(\frac{\textrm{Var}[X_2|X_1]a_{12}^2}{\sigma_{Y_1}^2 2^{-2(\mu_1-\epsilon)}-\sigma_{Z_1}^2}\right)}.
\end{equation*}
Using the correlation structure implied by (\ref{eq:model_gaussian_TWI}) we can obtain:
\begin{equation*}
R_2+\epsilon\geq\frac{1}{2}\log\left(\frac{(1-\rho_{x_1x_2}^2)(1-\rho_{x_1y_1}^2)-\beta}{2^{-2(\mu_1-\epsilon)}(1-\rho_{x_1x_2}^2)-\beta}\right).
\end{equation*}
As $\epsilon>0$ is arbitrary we obtain the desired result. The results for $R_1$ and $\mu_2$ can be obtained similarly.

\textit{Achievability:} We propose the following choices for auxiliary random variables. Let $V_{1}^{[2:K]}=V_{2}^{[2:K]}=\emptyset$ and $V_{1,1}=X_1+P_1$ and  $V_{2,1}=X_2+P_2$,  where $V_{1,1}$ and $V_{2,1}$ are zero-mean Gaussian random variables with  variances:
\begin{IEEEeqnarray*}{rCl}
\mathbb{E}[V_{1,1}^2]&=&\sigma_{x_1}^2+\sigma_{p_1}^2\ ,\\
\mathbb{E}[V_{2,1}^2]&=&\sigma_{x_2}^2+\sigma_{p_2}^2\ ,\\
\sigma_{p_1}^2&=&\sigma_{x_1}^2\frac{2^{-2\mu_2}(1-\rho_{x_1x_2}^2)-\delta}{1-\rho_{x_2y_2}-2^{-2\mu_2}}\ ,\\
\sigma_{p_2}^2&=&\sigma_{x_2}^2\frac{2^{-2\mu_1}(1-\rho_{x_1x_2}^2)-\beta}{1-\rho_{x_1y_1}-2^{-2\mu_1}}\ ,
\end{IEEEeqnarray*}
and $P_1,P_2$ are Gaussian zero-mean random variables such that $P_1\perp(X_1,X_2,Y_1,Y_2,P_2)$ and $P_2\perp(X_1,X_2,Y_1,Y_2,P_1)$. It is clear these choices satisfies the appropriate Markov chain conditions. Although a bit cumbersome, it is straightforward to calculate the corresponding values of $I(Y_1;\mathcal{W}_{[1;K+1]}X_2)$, $I(Y_2;\mathcal{W}_{[1;K+1]}X_2)$, $I(X_1;\mathcal{W}_{[1;K+1]}|X_2)$ and $I(X_2;\mathcal{W}_{[1;K+1]}|X_1)$ and conclude the proof.
\end{IEEEproof}
\vspace{1mm}
\begin{remark}
Notice that the maximum values of $\mu_1$ and $\mu_2$  in \eqref{eq:R_2_mu_1}) and \eqref{eq:R_1_mu_2} correspond to $I(X_1X_2;Y_1)$ and $I(X_1X_2;Y_2)$  and are achievable when $R_2\rightarrow \infty$ and  $R_1\rightarrow \infty$ respectively. Besides that, it is clear from the achievability that only one round of interaction suffices to achieve optimality when the sources are jointly Gaussian. In perspective, this is not surprising, and derives from the Wyner-Ziv's result~\cite{wyner_ziv76} which states that for Gaussian random variables, the rate-distortion function with side information at the encoder and decoder is not larger than the one with side information only at the decoder. These two cases correspond to two extreme situations: one in which there is no interaction between both encoders and the other in which interaction is not needed because both encoders have access to both observable sources. This conclusion follows easily by noticing that any code for a Gaussian rate-distortion problem, where decoder 1 desires to reconstruct $Y_1$ with distortion $\mu^\prime_1\triangleq\sigma_{Y_1}^22^{-2\mu_1}$ and decoder 2 desires to recover $Y_2$ with distortion $\mu^\prime_2\triangleq\sigma_{Y_2}^22^{-2\mu_2}$, is also good for an equivalent CRL problem with desired relevances levels $\mu_1$ and $\mu_2$. 
\end{remark}
\begin{figure*}[b]
\normalsize
\vspace*{4pt}
\hrulefill
\begin{IEEEeqnarray}{rCl}
\mu&\leq&\frac{1}{2}\log\left(\frac{1}{1-\rho_{x_2y}^2+\rho_{x_2y}^22^{-2R_2}-\rho_{x_2y}^2\rho_{x_1x_2}^22^{-2R_2}+\rho_{x_2y}^2\rho_{x_1x_2}^22^{-2(R_1+R_2)}}\right).
\label{eq:mu_rate_relevance_1}
\end{IEEEeqnarray}
\end{figure*}
\subsection{Gaussian  CDIB model: $X_1\mkv X_2\mkv Y$ case}
We study the Gaussian case for the region $\mathcal{R}_{\mbox{\tiny CDIB}}(1)$ investigated  in Section~\ref{sec:alternate} when $X_1\mkv X_2\mkv Y$. Let $(X_1,X_2,Y)$ be Gaussian random variables with zero-mean. We will assume without loss of generality that we can write:
\begin{equation*}
Y=aX_2+Z_a\ ,\ \  \qquad X_2=bX_1+Z_b\ ,
\end{equation*}
where $Z_a\perp (X_1,X_2)$ and $Z_b\perp X_1$ are Gaussian and constants $a$ and $b$ are obtained from the correlation structure of the random variables. That is:
\begin{equation*}
a\triangleq\rho_{x_2y}\frac{\sigma_y}{\sigma_{x_2}}\ ,\qquad b\triangleq\rho_{x_1x_2}\frac{\sigma_{x_2}}{\sigma_{x_1}}\ .
\end{equation*}
It is easy to check that
\begin{equation*}
\sigma_{z_a}^2=\sigma_{y}^2(1-\rho_{x_2y}^2)\ ,\qquad\sigma_{z_b}^2=\sigma_{x_2}^2(1-\rho_{x_1x_2}^2)\ .
\end{equation*}
\begin{theorem}[Complexity-relevance region for the Gaussian model when $X_1\mkv X_2\mkv Y$] \label{gauss_posta2}
Let $(X_1,X_2,Y_2)$ be jointly Gaussian random variables satisfying $X_1\mkv X_2\mkv Y$. The complexity-relevance region $\mathcal{R}_{\mbox{\tiny CDIB}}(1)$ is given by \eqref{eq:mu_rate_relevance_1}, with $R_1\geq 0$, $R_2\geq 0$.
\end{theorem}
\begin{IEEEproof}
We begin with the converse.

\textit{Converse:} Assume $(R_1,R_2,\mu)\in\mathcal{R}_{\mbox{\tiny CDIB}}(1)$ and consider rate constraint $R_1$. Using the fact that $I_1$ is function of $X_1^n$:
\begin{equation}
R_1+\epsilon\geq\frac{1}{n}I\left(X_1^n;I_{1}\right)= h(X_1)-\frac{1}{n}h\left(X_1^n\big|I_{1}\right)\ .
\label{eq:R1_x1x2y}
\end{equation}
From rate $R_2$ and using the fact that $J_1$ is function of $X_2^n$ and $I_1$ it is not hard to obtain:
\begin{align}
R_2+\epsilon&\geq \frac{1}{n}I\left(J_1;X_2^n\big|I_1\right)\nonumber \\
&= \frac{1}{n}h\left(bX_1^n+Z_b^n\big|I_1\right)-\frac{1}{n}h\left(X_2^n\big|I_1J_1\right)\nonumber \\
&\overset{(a)}{\geq} \frac{1}{2}\log\left(b^22^{\frac{2}{n}h(X_1^n|I_1)}+2\pi e\sigma_{z_b}^2\right)\nonumber\\
&\qquad-\frac{1}{n}h\left(X_2^n\big|I_1J_1\right)\nonumber \\
&\overset{(b)}{\geq} \frac{1}{2}\log\left(2\pi e\sigma_{x_1}^2b^22^{-2(R_1+\epsilon)}+2\pi e\sigma_{z_b}^2\right)\nonumber\\
&\qquad-\frac{1}{n}h\left(X_2^n\big|I_1J_1\right),
\label{eq:R2_x1x2y}
\end{align}
where $(a)$ uses the conditional EPI because $Z_b^n\perp I_1$, and $(b)$ use Eq. \eqref{eq:R1_x1x2y}. \par
From relevance condition we use the same idea:
\begin{align*}
\mu-\epsilon&\leq \frac{1}{n}I\left(Y^n;I_1J_1\right)\\
&= h(Y)-\frac{1}{n}h\left(aX_2^n+Z_a^n\big|I_1J_1\right)\\
&\overset{(c)}{\leq} \frac{1}{2}\log\left(2\pi e\sigma_y^2\right)-\frac{1}{2}\log\left(a^22^{\frac{2}{n}h(X_2^n|I_1J_1)}+2\pi e\sigma_{z_a}^2\right)
\end{align*}
where $(c)$ use the conditional EPI because $Z_a^n\perp\left(I_1,J_1\right)$. Then, \eqref{eq:mu_rate_relevance_1} is proved using Eq. \eqref{eq:R2_x1x2y} and the fact that $\epsilon>0$ is arbitrary.
\begin{figure}[t]
	\centering{\includegraphics[width=0.48\textwidth]{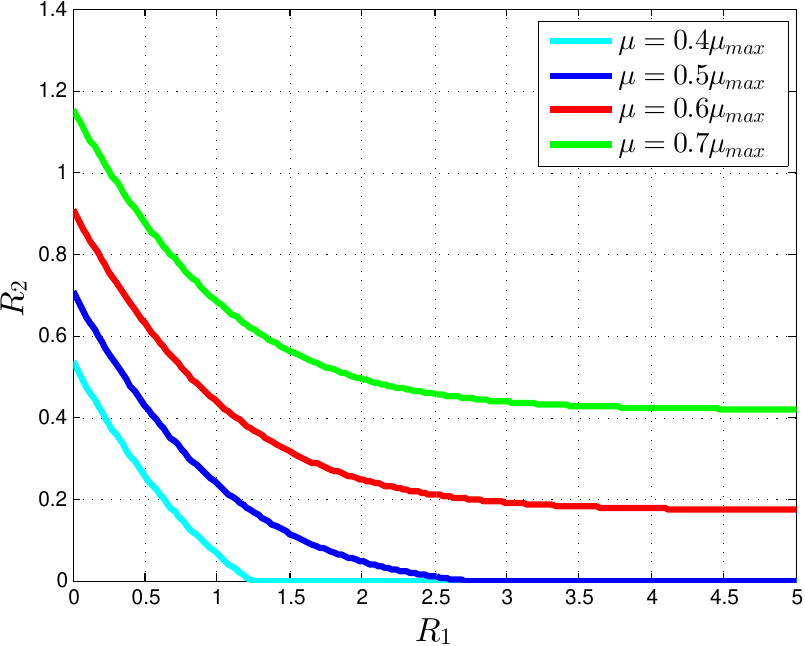}}
	\caption{Achievable rates $R_1$ and $R_2$ for the Gaussian case with $X_1\mkv X_2\mkv Y$ for several values of relevance $\mu$. The correlation coefficients are $\rho_{x_1x_2}=\rho_{x_2y}=0.8$.}
\label{fig:x1x2y}
\end{figure}

\textit{Achievability:} We propose the following choices for auxiliary random variables. Let $V_{1}=X_1+P_1$ and  $V_{2}=X_2+P_2$,  where $V_{1}$ and $V_{2}$ are zero-mean Gaussian random variables with  variances:
\begin{IEEEeqnarray*}{rCl}
\mathbb{E}[V_1^2]&=&\sigma_{x_1}^2+\sigma_{p_1}^2\ ,\qquad\mathbb{E}[V_2^2]=\sigma_{x_2}^2+\sigma_{p_2}^2\ ,\\
\sigma_{p_1}^2&=&\sigma_{x_1}^2\frac{2^{-2R_1}}{1-2^{-2R_1}}\ ,\\
\sigma_{p_2}^2&=&\sigma_{x_2}^2\frac{2^{-2R_2}}{1-2^{-2R_2}}\left(1-\rho_{x_1x_2}^2+\rho_{x_1x_2}^22^{-2R_1}\right)\ ,
\end{IEEEeqnarray*}
and $P_1,P_2$ are Gaussian zero-mean random variables such that $P_1\perp(X_1,X_2,Y,P_2)$ and $P_2\perp(X_1,X_2,$ $Y,P_1)$. It is clear these choices satisfy the appropriate Markov chain conditions. Although a bit cumbersome, it is straightforward to calculate the corresponding values of $I(X_1;V_1)$, $I(X_2;V_2|V_1)$ and $I(Y;V_1,V_2)$. This  concludes the proof.
\end{IEEEproof}
This region can also be written as:
\begin{align*}
R_1&\geq 0, \\
R_2&\geq \frac{1}{2}\left[\log{\left(\frac{\rho_{x_2y}^2\rho_{x_1x_2}^22^{-2R_1}+\rho_{x_2y}^2(1-\rho_{x_1x_2}^2)}{2^{-2\mu}-(1-\rho_{x_2y}^2)}\right)}\right]^{+}
\end{align*}
In Fig. \ref{fig:x1x2y} we plot this alternative parametrization for different values of $\mu$. Taking into account that $\mu_{\max}=I(Y;X_2)$ it is seen how when increasing $R_1$ the value of $R_2$ tends to saturate. If the value of $\mu$ required is small enough, after increasing sufficiently $R_1$, the information about $Y$ provided by the second encoder would be not useful. In fact, it can be proved that the critical value of $R_1$ (if exists) for which $R_2=0$  satisfy:
\begin{equation*}
R_1=\frac{1}{2}\log{\left(\frac{\rho_{x_1x_2}^2\rho_{x_2y}^2}{2^{-2\mu}-(1-\rho_{x_1x_2}^2\rho_{x_2y}^2)}\right)}. 
\end{equation*}
Moreover, it can be proved that there will be a critical value for $R_1$ if and only if the required level of relevance satisfy $\mu\leq I(Y;X_1)$. If the value for $\mu$ is greater than this quantity, it is not possible to have $R_2=0$ independently of the value of $R_1$. This not a surprise because it means, that for the level of relevance required, encoding of only $X_1$ is sufficient. If $\mu> I(Y;X_1)$ s required node 3 will require information from $X_2$ (remember that $X_1\mkv X_2\mkv Y$) which leads to $R_2>0$. 

\subsection{Gaussian CDIB model: $X_1\mkv Y\mkv X_2$ case}
\label{gauss_dcrl_x1yx2}
\begin{figure*}[b]
\normalsize
\vspace*{4pt}
\hrulefill
\begin{align}
R_2&\geq r_2-\frac{1}{2}\log\left(\frac{1-\rho_{x_1y}^2\rho_{x_2y}^2-\rho_{x_1y}^2(1-\rho_{x_2y}^2)2^{-2r_1}-\rho_{x_2y}^2(1-\rho_{x_1y}^2)}{(1-\rho_{x_1y}^2)(1-\rho_{x_2y}^2)}\right)+\mu,\label{eq:r2condition}\\
\mu&\leq\frac{1}{2}\log\left(\frac{1-\rho_{x_1y}^2\rho_{x_2y}^2-\rho_{x_1y}^2(1-\rho_{x_2y}^2)2^{-2r_1}-\rho_{x_2y}^2(1-\rho_{x_1y}^2)2^{-2r_2}}{(1-\rho_{x_1y}^2)(1-\rho_{x_2y}^2)}\right).\label{eq:mucondition}
\end{align}
\end{figure*}

We will consider the Gaussian case for the region $\mathcal{R}_{\mbox{\tiny CDIB}}(1)$  when $X_1\mkv Y\mkv X_2$, studied in section \ref{sec:case_l=1}. Let $(X_1,X_2,Y)$ be Gaussian random variables with zero-mean. We will assume without loss of generality, that we can write:
\begin{equation*}
Y=a_1X_1+a_2X_2+Z,
\end{equation*}
where $Z\perp (X_1,X_2)$ is Gaussian and constants $a_1$ and $a_2$ can be obtained from the correlation structure of the random variables, using the Markov chain $X_1\mkv Y\mkv X_2$. This is:
\begin{align*}
a_1&\triangleq\frac{\sigma_y}{\sigma_{x_1}}\frac{\rho_{x_1y}(1-\rho_{x_2y}^2)}{1-\rho_{x_1y}^2\rho_{x_2y}^2}\\ a_2&\triangleq\frac{\sigma_y}{\sigma_{x_2}}\frac{\rho_{x_2y}(1-\rho_{x_1y}^2)}{1-\rho_{x_1y}^2\rho_{x_2y}^2}\ .
\end{align*}
It is not difficult to check that
\begin{equation*}
\sigma_{z}^2=\sigma_{y}^2\frac{(1-\rho_{x_1y}^2)(1-\rho_{x_2y}^2)}{1-\rho_{x_1y}^2\rho_{x_2y}^2}\ .
\end{equation*}
\begin{theorem}[Outer bound to $\mathcal{R}_{\mbox{\tiny CDIB}}(1)$ for the Gaussian model when $X_1\mkv Y\mkv X_2$] \label{gauss_posta3}
If $(X_1,X_2,Y)$ are jointly Gaussian with $X_1\mkv Y\mkv X_2$ and if $(R_1,R_2,\mu)\in\mathcal{R}_{\mbox{\tiny CDIB}}(1)$, then there exists $r_1\geq0$ and $r_2\geq0$ such that they meet \eqref{eq:r2condition}, \eqref{eq:mucondition} and
\begin{align*}
R_1&\geq r_1-\frac{1}{2}\log\left(\frac{1}{1-\rho_{x_2y}^2}\right)+\mu,\\
R_1+R_2&\geq r_1+r_2+\mu,
\end{align*}
\end{theorem}

\begin{IEEEproof}
First of all, we define
\begin{equation*}
r_1\triangleq\frac{1}{n}I\left(X_1^n;I_1|Y^n\right)\ ,\ \quad r_2\triangleq\frac{1}{n}I\left(X_2^n;J_1|I_1Y^n\right)
\end{equation*}
Consider the constraint on $R_1-\mu$, using the Markov chain $X_1\mkv Y\mkv X_2$ we can write:
\begin{align*}
R_1-\mu&+2\epsilon\geq\frac{1}{n}H\left(I_1\right)- \frac{1}{n}I\left(Y^n;I_1J_1\right),\\
&\geq\frac{1}{n}I\left(X_1^nY^n;I_1\big|X_2^n\right)- \frac{1}{n}I\left(Y^n;I_1X_2^n\right),\\
&=\frac{1}{n}I\left(X_1^n;I_1\big|X_2^nY^n\right)- \frac{1}{n}I\left(Y^n;X_2^n\right),\\
&=r_1-\frac{1}{2}\log\left(\frac{\sigma_y^2}{\textrm{Var}[Y|X_2]}\right),\\
&=r_1-\frac{1}{2}\log\left(\frac{1}{1-\rho_{x_2y}^2}\right).
\end{align*}
For $R_1+R_2-\mu$, using Markov chain $X_1\mkv Y\mkv X_2$ again, it is not difficult  to obtain:
\begin{align*}
R_1&+R_2-\mu+3\epsilon\geq\frac{1}{n}H\left(I_1,J_1\right)- \frac{1}{n}I\left(Y^n;I_1J_1\right),\\
&=\frac{1}{n}H\left(I_1J_1\big|Y^n\right),\\
&=\frac{1}{n}I\left(X_1^nX_2^n;I_1J_1\big|Y^n\right),\\
&=\frac{1}{n}I\left(X_1^n;I_1J_1\big|Y^n\right)+\frac{1}{n}I\left(X_2^n;I_1J_1\big|X_1^nY^n\right),\\
&=r_1+r_2.
\end{align*}
For $R_2-\mu$, doing a similar analysis:
\begin{align*}
R_2-\mu&+2\epsilon\geq\frac{1}{n}H\left(J_1\right)- \frac{1}{n}I\left(Y^n;I_1J_1\right),\\
&\geq\frac{1}{n}I\left(X_2^nY^n;J_1\big|I_1\right)- \frac{1}{n}I\left(Y^n;I_1J_1\right),\\
&=\frac{1}{n}I\left(X_2^n;J_1\big|Y^nI_1\right)- \frac{1}{n}I\left(Y^n;I_1\right),\\
&=r_2-h\left(Y\right)+\frac{1}{n}h\left(Y^n\big|I_1\right),
\end{align*}
where the term $h\left(Y^n|I_1\right)$ can be bounded using the conditional EPI two times (in a similar fashion as in \cite{oohama_rate-distortion_2005}): firstly because of $Z^n\perp I_1$ and secondly because of $X_1^n\mkv\left(I_1,Y^n\right)\mkv X_2^n$,
\begin{align*}
&2^{\frac{2}{n}h\left(Y^n|I_1\right)}\geq 2^{\frac{2}{n}h\left(a_1X_1^n+a_2X_2^n|I_1\right)}+2\pi e\sigma_z^2,\\
&\;=\frac{2^{\frac{2}{n}h\left(a_1X_1^n+a_2X_2^n|Y^nI_1\right)}2^{\frac{2}{n}h\left(Y^n|I_1\right)}}{2\pi e\sigma_z^2}\nonumber\\
&\qquad\quad+2\pi e\sigma_z^2,\\
&\;\geq\frac{\left[a_1^22^{\frac{2}{n}h\left(X_1^n|Y^nI_1\right)}+a_2^22^{\frac{2}{n}h\left(X_2^n|Y^nI_1\right)}\right]2^{\frac{2}{n}h\left(Y^n|I_1\right)}}{2\pi e\sigma_z^2}\nonumber\\
&\qquad\quad+2\pi e\sigma_z^2,\\
&\;=\frac{\left[a_1^22^{\frac{2}{n}h\left(X_1^n|Y^n\right)}2^{-2r_1}+a_2^22^{\frac{2}{n}h\left(X_2^n|Y^n\right)}\right]2^{\frac{2}{n}h\left(Y^n|I_1\right)}}{2\pi e\sigma_z^2}\nonumber\\
&\qquad\quad+2\pi e\sigma_z^2,\\
&\;=\frac{\left(a_1^2\textrm{Var}[X_1|Y]2^{-2r_1}+a_2^2\textrm{Var}[X_2|Y]\right)2^{\frac{2}{n}h\left(Y^n|I_1\right)}}{\sigma_z^2}\nonumber\\
&\qquad\quad+2\pi e\sigma_z^2.
\end{align*}
Finally, this term is bounded by
\begin{equation*}
2^{\frac{2}{n}h\left(Y^n|I_1\right)}\geq\frac{2\pi e\sigma_z^4}{\sigma_z^2-\left(a_1^2\textrm{Var}[X_1|Y]2^{-2r_1}+a_2^2\textrm{Var}[X_2|Y]\right)}.
\end{equation*}
Then, the bound of $R_2-\mu$ can be written as:
\begin{align*}
&R_2-\mu+2\epsilon\geq r_2\nonumber\\
&-\frac{1}{2}\log\left(\frac{\sigma_y^2\left[\sigma_z^2-\left(a_1^2\textrm{Var}[X_1|Y]2^{-2r_1}+a_2^2\textrm{Var}[X_2|Y]\right)\right]}{\sigma_z^4}\right)
\end{align*}
and then \eqref{eq:r2condition} is proved because $\epsilon>0$ is arbitrary. The analysis is similar to the case for $h\left(Y^n|I_1J_1\right)$, because $Z^n\perp(I_1,J_1)$ and $X_1^n\mkv\left(I_1,J_1,Y^n\right)\mkv X_2^n$:
\begin{align*}
&2^{\frac{2}{n}h\left(Y^n|I_1J_1\right)}\geq\nonumber\\
&\qquad\frac{2\pi e\sigma_z^4}{\sigma_z^2-\left(a_1^2\textrm{Var}[X_1|Y]2^{-2r_1}+a_2^2\textrm{Var}[X_2|Y]2^{-2r_2}\right)}.
\end{align*}
Finally, the relevance condition can be bounded as:
\begin{align*}
&\mu-\epsilon\leq \frac{1}{n}I\left(Y^n;I_1J_1\right), \\
&=h\left(Y\right)-\frac{1}{n}h\left(Y^n\big|I_1J_1\right), \\
&=\frac{1}{2}\log\left(\frac{\sigma_y^2}{\sigma_z^4}\left[\sigma_z^2-\left(a_1^2\textrm{Var}[X_1|Y]2^{-2r_1}\right.\right.\right.\nonumber\\
&\qquad\qquad\qquad\qquad\qquad\quad\left.\left.\left.+a_2^2\textrm{Var}[X_2|Y]2^{-2r_2}\right)\right]\right),
\end{align*}
As $\epsilon>0$ is arbitrary, we obtain \eqref{eq:mucondition}.
\end{IEEEproof}

\begin{figure}[t]
	\centering{\includegraphics[width=0.48\textwidth]{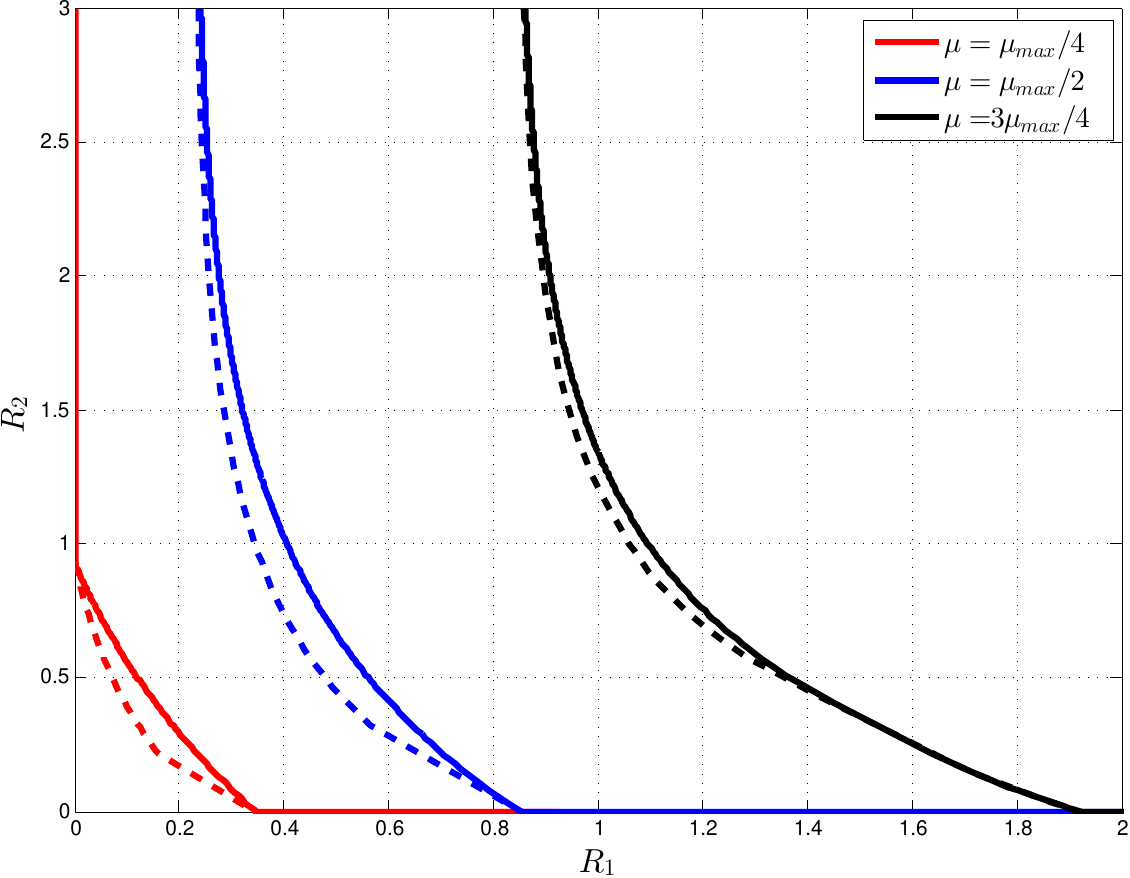}}
	\caption{Comparison between the outer bound (dashed) and the inner bound (solid) for $\mathcal{R}_{\mbox{\tiny CDIB}}(1)$ when $\rho_{x_1y}=0.8$ and $\rho_{x_2y}=0.6$. In this case $\mu_{\max}=I(Y;X_1X_2)$.}
\label{fig:cvr_gauss}
\end{figure}

An inner bound for $\mathcal{R}_{\mbox{\tiny CDIB}}(1)$ can be obtained defining $V_1=X_1+P_1$ and $V_2=X_2+V_1+P_2$, where $P_1$ and $P_2$ are Gaussian variables with $P_1\perp(X_1,X_2,Y,P_2)$ and $P_2\perp(X_1,X_2,$ $Y,P_1)$ and variances $\sigma_{P_1}^2$ and $\sigma_{P_2}^2$. Numerically choosing  $\sigma_{P_1}^2$ and $\sigma_{P_2}^2$ to satisfy the relevance condition we can plot the resulting inner bound and compare with the obtained outer bound. The results are showed in Fig. \ref{fig:cvr_gauss}.  We observe that there is small gap between both regions (the parameters $\rho_{x_1y}$ and $\rho_{x_2y}$ were chosen to maximize the observed difference).  Although we were unable to prove it, we suspect that in this special Gaussian case there is no gain from cooperation and that the observed gap is indeed not achievable. This suspicion is motivated by the fact that the non-cooperative Gaussian CEO region for this problem, which can be easily obtained from the corresponding CEO result with Gaussian inputs and quadratic distortion in \cite{prabhakaran_rate_2004}, was also numerically shown to be equal to the above inner bound for the cooperative case. It is interesting to observe that, if true, the conclusion that cooperation is not helpful should hold for the cooperative Gaussian CEO problem with quadratic distortion as well. In the case the gap were achievable, this would be due to possible gains in the individuals rates $R_1$ and $R_2$. The sum-rate and relevance do not increase when cooperation is allowed. This is rooted in the well-known result that when cooperation is in force there is no gain in the sum-rate for a two encoder rate-distortion problem with Gaussian inputs and quadratic distortions \cite{Wagner_2008b}. The same result for our setting with log-loss distortion can be obtained easily.

\section{Binary Source Model}
\label{sec:binary}

In this section, we will consider a binary example for the region obtained related to the TW-CIB problem. The study of the binary examples with multiple rounds proves to be a rather challenging problem for which closed forms remain elusive  to obtain. Our approach to the problem will be the following. We will consider the problem in which decoder 1 is intended  to learn a hidden variable $Y_1$ while decoder 2 desires to learn $Y_2$. Exchanges between encoder 1 and the decoder 2,  and between encoder 2 and decoder 1, are through a two decoupled half-rounds as will we explained below. First we will consider the problem where both encoders know $X_1$ and $X_2$. From the perspective of each encoder-decoder pair this is reminiscent of a noisy rate-distortion problem with side information at both the encoder and the decoder where the metric of interest is given by the relevance $\frac{1}{n}I\left(Y_1^n;J_1X^n_1\right)$ and $\frac{1}{n}I\left(Y_2^n;I_1X^n_2\right)$, respectively.  Let us refer this region to as  $\mathcal{R}_{\mbox{\tiny TW-CIB}}^{\mbox{\tiny ED}}(1/2)$. Secondly, we will consider the more interesting problem in which $X_2$ is not known at encoder 1 and $X_1$ is not known at encoder 2. This is reminiscent of a noisy rate-distortion problem with side information only at the decoder. We refer to this region to as $\mathcal{R}_{\mbox{\tiny TW-CIB}}^{\mbox{\tiny D}}(1/2)$. Notice that in these two regions there is not interaction between the encoders. In the first case, interaction is not needed because each node has full knowledge of the side information of the other node.  In the second case, we neglect any interaction. Encoder 1 sends its description to decoder 2 who uses its side information $X_2$ for decoding. Similar, and without consider the previous description received from node 1, encoder 2 sends its own description to decoder 1 who recover it with its side information $X_1$. It is clear that we have the following:
\begin{equation*}
\mathcal{R}_{\mbox{\tiny TW-CIB}}^{\mbox{\tiny D}}(1/2)\subseteq\mathcal{R}_{\mbox{\tiny TW-CIB}}\subseteq\mathcal{R}_{\mbox{\tiny TW-CIB}}^{\mbox{\tiny ED}}(1/2).
\end{equation*}
As a consequence, the existent gap  between $\mathcal{R}_{\mbox{\tiny TW-CIB}}^{\mbox{\tiny ED}}(1/2)$ and  $\mathcal{R}_{\mbox{\tiny TW-CIB}}^{\mbox{\tiny D}}(1/2)$ can be thought to be an upper bound to the potential gain to be obtained from multiple interactions. In more specific terms, each of the above regions can be characterized by two \emph{relevance-rate} functions (one for each encoder-decoder pair). For instance, for the encoder 1-decoder 2 pair, we have:
\begin{IEEEeqnarray*}{rCl}
\mu_{\mbox{\tiny TW-CIB}}^{\mbox{\tiny ED}}(R_1)&=&\sup{\Big\{\mu_2:(R_1,\mu_2)\in\mathcal{R}_{\mbox{\tiny TW-CIB}}^{\mbox{\tiny ED}}(1/2)\Big\}},\\
\mu_{\mbox{\tiny TW-CIB}}^{\mbox{\tiny D}}(R_1)&=&\sup{\Big\{\mu_2:(R_1,\mu_2)\in\mathcal{R}_{\mbox{\tiny TW-CIB}}^{\mbox{\tiny D}}(1/2)\Big\}}.
\end{IEEEeqnarray*}
Similar definitions are valid for the relevance-rate functions $\mu_{\mbox{\tiny TW-CIB}}^{\mbox{\tiny ED}}(R_2)$, $\mu_{\mbox{\tiny TW-CIB}}^{\mbox{\tiny D}}(R_2)$ corresponding to the encoder 2-decoder 1 pair. It is also clear that as the encoding and decoding of the encoders and decoders are decoupled, a full characterization of these functions for the encoder 1-decoder 2 pair also leads to the full characterization of the functions for the other pair. These functions which are concave (see Appendix~\ref{app:mu_D}) are to be computed when $(X_1,X_2,Y_1,Y_2)$ satisfy  $(X_1,X_2, Y_1,Y_2)\sim \mbox{Bern}(1/2)$ and subject to $Y_1\mkv X_2\mkv X_1\mkv Y_2$. This implies that $X_1=X_2\oplus Z$ with $Z\sim\mbox{Bern}(q)$, $q\in(0,1/2)$, $Z\bot X_2$,  $Y_2=X_1\oplus W_1$ and $Y_1=X_2\oplus W_2$ with $W_i\sim\mbox{Bern}(p_i)$, $p_i\in(0,1/2)$, $W_i\bot (X_1,X_2)$ for $i=1,2$. In the following, we will assume that $p_1=p_2$. In this way the above relevance-rate functions for both pairs of encoders and decoders are the same and we can work with only one encoder-decoder pair satisfying  $X_2\mkv X_1\mkv Y$, where the decoder has access to $X_2$ and wishes  to learn $Y$. With this in mind, we begin with the characterization of $\mu_{\mbox{\tiny TW-CIB}}^{\mbox{\tiny ED}}(R)$. We have the following result.

\begin{theorem}[Relevance-rate function for binary sources with side information to the encoder and the decoder] 
Consider random binary sources  $(X_1,X_2,Y)\sim \mbox{Bern}(1/2)$ with $X_2\mkv X_1\mkv Y$ such that  $X_1=X_2\oplus Z$ with $Z\sim\mbox{Bern}(q)$, $q\in(0,1/2)$, $Z\bot X_2$ and $Y=X_1\oplus W$ with $W\sim\mbox{Bern}(p)$, $p\in(0,1/2)$, $W\bot (X_1,X_2)$. The relevance-rate function $\mu_{\mbox{\tiny TW-CIB}}^{\mbox{\tiny ED}}(R)$ can be put as:
\begin{equation*}
\mu_{\mbox{\tiny TW-CIB}}^{\mbox{\tiny ED}}(R)=1-h_2\left(h^{-1}_2\left(\left[h_2(q)-R\right]^{+}\right)\ast p\right).
\end{equation*}
\label{theo:binary_ED}
\end{theorem}
\begin{IEEEproof}
For the converse, we can without loss of generality begin from a single letter description. If $(R,\mu)$ is achievable, it is clear that there exists $U$ such that $U\mkv (X_1,X_2)\mkv Y$ and
\begin{equation*}
R\geq I(X_1;U|X_2)\ ,\ \ \mu\leq I(Y;UX_2).
\end{equation*}
 is straightforward to obtain:
\begin{equation*}
H(X_1|X_2U)\geq \left[h_2(q)-R\right]^{+},\ \ \ \mu\leq 1-H(Y|X_2U).
\end{equation*}
As $Y=X_1\oplus W$ with $W\sim\mbox{Bern}(p)$ and $W\bot (X_1,X_2)$ it is clear that $W\bot (U,X_1)$. This allows us to use Mrs. Gerber lemma \cite{wyner_ziv73} to obtain:
\begin{align*}
H(Y|X_2U)&\geq h_2\left(h_2^{-1}\left(H(X_1|X_2U)\right)*p\right)\\
&\geq h_2\left(h_2^{-1}\left(\left[h_2(q)-R\right]^{+}\right)\ast p\right),
\end{align*}
which implies 
\begin{equation*}
\mu_{\mbox{\tiny TW-CIB}}^{\mbox{\tiny ED}}(R)\leq 1-h_2\left(h^{-1}_2\left(\left[h_2(q)-R\right]^{+}\right)\ast p\right).
\end{equation*}
The achievability is straightforward choosing $U=U_0\mathds{1}\left\{X_2=0\right\}+U_1\mathds{1}\left\{X_2=1\right\}$,
where $U_i$, $i=0,1$ are binary random variables which are schematized in Fig. \ref{fig:U_ED}
\begin{figure}[t]
	\centering{\includegraphics[width=0.45\textwidth]{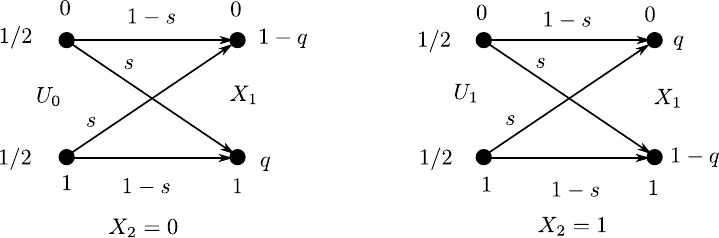}}
	\caption{Optimal choice of the random variable $U$  exhausting $\mu_{\mbox{\tiny TW-CIB}}^{\mbox{\tiny ED}}(R)$.}
\label{fig:U_ED}
\end{figure}
and the value of $s$ is given by $
s=h_2^{-1}\left(\left[h_2(q)-R\right]^{+}\right).$
\end{IEEEproof}
Now we consider the problem of obtaining $\mu_{\mbox{\tiny TW-CIB}}^{\mbox{\tiny D}}(R)$. Unfortunately in this case, as $U$ should depend only on $X_1$ (and not on $X_2$) we cannot apply Mrs. Gerber Lemma to obtain a tight upper bound to $\mu_{\mbox{\tiny TW-CIB}}^{\mbox{\tiny D}}(R)$. The converse and achievability in this case are more involved requiring the use of convex analysis. The following theorem provides the characterization of $\mu_{\mbox{\tiny TW-CIB}}^{\mbox{\tiny D}}(R)$ and its proof is deferred to Appendix~\ref{app:mu_D}.

\begin{theorem}[Relevance-rate function for binary sources with side information only to the decoder] 
\label{theo:mu_D}
Consider random Binary sources  $(X_1,X_2,Y)\sim \mbox{Bern}(1/2)$ with $X_2\mkv X_1\mkv Y$ such that  $X_1=X_2\oplus Z$ with $Z\sim\mbox{Bern}(q)$, $q\in(0,1/2)$, $Z\bot X_2$ and $Y=X_1\oplus W$ with $W\sim\mbox{Bern}(p)$, $p\in(0,1/2)$, $W\bot (X_1,X_2)$. The relevance-rate function $\mu_{\mbox{\tiny TW-CIB}}^{\mbox{\tiny D}}(R)$ can be put as:
\begin{align*}
&\mu_{\mbox{\tiny TW-CIB}}^{\mbox{\tiny D}}(R)=\\
&\left\{\begin{array}{cc}
1-h_2(p\ast q)+\displaystyle \frac{f\left(g^{-1}(R_c)\right)}{R_c}R & 0\leq R\leq R_c,\\
1-h_2(p\ast q)+f\left(g^{-1}(R)\right) & R_c<R\leq h_2(q),\\
1-h_2(p) & R>h_2(q),
\end{array}\right.\nonumber
\end{align*}
where $R_{c}$ is given by:
\begin{equation*}
\frac{f'\left(g^{-1}(R_c)\right)}{g'\left(g^{-1}(R_c)\right)}=\frac{f\left(g^{-1}(R_c)\right)}{R_c},
\end{equation*}
and $g(\cdot)$ and $f(\cdot)$ are defined in (\ref{eq:g_fun}) and (\ref{eq:f_fun}).
\end{theorem}

It is important to emphasize, as it is discussed in Appendix~\ref{app:mu_D},  that this region is achieved by time-sharing. This is similar to the Wyner-Ziv problem for binary sources \cite{wyner_ziv76}.
\begin{remark}
The proof in Appendix~\ref{app:mu_D} can be generalized to the cases in which $X_1$, $X_2$ and $Y$ are \emph{Bernoulli} random variables with other parameters than $1/2$. Moreover, a similar (but even more cumbersome) analysis can be carried over for arbitrary discrete random sources that satisfy the above Markov chain.
\end{remark}
\begin{figure}[t]
	\centering{\includegraphics[width=0.48\textwidth]{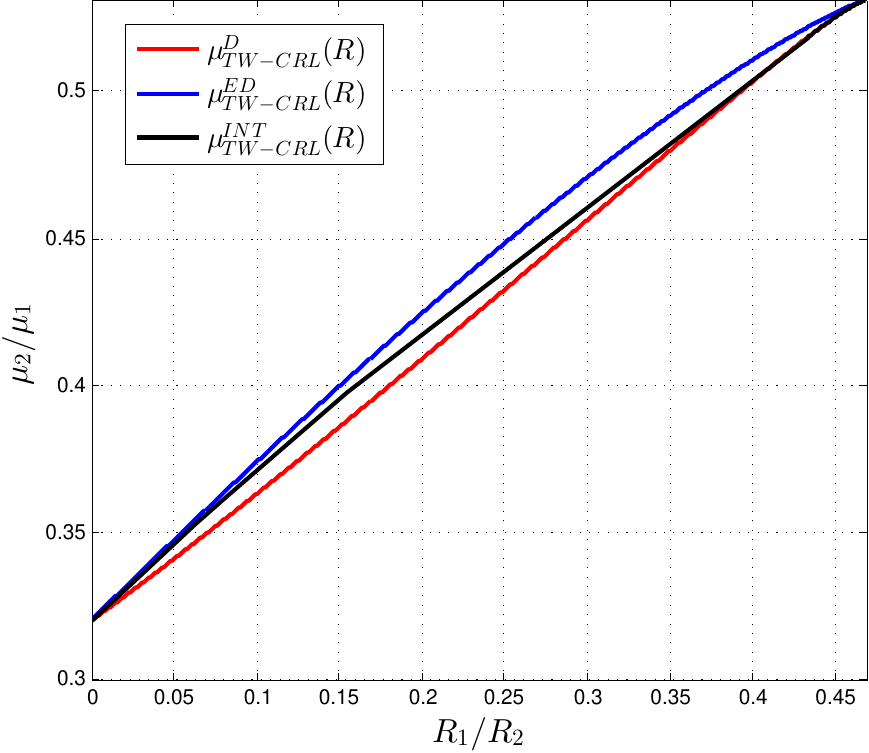}}
	\caption{$\mu_{\mbox{\tiny TW-CIB}}^{\mbox{\tiny D}}(R_1)$,$\mu_{\mbox{\tiny TW-CIB}}^{\mbox{\tiny D}}(R_2)$, $\mu_{\mbox{\tiny TW-CIB}}^{\mbox{\tiny ED}}(R_1)$, $\mu_{\mbox{\tiny TW-CIB}}^{\mbox{\tiny ED}}(R_2)$ and $\mu_{\mbox{\tiny TW-CIB}}^{\mbox{\tiny INT}}(R_2)$ as functions of $R_1$ and $R_2$ respectively and when $p_1=p_2=0.1$ and $q=0.1$.}
\label{fig:mu_binary_comp_2}
\end{figure}

In order to compare these two extreme cases, where there is no interaction with an example where there is some coupling between the two pairs of encoder-decoder, we study the full interactive case with one round for random binary sources that satisfy $Y_1\mkv X_2\mkv X_1\mkv Y_2$ with $p_1=p_2$. Assume that in the first half round, encoder 1 transmits a description to decoder 2, who wants to learn hidden variable $Y_2$. After that, encoder 2 sends a description to decoder 1. In this case, and according to Theorem~\ref{theo:p1}, encoder 2 should transmit with rates and relevances satisfying:
\begin{align*}
R_1&\geq I(X_1;V_1|X_2),\\
R_2&\geq I(X_2;V_2|X_1V_1),\\
\mu_1&\leq I(Y_1;V_1V_2X_1)=I(Y_1;V_2X_1)\\
\mu_2&\leq I(Y_2;V_1X_2),
\end{align*}
where $V_1\mkv X_1\mkv (X_2,Y_1,Y_2)$ and $V_2\mkv (V_1,X_2) \mkv (X_1,Y_1,Y_2)$. It is clear that the tradeoff between $R_1$ and $\mu_2$ is given by the function $\mu_{\mbox{\tiny TW-CIB}}^{\mbox{\tiny D}}(R_1)$, with the optimal choice of $V_1$ given in Appendix~\ref{app:mu_D}. Regarding  the choice of $V_2$, we should consider the following problem (with $V_1$ fixed with the mentioned optimal choice):
\begin{equation}
\max_{p(v_2|x_2v_1)}I(Y_1;V_2X_1)\ \ \mbox{s.t.}\ \ R_2\geq I(X_2;V_2|X_1V_1).
\label{eq:second_problem}
\end{equation}
This problem is similar to the one considered for $\mu_{\mbox{\tiny TW-CIB}}^{\mbox{\tiny D}}(R_1)$. It is however, a little more subtle and difficult to solve. It can be seen that it corresponds to a source coding problem where both encoder and decoder have side information ($V_1$ and $X_1$ respectively), but the side information of the encoder is degraded with respect to that of the decoder. We simply evaluated the resulting rate region by  numerically generating random probability distributions of $V_2$ with cardinalities no lower than $7$, as indicated in Theorem~\ref{theo:p1}, for each value of $R_2$. Taking the maximum\footnote{It is clear that with this approach we cannot guarantee the  solution to the optimization problem in~\eqref{eq:second_problem}, but that it is not necessary as we only aim at generating an achievable region which shows that interaction helps.} of the generated value of $\mu_1$ for each value of $R_2$ and considering the concave envelope of the resulting curve (allowing for time-sharing between different points in the curve), we obtained the function $\mu_{\mbox{\tiny TW-CIB}}^{\mbox{\tiny INT}}(R_2)$ which is clearly achievable. In Fig.~\ref{fig:mu_binary_comp_2}, we plot this function along with $\mu_{\mbox{\tiny TW-CIB}}^{\mbox{\tiny D}}(R_1)$, $\mu_{\mbox{\tiny TW-CIB}}^{\mbox{\tiny D}}(R_2)$ (plotted as only one curve, as these are equivalent) and  $\mu_{\mbox{\tiny TW-CIB}}^{\mbox{\tiny ED}}(R_1)$, $\mu_{\mbox{\tiny TW-CIB}}^{\mbox{\tiny ED}}(R_2)$ (again, plotted together because they are equivalent). It is seen that, in contrast with the corresponding Gaussian TW-CIB case analyzed in Section~\ref{sec:gaussian} where interaction does not help and both encoder-decoder pairs operate in a complete decoupled manner, interaction clearly helps in this binary setting.  Actually, during the second half round, the first description sent by encoder 1 is useful for encoder 2 and decoder 1 in the task of learning $Y_1$.

\section{Summary and Discussion}
\label{sec:summary}

We investigated a multi-terminal collaborative source coding problem with a non-additive logarithmic distortion. This work intended to characterize tradeoffs between rates of complexity and relevance to the source-coding problem of cooperatively extracting information about hidden variables from some observed and physically distributed ones. Two different scenarios are distinguished: the so-called Two-way Collaborative Information Bottleneck (TW-CIB)  and the Collaborative Distributed Information Bottleneck (CDIB). These problems differ from each other in the fact that the decoder may or may not be one of the encoders, necessitating  fundamentally different approaches.  Inner and outer bounds to the complexity-relevance region of these problems are derived and optimality is characterized for several cases of interest. 

Specific applications of our results to binary symmetric and Gaussian statistical models were also considered and optimality is characterized for most of the cases. These results show that cooperation does not improve the rates of relevance in presence of  Gaussian statistical models in most cases. This can be explained from the well-known result by Wyner-Ziv~\cite{wyner_ziv76} which implies that side information at the encoder does not improve the quadratic distortion in presence of Gaussian sources. In contrast, we have shown that cooperation clearly helps in the TW-CIB scenario. In particular, the converse to the complexity-relevance region of the binary model appears to be rather involved and required the use of tools of convex analysis. It will be the purpose of future work to study the binary source model within  the CDIB framework for which we conjecture that cooperation also helps. 


 


\appendices

\section{Strongly typical sequences and related results}
  \label{app:strongly}
  In this appendix we introduce standard notions in information theory but suited for the mathematical developments and proof needed in this work. The results presented can be easily derived from the standard formulations provided in~\cite{ELGamal-Kim-book} and~\cite{Csiszar81}. Be $\mathcal{X}$ and $\mathcal{Y} $  finite alphabets and $(x^n,y^n)\in\mathcal{X}^n\times \mathcal{Y}^n$.  With $\mathcal{P}(\mathcal{X}\times \mathcal{Y})$ we denote the set of all probability distributions on $\mathcal{X}\times\mathcal{Y}$. We define the \emph{strongly $\delta$-typical} sets as:
 
 \begin{definition}[Strongly typical set]
 Consider $p\in\mathcal{P}(\mathcal{X})$ and $\delta>0$. We say that $x^n\in\mathcal{X}^n$ is $p\delta$- strongly typical if $x^n\in\mathcal{T}_{[p]\delta}^n$ with:
 \begin{align*}
 &\mathcal{T}_{[p]\delta}^n=\left\{x^n\in\mathcal{X}^n:\Big|\frac{N(a|x^n)}{n}-p(a)\Big|\leq\frac{\delta}{|\mathcal{X}|},\right.\nonumber\\ &\qquad\qquad\left.\forall a\in\mathcal{X}\ \mbox{such that}\ p(a)\neq 0\right\}\ ,
 \end{align*}
 where $N(a|x^n)$ denotes de number of occurrences of $a\in\mathcal{X}$ in $x^n$ and $p\in\mathcal{P}(\mathcal{X})$.  When $X\sim p_X$ we can denote the corresponding set of strongly typical sequences as $\mathcal{T}_{[X]\delta}^n$. 
  \end{definition}
 Similarly, given $p_{XY}\in\mathcal{P}\left(\mathcal{X}\times\mathcal{Y}\right)$ we can construct the set of $\delta$-jointly typical sequences as:
  \begin{align*}
  &\mathcal{T}_{[XY]\delta}^n=\left\{(x^n,y^n)\in\mathcal{X}^n\times\mathcal{Y}^n:\right.\nonumber\\
  &\qquad\Big|\frac{N(a,b|x^n,y^n)}{n}-p_{XY}(a,b)\Big|\leq\frac{\delta}{|\mathcal{X}||\mathcal{Y}|},\nonumber\\
  &\qquad\left.\ \forall (a,b)\in\mathcal{X}\times\mathcal{Y}\ \mbox{such that}\ p_{XY}(a,b)\neq 0\right\}\ .
  \end{align*}
 We also define the \emph{conditional} typical sequences. In precise terms, given $x^n\in\mathcal{X}^n$ we consider the set:
 \begin{align*}
&\mathcal{T}_{[Y|X]\delta}^n(x^n)=\Big\{y^n\in\mathcal{Y}^n:\nonumber\\
&\qquad\Big|\frac{N(a,b|x^n,y^n)}{n}-p_{XY}(a,b)\Big|\leq\frac{\delta}{|\mathcal{X}||\mathcal{Y}|},\nonumber\\
&\qquad\forall (a,b)\in\mathcal{X}\times\mathcal{Y}\ \mbox{such that}\ p_{XY}(a,b)\neq 0\Big\}\ .
 \end{align*} 
 Notice that we the following is an alternative writing of this set:
 \begin{equation*} 
 \mathcal{T}_{[Y|X]\delta}^n(x^n)=\left\{y^n\in\mathcal{Y}^n:(x^n,y^n)\in\mathcal{T}_{[XY]\delta}^n\right\} \ .
 \end{equation*}
 
 We have several useful and standard lemmas, which will be presented without proof:
 
 
 \begin{lemma}[Conditional typicality lemma~\cite{Csiszar81}]
 Consider de product measure $\prod_{t=1}^np_{XY}(x_t,y_t)$. Using that measure, we have the following 
 \begin{equation*}
 \operatorname{Pr}\left\{\mathcal{T}_{[X]\epsilon}^n\right\}\geq 1-\mathcal{O}\left(c_1^{-nf(\epsilon)}\right),\ \ c_1>1
 \end{equation*}
 where $f(\epsilon)\rightarrow 0$ when $\epsilon\rightarrow 0$. In addition, for every  $x^n\in\mathcal{T}^n_{[X]\epsilon'}$ with $\epsilon'<\frac{\epsilon}{|\mathcal{Y}|}$ we have:
 \begin{equation*}
\operatorname{Pr}\left\{\mathcal{T}_{[Y|X]\epsilon}^n(x^n)|x^n\right\}\geq 1-\mathcal{O}\left(c_2^{-ng(\epsilon,\epsilon')}\right),\ \ c_2>1
 \end{equation*}
 where $g(\epsilon,\epsilon')\rightarrow 0$ when $\epsilon,\epsilon'\rightarrow 0$.
 \label{lemma:prob_lim}
 \end{lemma}

\begin{lemma}[Covering Lemma~\cite{ELGamal-Kim-book}]
\label{lemma:covering}
Be $(U,V,X)\sim p_{UVX}$ and $(x^n,u^n)\in\mathcal{T}_{[XU]\epsilon'}^n$, $\epsilon'<\frac{\epsilon}{|\mathcal{V}|}$ and $\epsilon<\epsilon''$. Consider also $\left\{V^n(m)\right\}_{m=1}^{2^{nR}}$ random vectors  which are independently generated according to $\frac{\mathds{1}\left\{v^n\in\mathcal{T}_{[V|U]\epsilon''}^n(u^n)\right\}}{|\mathcal{T}_{[V|U]\epsilon''}(u^n)|}$. Then:
\begin{equation*}
\mbox{Pr}\left\{V^n(m)\notin\mathcal{T}_{[V|UX]\epsilon}^n(x^n,u^n)\ \mbox{for all}\ m\right\}\xrightarrow[n\rightarrow\infty]{}0
\end{equation*}
uniformly for every  $(x^n,u^n)\in\mathcal{T}^n_{[XU]\epsilon'}$  if: 
\begin{equation}
R>I\left(V;X|U\right)+\delta(\epsilon,\epsilon',\epsilon'',n)
\label{eq:covering_lemma}
\end{equation}
where $\delta(\epsilon,\epsilon',\epsilon'',n)\rightarrow 0$ when $\epsilon,\epsilon',\epsilon''\rightarrow 0$ and $n\rightarrow \infty$.
\end{lemma}
\begin{corollary}
\label{coro:covering}
Assume the conditions in Lemma~\ref{lemma:covering}, and  also:
\begin{equation*}
\mbox{Pr}\left\{\left(X^n,U^n\right)\in\mathcal{T}^n_{[XU]\epsilon'}\right\}\xrightarrow[n\rightarrow\infty]{}1\ .
\end{equation*}
Then:
\begin{equation*}
\mbox{Pr}\left\{\left(U^n,X^n,V^n(m))\right)\notin\mathcal{T}_{[UXV]\epsilon}^n\ \mbox{for all}\ m\right\}\xrightarrow[n\rightarrow\infty]{}0
\end{equation*}
when (\ref{eq:covering_lemma}) is satisfied.
\end{corollary}

\begin{lemma}[Packing Lemma~\cite{ELGamal-Kim-book}]
\label{lemma:packing}
Be $(U_1U_2WV_1V_2X)\sim p_{U_1U_2WV_1V_2X}$, $(x^n,w^n,v_1^n,v_2^n)\in\mathcal{T}^n_{[XWV_1V_2]\epsilon'}$ and  $\epsilon'<\frac{\epsilon}{|\mathcal{U}_1||\mathcal{U}_2|}$ and $\epsilon<\min{\left\{\epsilon_1,\epsilon_2\right\}}$. Consider random vectors $\left\{U_1^n(m_1)\right\}_{m_1=1}^{\mathcal{A}_1}$ and $\left\{U_2^n(m_2)\right\}_{m_2=1}^{\mathcal{A}_2}$ which are independently generated according to 
\begin{equation*}
\frac{\mathds{1}\left\{u_i^n\in\mathcal{T}_{[U_i|V_iW]\epsilon_i}^n(v_i^n,w^n)\right\}}{|\mathcal{T}_{[U_i|V_iW]\epsilon_i}(w^n,v_i^n)|},\ i=1,2\ ,
\end{equation*}
and $\mathcal{A}_1,\mathcal{A}_2$ are positive random variables independent of everything else. Then
\begin{align*}
&\mbox{Pr}\left\{\left(U_1^n(m_1),U_2^n(m_2)\right)\in\mathcal{T}_{[U_1U_1|XWV_1V_2]\epsilon}^n(x^n,w^n,v_1^n,v_2^n)\right.\nonumber\\
&\qquad\left.\mbox{for some}\ (m_1,m_2)\right\}\xrightarrow[n\rightarrow\infty]{}0
\end{align*}
uniformly for every  $(x^n,w^n,v_1^n,v_2^n)\in\mathcal{T}^n_{[XWV_1V_2]\epsilon'}$  provided that: 

\begin{align}
\frac{\log{\mathbb{E}\left[\mathcal{A}_1\mathcal{A}_2\right]}}{n}<&I\left(U_1;XV_2U_2|WV_1\right)+I\left(U_2;XV_1U_1|WV_2\right)\nonumber\\
&-I\left(U_1;U_2|XWV_1V_2\right)-\delta
\label{eq:packing_lemma}
\end{align}
where $\delta\triangleq \delta(\epsilon,\epsilon',\epsilon_1,\epsilon_2,n)\rightarrow 0$ when $\epsilon,\epsilon',\epsilon_1,\epsilon_2\rightarrow 0$ and $n\rightarrow \infty$.
\end{lemma}

\begin{corollary}
\label{coro:packing}
Assume the conditions in Lemma~\ref{lemma:packing}, and  also:
\begin{equation*}
\mbox{Pr}\left\{\left(X^n,W^n,V_1^n,V_2^n\right)\in\mathcal{T}^n_{[XWV_1V_2]\epsilon'}\right\}\xrightarrow[n\rightarrow\infty]{}1.
\end{equation*}
Then:
\begin{align*}
&\mbox{Pr}\left\{\left(U_1^n(m_1),U_2^n(m_2),X^n,W^n,V_1^n,V_2^n)\right)\in\mathcal{T}_{[U_1U_1XWV_1V_2]\epsilon}^n\right.\nonumber\\
&\left.\qquad\mbox{for some}\ (m_1,m_2)\right\}\xrightarrow[n\rightarrow\infty]{}0
\end{align*}
when \eqref{eq:packing_lemma} is satisfied.
\end{corollary}

\begin{lemma}[Generalized Markov Lemma~\cite{GML_ours_2014} ]
Consider a pmf $p_{UXY}$ belonging to $\mathcal{P}\left(\mathcal{X}\times\mathcal{Y}\times\mathcal{U}\right)$ and that satisfies de following: $Y\mkv X\mkv U$.\par
Consider $(x^n,y^n)\in\mathcal{T}^n_{[XY]\epsilon'}$ and random vectors $U^n$  generated according to:

\begin{align}
\mbox{Pr}&\left\{U^n=u^n\big|x^n,y^n,U^n\in\mathcal{T}^n_{[U|X]\epsilon''}(x^n)\right\}\nonumber\\
&\qquad=\frac{\mathds{1}\left\{u^n\in\mathcal{T}_{[U|X]\epsilon''}^n(x^n)\right\}}{|\mathcal{T}_{[U|X]\epsilon''}^n(x^n)|}.\label{eq:u_dist}
\end{align}
For sufficiently small $\epsilon,\epsilon',\epsilon''$ the following holds uniformly for every  $(x^n,y^n)\in\mathcal{T}^n_{[XY]\epsilon'}$:

\begin{align*}
\mbox{Pr}&\left\{U^n\notin\mathcal{T}^n_{[U|XY]\epsilon}(x^n,y^n)\big|x^n,y^n,U^n\in\mathcal{T}^n_{[U|X]\epsilon''}(x^n)\right\}\nonumber\\
&\qquad=\mathcal{O}\left(c^{-n}\right)
\end{align*}
where $c>1$.
  \label{lemma:markov}
  \end{lemma}

\begin{corollary}
\label{coro:markov}
Assume the conditions in Lemma~\ref{lemma:markov}, and also:
\begin{equation*}
\mbox{Pr}\left\{\left(X^n,Y^n\right)\in\mathcal{T}^n_{[XY]\epsilon'}\right\}\xrightarrow[n\rightarrow\infty]{}1
\end{equation*}
and that uniformly for every $(x^n,y^n)\in\mathcal{T}^n_{[XY]\epsilon'}$:
\begin{equation*}
\mbox{Pr}\left\{U^n\notin\mathcal{T}^n_{[U|X]\epsilon''}(x^n)\big| x^n,y^n\right\}\xrightarrow[n\rightarrow\infty]{}0
\end{equation*}
we obtain:
\begin{equation*}
\mbox{Pr}\left\{(U^n,X^n,Y^n)\in\mathcal{T}^n_{[UXY]\epsilon}\right\}\xrightarrow[n\rightarrow\infty]{}1\ .
\end{equation*}
\end{corollary}

We next present a result which will be very useful to us. In order to use the Markov lemma we need to show that the descriptions induced by the encoding procedure in each encoder satisfies (\ref{eq:u_dist}). A proof of this result can be found in \cite{ours_2015}
\begin{lemma}[Encoding induced distribution]   \label{lemma:induced}
 Consider a pmf $p_{UXW}$ belonging to $\mathcal{P}\left(\mathcal{U}\times\mathcal{X}\times\mathcal{W}\right)$ and $\epsilon'\geq\epsilon$.  Be $\left\{U^n(m)\right\}_{m=1}^{S}$ random vectors independently generated according to 
\begin{equation*}
 \frac{\mathds{1}\left\{u^n\in\mathcal{T}_{[U|W]\epsilon'}^n(w^n)\right\}}{|\mathcal{T}_{[U|W]\epsilon'}(w^n)|}
\end{equation*}
 and where $(W^n,X^n)$ are generated with an arbitrary distribution. Once these vectors are generated, and given $x^n$ and $w^n$, we choose one of them if:
  \begin{equation*}
  \left(u^n(m),w^n,x^n\right)\in\mathcal{T}^n_{[UWX]\epsilon},\ \mbox{for some}\ m\in[1:S] \ .\end{equation*}
 If there are various vectors $u^n$ that satisfies this we choose the one with smallest index. If there are none we choose an arbitrary one. Let $M$ denote the index chosen. 
 Then we have that:
  \begin{align*}
 \mbox{Pr}&\left\{U^n(M)=u^n\big|x^n,w^n,U^n(M)\in\mathcal{T}^n_{[U|XW]\epsilon}(x^n,w^n)\right\}\nonumber\\
 &\qquad=
 \frac{\mathds{1}\left\{u^n\in\mathcal{T}_{[U|XW]\epsilon}^n(x^n,w^n)\right\}}{|\mathcal{T}_{[U|XW]\epsilon}(x^n,w^n)|} \ .
 \end{align*}
   \end{lemma}

\section{Achievability proofs}
\label{sec:achiev}

We will begin with the proof of Theorem~\ref{theo:inner_bound}. The proof of Theorem~\ref{theo:p1} can be seen as a simple extension with some minor differences to be discussed next.

\subsection{Proof of Theorem~\ref{theo:inner_bound}}

Let us describe the coding generation, encoding and decoding procedures. We will consider the following notation. With $m_{i,l}$ we will generically denote the indices used for the descriptions $V^n_{i,l}$ generated at encoder $i$ at round $l$.  With $M_{i,l}$ we will denote the actual index corresponding to the actual description $V^n_{i,l}$ generated at encoder $i$ at round $l$. With $m_{W_{i,l}}$ we denote the indices used for the sets of descriptions generated just before encoder $i$ generated its own description at round $l$ and with $M_{W_{i,l}}$ the actual corresponding indices generated. Similarly, $p_{i,l}$ will denote the bin indices used at encoder $i$ at round $l$ and $P_{i,l}$ will denote the actual bin index generated at encoder $i$ at round $l$. With $\hat{M}_{i,l}(j)$ where $i\neq j$ we denote the  estimation at encoder $j$ of the actual index generated at encoder $i$ at round $l$, where $i\in\{1,2\}$ and $j\in\{1,2,3\}$. We will fix codeword length $n$ and a distribution which satisfies the Markov chains \eqref{eq:markv1-p2} and \eqref{eq:markv2-p2}. We will describe the coding procedure to be used.

\textit{Coding generation}: Consider the round $l\in [1:K]$. For each $m_{W_{1,l}}$,  we generate $2^{n\hat{R}_{1,l}}$ independent and identically distributed $n$-length codewords $V_{1,l}^n(m_{1,l},m_{W_{1,l}})$ according to:
\begin{align*}
\mbox{Pr}&\left\{V^n_{1,l}(m_{1,l},m_{W_{1,l}})=v_{1,l}^n\right\}=\\
&\qquad\frac{\mathds{1}\left\{v^n_{1,l}\in\mathcal{T}_{[V_{1,l}|W_{1,l}]\epsilon(1,l)}^n\left(w_{1,l}^n\right)\right\}}{\left|\mathcal{T}_{[V_{1,l}|W_{1,l}]\epsilon(1,l)}^n\left(w_{1,l}^n\right)\right|},\  \epsilon(1,l)>0\nonumber
\end{align*}
where $m_{1,l}\in[1:2^{n\hat{R}_{1,l}}]$ and let $m_{W_{1,l}}$ denote the indices of the descriptions $W^n_{1,l}$ generated at encoders 1 and 2 in the past rounds $t\in[1:l-1]$ as explained above. For example, $m_{W_{1,l}}=\left\{m_{1,t}, m_{2,t}\right\}_{t=1}^{l-1}$. With $w_{1,l}^n$ we denote the set of $n$-length  codewords (which are realizations of $W^n_{1,l}$) from previous rounds corresponding to the indices  $m_{W_{1,l}}$. Constant $ \epsilon(1,l)$ is  chosen to be sufficiently small. It is clear that there exists $2^{n(\hat{R}_{1,l}+\sum_{k=1}^{l-1}\hat{R}_{1,k}+\hat{R}_{2,k})}$ $n$-length codewords   $V^n_{1,l}(m_{1,l},m_{W_{1,l}})$. These codewords are distributed independently and uniformly over $2^{nR_{1,l}}$ bins which are denoted as $\mathcal{B}_{1,l}(p_{1,l})$ with $p_{1,l}\in[1:2^{nR_{1,l}}]$. Similarly, for encoder 2 and for each $m_{W_{2,l}}$ we generate $2^{n\hat{R}_{2,l}}$ independent and identically distributed $n$-length codewords $V_{1,l}^n(m_{1,l},m_{W_{1,l}})$ according to:
  \begin{align*}
\mbox{Pr}&\left\{V^n_{2,l}(m_{1,l},m_{W_{2,l}})=v_{2,l}^n\right\}=\\
&\qquad\frac{\mathds{1}\left\{v^n_{2,l}\in\mathcal{T}_{[V_{2,l}|W_{2,l}]\epsilon(2,l)}^n\left(w_{2,l}^n\right)\right\}}{\left|\mathcal{T}_{[V_{2,l}|W_{2,l}]\epsilon(2,l)}^n\left(w_{2,l}^n\right)\right|},\  \epsilon(2,l)>0\nonumber
  \end{align*}
These $2^{n(\hat{R}_{2,l}+\hat{R}_{1,l}+\sum_{k=1}^{l-1}\hat{R}_{1,k}+\hat{R}_{2,k})}$ $n$-length codewords  are distributed independently and uniformly over $2^{nR_{2,l}}$ bins which are denoted as $\mathcal{B}_{2,l}(p_{2,l})$ with $p_{2,l}\in[1:2^{nR_{2,l}}]$. It is clear that we should impose that
\begin{IEEEeqnarray}{rCl}
R_{1,l}& <& \hat{R}_{1,l}+\sum_{k=1}^{l-1}\hat{R}_{1,k}+\hat{R}_{2,k},\nonumber\\
R_{2,l}&<& \hat{R}_{2,l}+\hat{R}_{1,l}+\sum_{k=1}^{l-1}\hat{R}_{1,k}+\hat{R}_{2,k},
\label{eq:first_rate_cond}
\end{IEEEeqnarray}
for each $l\in[1:K]$. 

This procedure for the codebooks generation is done sequentially beginning at encoder 1 and round 1 and terminated at encoder 2 and round $K$. After that, the codebooks are revealed to all parties. 

\textit{Encoding}: Consider encoder $1$ at round $l\in [1:K]$. Upon observing $x_1^n$ and given all of its encoding and decoding history up to round $l$, encoder $1$ first looks for a codeword $v_{1,l}^n(m_{1,l},\hat{m}_{W_{1,l}}(1))$ such that $\left(x_1^n,w_{1,l}^n(\hat{m}_{W_{1,l}}(1)),v_{1,l}^n(m_{1,l},\hat{m}_{W_{1,l}}(1)) \right)
 \in\mathcal{T}^n_{[V_{1,l}X_1W_{1,l}]\epsilon_c(1,l)}$, where $\epsilon_c(1,l)>0$. Notice that some components in $\hat{m}_{W_{1,l}}(1)$  are generated at encoder 1 and are perfectly known. If more than one codeword satisfies this condition, then we choose the one with the smallest index. Otherwise, if no such codeword exists,  we choose an arbitrary index and declare an error. With the chosen index $m_{1,l}$, and with $\hat{m}_{W_{1,l}}(1)$, we determine the index $p_{1,l}$ of the bin $\mathcal{B}_{1,l}\big(p_{1,l}\big)$ to which $v_{1,l}^n(m_{1,l},\hat{m}_{W_{1,l}}(1))$ belongs. Then, the index $p_{1,l}$ is transmitted to encoder 2 and 3. Similarly, encoder 2 looks for a codeword d $v_{2,l}^n(m_{2,l},\hat{m}_{W_{2,l}}(2))$ such that $\left(x_2^n,w_{2,l}^n(\hat{m}_{W_{2,l}}(1)),v_{2,l}^n(m_{2,l},\hat{m}_{W_{2,l}}(2)) \right)
 \in\mathcal{T}^n_{[V_{2,l}X_2W_{2,l}]\epsilon_c(2,l)}$, where $\epsilon_c(2,l)>0$. If more than one codeword satisfies this condition, then we choose the one with the smallest index. Otherwise, if no such codeword exists,  we choose an arbitrary index and declare an error. With the chosen index $m_{2,l}$, and with $\hat{m}_{W_{2,l}}(2)$, we determine the index $p_{2,l}$ of the bin $\mathcal{B}_{2,l}\big(p_{2,l}\big)$ to which $v_{2,l}^n(m_{2,l},\hat{m}_{W_{2,l}}(2))$ belongs. Then, the index $p_{2,l}$ is transmitted to encoder 1 and the decoder. 

\textit{Decoding:}  At round $l\in[1:K]$ encoder 1, after receiving $p_{2,l-1}$ looks for $m_{2,l-1}$ such that $\left(x_1^n,w^n_{[2,l-1]}(\hat{m}_{W_{2,l-1}}(1)),v_{2,l-1}^n(m_{2,l-1},\hat{m}_{W_{2,l-1}}(1)) \right)
 \in\mathcal{T}^n_{[V_{2,l-1}X_1W_{2,l-1}]\epsilon_d(1,l)}$ and such that $(m_{2,l},\hat{m}_{W_{2,l-1}}(1)) \in\mathcal{B}(p_{2,l-1})$. If there are more than one pair of codewords, or none that satisfies this, we choose a predefined one and declare an error. Similarly, at round $l$ and after receiving $p_{1,l}$ encoder 2 looks for $m_{1,l}$ such that $\left(x_2^n,w_{1,l}^n(\hat{m}_{W_{1,l}}(1)),v_{1,l}^n(m_{1,l},\hat{m}_{W_{1,l}}(2)) \right)
 \in\mathcal{T}^n_{[V_{1,l}X_2W_{1,l}]\epsilon_d(2,l)}$ and such that $(m_{1,l},\hat{m}_{W_{1,l}}(2)) \in\mathcal{B}(p_{1,l})$. If there are more than one pair of codewords, or none that satisfies this, we choose a predefined one and declare an error. After all exchanges are done it is the turn of encoder 3 to recover the descriptions generated at encoder 1 and 2. After receiving $\left\{p_{1,l},p_{2,l}\right\}_{l=1}^K$, the encoder $3$ looks for a codeword $W^n_{1,K+1}(m_{W_{1,K+1}})$ such that $(m_{1,l},m_{W_{1,l}})\in\mathcal{B}(p_{1,l})$ and $(m_{2,l},m_{W_{2,l}})\in\mathcal{B}(p_{2,l})$ for all $l\in[1:K]$. Notice that as encoder 3 has no side information, it is not needed to employ joint decoding. It suffices to search over the codebooks and bins for appropriate indices. The coding guarantee that with high probability only one set of indices will be compatible with the above conditions.

We are now ready to analyze the error probability and relevance averaged over all random codes. We will explain, without a detailed mathematical treatment, the basic idea of the achievability for the case of only one round ($K=1$). Following this, the analysis for a generic $K$ would be done in precise and rigorous mathematical terms.

In the one round scenario, after observing $X_1^n$, node 1 choose $V_1^n(M_1)$ with $M_1\in[1:2^{n\hat{R}_1}]$ such that $(X_1^n,V_1^n(M_1))$ are typical. This would the case with high probability if
\begin{equation*}
\hat{R}_1>I(V_1;X_1)
\end{equation*}
The index $M_{1}$ of $V_1^n$ belongs to a given bin whose index ($P_1$) is sent to node 2 and 3.  The numbers of bins ($2^{nR_1}$, $R_1< \hat{R}_1$) in node 1 is chosen such that the use of  side information ($X_2^n$) in node 2 allows for the recovery of the index of $V_1^n(M_1)$. The joint-typicality decoding at node 2 would be successful with high probability if
\begin{equation*}
\hat{R}_1-R_1<I(V_1;X_2)
\end{equation*}
 Of course, it is not guaranteed that node 3 could recover that index because it does not have side information. In this way, the information sent by node 2 should provide something to be used by node 3 to recover not only the index generated at node 2 but also the index generated at node 1. First, node 2 choose $V_2^n(M_2,\hat{M}_1(2))$ with $M_2\in[1:2^{n\hat{R}_2}]$ such that  $(X_2^n,V_1^n(\hat{M}_1(2)),V_2^n(M_2,\hat{m}_1))$ is typical, where $\hat{M}_1(2)$ is estimation of $M_1$ at node 2 (which with probability close to one will be equal to the true $M_1$). In order to achieves this with high probability:
 \begin{equation*}
 \hat{R}_2>I(V_2;X_2|V_1)
 \end{equation*}
 After that, node 2 look for the bin index where {\bf both} $(\hat{M_1}(2),M_2)$ live ($P_2$) and send it to node 3. Notice that as explained before, at node 2 the bins contain all possible pairs $(m_1,m_2)$ (distributed in uniform fashion). This is the key fact. Node 2 bins both indices: the one recovered from node 1 and the one it generates. In this way an explicit cooperation is achieved between node 1 and 2 through binning in order to help the decoder in node 3 to recover both $M_1$ and $M_2$. Clearly, the number of bins in node 3 should satisfy:
 \begin{equation*}
R_2<\hat{R}_2 +\hat{R}_1.
 \end{equation*}
 Finally, node 3 should recover $M_1$ and $M_2$ from the bin indices $P_1$ and $P_2$. This is simply done by looking for $(m_1,m_2)$ such that $m_1\in\mathcal{B}(P_1)$ and $(m_1,m_2)\in\mathcal{B}(P_2)$ and $,V_1^n(m_1),V_2^n(m_2,m_1))$ are jointly typical. As the bins formations in node 1 and 2 are done with uniform distributions over the indices sets, the probability of failure of this procedure is shown to go to zero exponentially fast if:
  \begin{IEEEeqnarray*}{rCl}
\hat{R}_1&<&R_1+R_2,\\
\hat{R}_2&<&R_2,\\
\hat{R}_1+\hat{R}_2&<&R_1+R_2.
   \end{IEEEeqnarray*}
The mathematical details of the proof of this fact can be found  in appendix B in \cite{ours_2015} (setting $X_3=V_1=V_2=\varnothing$).  Eliminating $\hat{R}_1$ and $\hat{R}_2$ through a Fourier-Motzkin elimination procedure we obtain:
\begin{align*}
R_1\;&\geq\;I(X_1;V_1|X_2),\\
R_2\;&\geq\;I(X_2;V_2|V_1),\\ 
R_1+R_2\;&\geq\;I(X_1X_2;V_1V_2),
\end{align*}
In the following, we will provide the detailed mathematical proof for case with arbitrary $K$. In order to maintain expressions simple, in the following when we denote a description without the corresponding index, i.e. $V^n_{i,l}$  or $W_{i,l}^n$ for $i\in\{1,2\}$, we will assume that the corresponding index is the true one generated in the corresponding encoders through the detailed  encoding procedure. Consider round $l$ and the event $\mathcal{D}_l=\mathcal{G}_l\cap\mathcal{F}_l$, where for $\epsilon_l>0$,
\begin{equation*}
\mathcal{G}_l=\left\{(X_1^n,X_2^n,Y^n,W^n_{1,l})\in\mathcal{T}^n_{[X_1X_2YW_{1,l}]\epsilon_l}\right\},
\end{equation*}
for all $l\in[1:K+1]$ and
\begin{equation*}
\mathcal{F}_l=\left\{\hat{M}_{1,t}(2)=M_{1,t},\ \hat{M}_{2,t}(1)=M_{2,t},\ t\in[1:l-1]\right\},
\end{equation*}
for all $l\in[1:K]$. We also define
\begin{equation*}
\mathcal{F}_{K+1}=\left\{\hat{M}_{1,t}(3)=M_{1,t},\ \hat{M}_{2,t}(3)=M_{2,t},\ t\in[1:K]\right\}.
\end{equation*}
Sets $\mathcal {G}_l$ tell us that all the descriptions generated up to round $l$ are jointly typical with the sources $X_1,X_2,Y$. This is an event that clearly depend on the encoding procedure at encoders 1 and 2. Sets $\mathcal{F}_l$ indicate that encoders are able to recover without error the indices generated in the other encoders. Clearly, this event depends on the decoding procedure employed. The occurrence of event $\mathcal{D}_l$ guarantees that encoders 1 and 2 share a common path of descriptions $W_{1,l}^n$ which are typical with $(X_1^n,X_2^n,Y^n)$. Similarly, $\mathcal{D}_{K+1}=\mathcal{F}_{K+1}\cap\mathcal{G}_{K+1}$ guarantees that all the generated descriptions are typical with $(X_1^n,X_2^n,Y^n)$ and are perfectly recovered at the decoder. Let us also define the event $\mathcal{E}_l=\left\{\right.$there exists at least an error at the encoding or decoding in a encoder during round $\left.l\right\}$
\begin{equation}
\label{eq:round_error}
\mathcal{E}_l=\mathcal{E}_{enc}(1,l)\cup\mathcal{E}_{dec}(2,l)\cup\mathcal{E}_{enc}(2,l)\cup\mathcal{E}_{dec}(1,l),
\end{equation}
where $\mathcal{E}_{dec}(i,l)$ considers the event that at encoder $i$ during round $l$ there is a failure at recovering an index generated previously in the other encoder and $\mathcal{E}_{enc}(i,l)$ contains the errors at the encoding in encoder $i$ during round $l$. In precise terms:
\begin{align*}
\mathcal{E}_{enc}(1,l)&= \left\{(X_1^n,W^n_{1,l}(\hat{M}_{W_{1,l}}(1)),V_{1,l}^n(m_{1,l},\hat{M}_{W_{1,l}}(1))\right.\nonumber\\
&\;\left.\notin\mathcal{T}^n_{[V_{1,l}W_{1,l}X_1]\epsilon_{c}(1,l)}\ \forall m_{1,l}\in[1:2^{n\hat{R}_{1,l}}]\right\}\\
\mathcal{E}_{enc}(2,l)&= \left\{(X_2^n,W^n_{2,l}(\hat{M}_{W_{2,l}}(2)),V_{2,l}^n(m_{2,l},\hat{M}_{W_{2,l}}(2))\right.\nonumber\\
&\;\left.\notin\mathcal{T}^n_{[V_{2,l}W_{2,l}X_2]\epsilon_{c}(2,l)}\ \forall m_{2,l}\in[1:2^{n\hat{R}_{2,l}}]\right\}\\
\mathcal{E}_{dec}(1,l)& = \left\{\hat{M}_{2,l}(1)\neq M_{2,l}\right\},\ \\
\mathcal{E}_{dec}(2,l) &= \left\{\hat{M}_{1,l}(2)\neq M_{1,l}\right\},
\end{align*}
for all $l\in[1:K]$. Defining the fictitious round $K+1$, where there are not descriptions generation and exchanges but only a decoding procedure at encoder 3, we can write:
\begin{align*}
\mathcal{E}_{K+1}\triangleq& \mathcal{F}^{c}_{K+1}=\left\{\hat{M}_{1,l}(3)\neq M_{1,l},\ \hat{M}_{2,l}(3)\neq M_{2,l},\right.\nonumber\\
&\qquad\left.\mbox{for some}\  l\in[1:K]\right\}.
\end{align*}
The main goal is to prove the occurrence of $\mathcal{D}_{K+1}$ (with high probabilty) which guarantees that the descriptions generated at each encoder are jointly typical with the sources and are perfectly recovered at the decoder 3. We can write:
\begin{align*}
 \mbox{Pr}\left\{\mathcal{D}^{c}_{K+1}\right\}&=\mbox{Pr}\left\{\mathcal{D}^{c}_{K+1}\cap \mathcal{D}_{K}\right\}+\mbox{Pr}\left\{\mathcal{D}^{c}_{K+1}\cap\mathcal{D}^{c}_{K}\right\}\nonumber\\
 &\leq \mbox{Pr}\left\{\mathcal{D}^{c}_{K+1}\cap \mathcal{D}_{K}\right\}+\mbox{Pr}\left\{\mathcal{D}  ^{c}_{K}\right\} \\
 &\leq \mbox{Pr}\left\{\mathcal{D}^{c}_{K}\right\}+\mbox{Pr}\left\{\mathcal{D}^{c}_{K+1}\cap\left(\mathcal{D}_K\cap \mathcal{E}^{c}_K\right)\right\}\nonumber\\
 &\qquad+
 \mbox{Pr}\left\{\mathcal{D}^{c}_{K+1}\cap\left(\mathcal{D}_K\cap \mathcal{E}_K\right)\right\} \\
 &\leq \mbox{Pr}\left\{\mathcal{D}^{c}_{K}\right\}+\mbox{Pr}\left\{\mathcal{D}^{c}_{K+1}\cap\left(\mathcal{D}_K\cap \mathcal{E}^{c}_K\right)\right\}\nonumber\\
  &\qquad+
  \mbox{Pr}\left\{\mathcal{D}_K\cap \mathcal{E}_K\right\} \\
 &\leq \mbox{Pr}\left\{\mathcal{D}^{c}_{1}\right\}+\sum_{l=1}^{K}\mbox{Pr}\left\{\mathcal{D}_{l}\cap\mathcal{E}_{l}\right\}\nonumber\\
  &\qquad+
 \sum_{l=1}^K\mbox{Pr}\left\{\mathcal{D}^{c}_{l+1}\cap\left(\mathcal{D}_l\cap \mathcal{E}^{c}_l\right)\right\}.
 \end{align*}
Notice that
 \begin{equation*}
 \mathcal{D}_1=\left\{(X_1^n,X_2^n,Y^n)\in\mathcal{T}^n_{[X_1X_2Y]\epsilon_1}\right\}\ , \ \epsilon_1>0\ .
 \end{equation*}
 From Lemma~\ref{lemma:prob_lim}, we see that for every $\epsilon_1>0$, $\mbox{Pr}\left\{\mathcal{D}^{c}_{1}\right\}\xrightarrow[n\rightarrow\infty]{}0$. Then, it is easy to see that $\mbox{Pr}\left\{\mathcal{D}_{K+1}\right\}\xrightarrow[n\rightarrow\infty]{}1$ will hold if the coding generation, the encoding and decoding procedures described above allow us to have the following: 
 \begin{enumerate}
  \item If $\mbox{Pr}\left\{\mathcal{D}_{l}\right\}\xrightarrow[n\rightarrow\infty]{}1$ then $\mbox{Pr}\left\{\mathcal{D}_{l+1}\right\}\xrightarrow[n\rightarrow\infty]{}1$ $\forall l\in[1:K-1]$;
  \item $\mbox{Pr}\left\{\mathcal{D}_{l}\cap\mathcal{E}_{l}\right\}\xrightarrow[n\rightarrow\infty]{}0$ $\forall l\in[1:K]$;
  \item $\mbox{Pr}\left\{\mathcal{D}^{c}_{K+1}\cap\left(\mathcal{D}_K\cap \mathcal{E}^{c}_K\right)\right\}\xrightarrow[n\rightarrow\infty]{}0.$
  \end{enumerate}
  In the following we will prove these facts. Observe that, at round $l$ the encoders act sequentially: $\mbox{Encoding at encoder 1}\rightarrow\mbox{Decoding at decoder  2}\rightarrow\mbox{Encoding at encoder 2}\rightarrow\mbox{Decoding at decoder 1}$.  Then, using \eqref{eq:round_error} we can write condition 2) as:
    \begin{align*}
&\mbox{Pr}\left\{\mathcal{D}_{l}\cap\mathcal{E}_{l}\right\}=\mbox{Pr}\left\{\mathcal{D}_{l}\cap\mathcal{E}_{enc}(1,l)\right\}\\
       &+\mbox{Pr}\left\{\mathcal{D}_{l}\cap\mathcal{E}_{dec}(2,l)\cap\mathcal{E}^{c}_{enc}(1,l)\right\}\nonumber\\
   &+ \mbox{Pr}\left\{\mathcal{D}_{l}\cap\mathcal{E}_{enc}(2,l)\cap\mathcal{E}^{c}_{enc}(1,l)\cap\mathcal{E}^{c}_{dec}(2,l)\right\}+\nonumber\\
   &+
   \mbox{Pr}\left\{\mathcal{D}_{l}\cap\mathcal{E}_{dec}(1,l)\cap\mathcal{E}^{c}_{enc}(1,l)\cap\mathcal{E}^{c}_{dec}(2,l)\cap\mathcal{E}^{c}_{enc}(2,l)\right\}.\nonumber
    \end{align*}
Assume then that at the beginning of round $l$ we have $\mbox{Pr}\left\{\mathcal{D}_{l}\right\}\xrightarrow[n\rightarrow\infty]{}1$. This implies that $\mbox{Pr}\left\{\mathcal{G}_{l}\right\}$, $\mbox{Pr}\left\{\mathcal{F}_{l}\right\}\xrightarrow[n\rightarrow\infty]{}1$. Clearly, we have:
\begin{equation*}
\mbox{Pr}\left\{(X_1^n,W^n_{1,l})\in\mathcal{T}^n_{[X_1W_{1,l}]\epsilon_l}\right\}\xrightarrow[n\rightarrow\infty]{}1.
\end{equation*}
We can clearly write:
\begin{align*}
\mbox{Pr}&\left\{\mathcal{D}_l\cap\mathcal{E}_{enc}(1,l)\right\}\leq\mbox{Pr}\left\{(X_1^n,W_{1,l}^n,V_{1,l}^n(m_{1,l},M_{W_{1,l}}))\notin\right.\nonumber\\
&\qquad\quad\left.\mathcal{T}^n_{[X_1W_{1,l}V_{1,l}]\epsilon_c(1,l)}\ \ \forall m_{1,l}\in[1:2^{n\hat{R}_{1,l}}]\right\}.
\end{align*}
 We can use lemma \ref{lemma:covering}  to show that:
\begin{align*}
\mbox{Pr}&\left\{(X_1^n,W_{1,l}^n,V_{1,l}^n(m_{1,l},M_{W_{1,l}}))\notin\mathcal{T}^n_{[X_1W_{1,l}V_{1,l}]\epsilon_c(1,l)}\right.\nonumber\\
&\qquad\left. \ \forall m_{1,l}\in[1:2^{n\hat{R}_{1,l}}]\right\}\xrightarrow[n\rightarrow\infty]{}0,
\end{align*}
 if 
\begin{equation}
 \hat{R}_{1,l}>I(V_{1,l};X_1|W_{1,l})+\delta_{c,1},
 \label{eq:first_rate_condb}
\end{equation}
 where $\delta_{c,1}$ can be made arbitrarly small. In this situation we clearly guarantee that:
\begin{equation*}
\mbox{Pr}\left\{(X_1^n,W^n_{2,l})\in\mathcal{T}^n_{[X_1W_{2,l}]\epsilon_{c}(1,l)}\right\}\xrightarrow[n\rightarrow\infty]{}1.
\end{equation*}
The conditions in Lemma~\ref{lemma:induced} are also satisfied implying:
\begin{align*}
\mbox{Pr}&\left\{V^n_{1,l}(m_{1,l})=v^n_{1,l}\big|x_1^n,x_2^n,y^n,w_{1,l}^n,V^n_{1,l}(m_{1,l})\in\right.\nonumber\\
&\qquad\left.\mathcal{T}^n_{[V_{1,l}|X_1W_{1,l}]\epsilon_c(1,l)}(x_1^n,w_{1,l}^n)\right\}\nonumber\\
&\qquad\quad=\frac{\mathds{1}\left\{v^n_{1,l}\in\mathcal{T}_{[V_{1,l}|X_1W_{1,l}]\epsilon_{c}(1,l)}^n(x_1^n,w_{1,l}^n)\right\}}{|\mathcal{T}_{[V_{1,l}|X_1W_{1,l}]\epsilon_{c}(1,l)}^n(x_1^n,w_{1,l}^n)|}.
 \end{align*}
 Given that we imposed the Markov chain $V_{1,l}\mkv (X_1,W_{1,l}) \mkv (X_2,Y)$ we can use lemma \ref{lemma:markov} and its corresponding corollary to obtain:
\begin{equation*}\mbox{Pr}\left\{(X_1^n,X_2^n,Y^n,W^n_{2,l})\in\mathcal{T}^n_{[X_1X_2YW_{2,l}]\epsilon_{l}'}\right\}\xrightarrow[n\rightarrow\infty]{}1,
\end{equation*}
with $\epsilon_l'$ sufficiently small. At this point we have that all descriptions generated up to round $l$, including the one generated at encoder 1 at round $l$ are jointly typical with the sources $X_1^n,X_2^n,Y^n$ with probability arbitrarily close to 1. Next, we should analyze the decoding at encoder 2. We can write:
    \begin{align}
    \mbox{Pr}&\left\{\mathcal{D}_{l}\cap\mathcal{E}_{dec}(2,l)\cap\mathcal{E}^{c}_{enc}(1,l)\right\}\leq
    \mbox{Pr}\Big\{\mathcal{D}_{l}\cap\mathcal{E}_{dec}(2,l)\nonumber\\
    &  \cap\mathcal{E}^{c}_{enc}(1,l) \cap\left\{(X_1^n,X_2^n,Y^n,W^n_{2,l})\in\mathcal{T}^n_{[X_1X_2YW_{2,l}]\epsilon_{l}'}\right\}\Big\}\nonumber\\
    & +\mbox{Pr}\left\{(X_1^n,X_2^n,Y^n,W^n_{2,l})\notin\mathcal{T}^n_{[X_1X_2YW_{2,l}]\epsilon_{l}'}\right\}\nonumber\\
    &\leq \mbox{Pr}\left\{(X_2^n,W_{2,l}^n)\in\mathcal{T}^n_{[X_2W_{2,l}]\epsilon_l'}\cap\mathcal{F}_l\cap\mathcal{E}_{dec}(2,l)\right\}\nonumber\\
    & +\mbox{Pr}\left\{(X_1^n,X_2^n,Y^n,W^n_{2,l})\notin\mathcal{T}^n_{[X_1X_2YW_{2,l}]\epsilon_{l}'}\right\}.
    \label{eq:error_prob_decod_2}
    \end{align}
Clearly, the second term in the RHS of (\ref{eq:error_prob_decod_2}) goes to zero when $n\rightarrow\infty$. The first term is bounded by:
\begin{equation*}
\mbox{Pr}\left\{(X_2^n,W_{2,l}^n)\in\mathcal{T}^n_{[X_2W_{2,l}]\epsilon_l'}\cap\mathcal{F}_l\cap\mathcal{E}_{dec}(2,l)\right\}\leq \mbox{Pr}\left\{\mathcal{K}_{2,l}\right\},
\end{equation*}
where
\begin{align*}
\mathcal{K}_{2,l}=&\left\{\exists \tilde{m}_{1,l}\neq M_{1,l}: (\tilde{m}_{1,l},M_{W_{1,l}}\in\mathcal{B}(P_{1,l}),\right.\\
&\left.(X_2^n,W_{1,l}^n,V_{1,l}^n(\tilde{m}_{1,l},M_{W_{1,l}}))\in\mathcal{T}^n_{[X_2W_{1,l}V_{1,l}]\epsilon_d(2,l)}\right\}.\nonumber
\end{align*}
From lemma \ref{lemma:packing} we can easily obtain that:
\begin{equation*}
\mbox{Pr}\left\{\mathcal{K}_{2,l}\right\}\xrightarrow[n\rightarrow\infty]{}0,
\end{equation*}
if
\begin{align*}
\frac{1}{n}& \log{\mathbb{E}\left[|\tilde{m}_{1,l}:(\tilde{m}_{1,l}, M_{W_{1,l}})\in\mathcal{B}(P_{1,l})|\right]}\nonumber\\
&\qquad\leq I(X_2;V_{1,l}|W_{1,l})-\delta',
\end{align*}
where $\delta'$ can be made arbitrarly small. It is very easy to show that:
\begin{equation*}
\mathbb{E}\left[|\tilde{m}_{1,l}:(\tilde{m}_{1,l}, M_{W_{1,l}})\in\mathcal{B}(P_{1,l})|\right]=2^{n(\hat{R}_{1,l}-R_{1,l})},
\end{equation*}
which implies that:
\begin{equation}
\hat{R}_{1,l}-R_{1,l}<I(X_2;V_{1,l}|W_{1,l})-\delta'.
\label{eq:second_rate_cond}
\end{equation}
At this point, we should analyze the encoding at encoder 2. This is done along the same lines of thought used for the encoding at encoder 1. The same can be said of the decoding at encoder 1. In summary we obtain the following rate equations:
 \begin{IEEEeqnarray}{rCl}
 \hat{R}_{2,l} &>& I(V_{2,l};X_2|W_{2,l})+\delta_{c,2},
 \label{eq:third_rate_cond}\\
\hat{R}_{2,l}-R_{2,l} &<& I(X_1;V_{2,l}|W_{2,l})-\delta''.
\label{eq:fourth_rate_cond}
\end{IEEEeqnarray}
It is straightforward to see that $\mbox{Pr}\left\{\mathcal{D}_l\cap\mathcal{E}_l\right\}\xrightarrow[n\rightarrow\infty]{}0\ \ \forall l\in[1:K]$ . The analysis for the joint typicality of  all descriptions generated up to round $l$, including the one generated at encoder 1 at round $l$ are jointly typical with the sources $X_1^n,X_2^n,Y^n$ with probability arbitrarly close to 1, can be repeated at encoder 2 obtaining:
\begin{equation*}
\mbox{Pr}\left\{(X_1^n,X_2^n,Y^n,W^n_{1,l+1})\in\mathcal{T}^n_{[X_1X_2YW_{[1:l+1]}]\epsilon_{l+1}'}\right\}\xrightarrow[n\rightarrow\infty]{}1,
\end{equation*}
which  is a restatement of  $\mbox{Pr}\left\{ \mathcal{G}_{l+1}\right\}\xrightarrow[n\rightarrow\infty]{}1$. In conjunction with the fact the above rate conditions guarantee that there are not errors at the enconding and decoding at encoder 1 and 2 during round $l$ we have that $\mbox{Pr}\left\{\mathcal{D}_{l+1}\right\}\xrightarrow[n\rightarrow\infty]{}1$. In this manner we can conclude that  $\mbox{Pr}\left\{\mathcal{D}_l\right\}\xrightarrow[n\rightarrow\infty]{}1$ implies that  $\mbox{Pr}\left\{\mathcal{D}_{l+1}\right\}\xrightarrow[n\rightarrow\infty]{}1$ for $l\in[1:K-1]$. In order to complete the error probability analysis we need to prove that 
\begin{equation*}
\mbox{Pr}\left\{\mathcal{D}^{c}_{K+1}\cap\left(\mathcal{D}_K\cap \mathcal{E}^{c}_K\right)\right\}\xrightarrow[n\rightarrow\infty]{}0.
\end{equation*}
In order to do this we need to analyze the decoding at encoder 3. It is easy to show that:
\begin{align}
\mbox{Pr}&\left\{\mathcal{D}^{c}_{K+1}\cap\left(\mathcal{D}_K\cap \mathcal{E}^{c}_K\right)\right\}\nonumber\\
&\qquad\leq\mbox{Pr}\left\{\mathcal{G}_{K+1}\cap\mathcal{F}^{c}_{K+1}\right\}+\mbox{Pr}\left\{\mathcal{G}^{c}_{K+1}\right\},
\label{eq:decoding_3}
\end{align}
where
\begin{equation*}
\mathcal{G}_{K+1}=\left\{(X_1^n,X_2^n,Y^n,W_{1,K+1}^n)\in\mathcal{T}^n_{[X_1X_2YW_{1,K+1}]\epsilon_{K+1}}\right\}.
\end{equation*}
From the previous analysis it is easy to see that $\mbox{Pr}\left\{\mathcal{G}^{c}_{K+1}\right\}\xrightarrow[n\rightarrow\infty]{}0$. The first term in the RHS of (\ref{eq:decoding_3}) can be bounded as:
\begin{align*}
\mbox{Pr}&\left\{\mathcal{G}_{K+1}\cap\mathcal{F}^{c}_{K+1}\right\}\nonumber\\
&\qquad\leq\mbox{Pr}\left\{\left\{W_{1,K+1}^n\in\mathcal{T}^n_{W_{1,K+1}\epsilon_{K+1}}\right\}\cap\mathcal{F}^{c}_{K+1}\right\}\\ &\qquad\leq\mbox{Pr}\left\{ \mathcal{K}_{3}\right\},
\end{align*}
where
\begin{align*}
\mathcal{K}_3=&\left\{\exists \left\{\tilde{m}_{1,l},\tilde{m}_{2,l}\right\}_{l=1}^K\neq \left\{M_{1,l},M_{2,l}\right\}_{l=1}^K:\right.\nonumber\\
&\quad\left.(\tilde{m}_{1,l},\tilde{m}_{W_{1,l}})\in\mathcal{B}(P_{1,l}),\ (\tilde{m}_{2,l},\tilde{m}_{W_{2,l}})\in\mathcal{B}(P_{2,l}),\ \right.\nonumber\\
& \quad\left.\left\{V_{1,l}^n(\tilde{m}_{1,l},\tilde{m}_{W_{1,l}}),V_{2,l}^n(\tilde{m}_{2,l},\tilde{m}_{W_{2,l}})\right\}_{l=1}^K
\in\right.\nonumber\\
&\quad\left.\mathcal{T}^n_{\left[\left\{V_{1,l},V_{2,l}\right\}_{l=1}^K\right]\epsilon_F}\right\}.
\end{align*}
We can write:
\begin{align*}
&\mbox{Pr}\left\{\mathcal{K}_3\right\}=\mathbb{E}\left[\mbox{Pr}\left\{\mathcal{K}_3\big| \left\{M_{1,l},M_{2,l}\right\}_{l=1}^K= \left\{m_{1,l},m_{2,l}\right\}_{l=1}^K,\right.\right.\nonumber\\
&\qquad\qquad\left.\left.\left\{P_{1,l},P_{2,l}\right\}_{l=1}^K= \left\{p_{1,l},p_{2,l}\right\}_{l=1}^K\right\}\right]\\
&=\mathbb{E}\left[\mbox{Pr}\left\{\bigcup_{\left\{\tilde{m}_{1,l},\tilde{m}_{2,l}\right\}_{l=1}^K\in\mathcal{A}\left(\left\{m_{1,l},m_{2,l}\right\}_{l=1}^K\right)}\right.\right.\nonumber\\
&\quad\left.\left.
\left\{V_{1,l}^n(\tilde{m}_{1,l},\tilde{m}_{W_{1,l}}),V_{2,l}^n(\tilde{m}_{2,l},\tilde{m}_{W_{2,l}})\right\}_{l=1}^K
\right.\right.\\
&\quad\left.\left.\in\mathcal{T}^n_{\left[\left\{V_{1,l},V_{2,l}\right\}_{l=1}^K\right]\epsilon_F}\Bigg|\left\{M_{1,l},M_{2,l}\right\}_{l=1}^K=\left\{m_{1,l},m_{2,l}\right\}_{l=1}^K\right\}\right]\nonumber\\
&\leq\mathbb{E}\left[\sum_{\left\{\tilde{m}_{1,l},\tilde{m}_{2,l}\right\}_{l=1}^K\in\mathcal{A}\left(\left\{m_{1,l},m_{2,l}\right\}_{l=1}^K\right)}\right.\nonumber\\
&\quad\left.\mbox{Pr}\left\{\left\{V_{1,l}^n(\tilde{m}_{1,l},\tilde{m}_{W_{1,l}}),V_{2,l}^n(\tilde{m}_{2,l},\tilde{m}_{W_{2,l}})\right\}_{l=1}^K
\right.\right.\\
&\quad\left.\left.\in\mathcal{T}^n_{\left[\left\{V_{1,l},V_{2,l}\right\}_{l=1}^K\right]\epsilon_F} \Bigg|\left\{M_{1,l},M_{2,l}\right\}_{l=1}^K=\left\{m_{1,l},m_{2,l}\right\}_{l=1}^K\right\}\right]\nonumber\\
&\leq \mathbb{E}\left[\left|\mathcal{A}\left(\left\{M_{1,l},M_{2,l}\right\}_{l=1}^K\right)\right|\right],
    \end{align*}
where we used the fact that $\left\{P_{1,l},P_{2,l}\right\}_{l=1}^K$ are functions of $\left\{m_{1,l},m_{2,l}\right\}_{l=1}^K$, and that:
\begin{align*}
\mbox{Pr}&\left\{\left\{V_{1,l}^n(\tilde{m}_{1,l},\tilde{m}_{W_{1,l}}),V_{2,l}^n(\tilde{m}_{2,l},\tilde{m}_{W_{2,l}})\right\}_{l=1}^K\in\right.\nonumber\\
&\qquad\left.\mathcal{T}^n_{\left[\left\{V_{1,l},V_{2,l}\right\}_{l=1}^K\right]\epsilon_F}\right\}\leq1
\end{align*}
 and where we defined
\begin{align*}
\mathcal{A}&\left(\left\{m_{1,l},m_{2,l}\right\}_{l=1}^K\right)\nonumber\\
&\qquad=\left\{\begin{array}{c}
\left\{\tilde{m}_{1,l},\tilde{m}_{2,l}\right\}_{l=1}^K\neq \left\{m_{1,l},m_{2,l}\right\}_{l=1}^K,\\ 
(\tilde{m}_{1,l},\tilde{m}_{W_{1,l}})\in\mathcal{B}(p_{1,l}),\\
(\tilde{m}_{2,l},\tilde{m}_{W_{2,l}})\in\mathcal{B}(p_{2,l})\end{array}\right\},
\end{align*}
considering that $\left\{p_{1,l},p_{2,l}\right\}_{l=1}^K$ are the functions of $\left\{m_{1,l},m_{2,l}\right\}_{l=1}^K$ generated by the described encoding procedure. In order to compute  $\mathbb{E}\left[\left|\mathcal{A}\left(\left\{m_{1,l},m_{2,l}\right\}_{l=1}^K\right)\right|\right]$ we will consider a relabelling of the indices of the exchanged descriptions. We define for every $s\in[1:2K]$:
\begin{equation}
m_s=\left\{\begin{array}{cc}
 I_{\frac{s+1}{2}}& s\ \mbox{odd}\\
 J_{\frac{s}{2}} & s\ \mbox{even}
\end{array}\right.,\;\; p_s=\left\{\begin{array}{cc}
 p_{1,\frac{s+1}{2}}& s\ \mbox{odd}\\
p_{2,\frac{s}{2}} & s\ \mbox{even}.
\end{array}\right.
\label{eq:relabel} 
\end{equation}
Clearly, we can write:
\begin{align*}
\mathcal{A}\left(\left\{m_{1,l},m_{2,l}\right\}_{l=1}^K\right)&=\Big\{\{\tilde{m}_s\}_{s=1}^{2K}\neq\{m_s\}_{s=1}^{2K}: \{\tilde{m}_s\}_{s=1}^{2K}\nonumber\\
&\qquad\in\mathcal{B}(p_s),\ s\in[1:2K]\Big\}.
\end{align*}
Consider $\mathcal{M}=[1:2K]$ and its power set $2^{\mathcal{M}}$. It is straightforward to obtain:
\begin{align*}
\mathbb{E}&\left[\left|\mathcal{A}\left(\left\{m_{1,l},m_{2,l}\right\}_{l=1}^K\right)\right|\right]=\sum_{\mathcal{H}\in 2^{\mathcal{M}}}E\left[\left|\left\{
\{\tilde{m}_s\}_{s\in\mathcal{M}}:\right.\right.\right.\nonumber\\ 
&\;\left.\left.\left.\tilde{m}_s\neq M_s\ \forall s\in\mathcal{H}, \{\tilde{m}_l\}_{l=1}^s\in\mathcal{B}(P_s),\ s\in\mathcal{M}\right\}\right|\right].
\end{align*}
Let us analyze each term in the above sumation. Consider $\mathcal{H}\in 2^{\mathcal{M}}$ and $s_{\rm min}(\mathcal{H})=\min{\left\{s:s\in\mathcal{H}\right\}}$. We are interested in computing: $E\left[\left|\left\{
\{\tilde{m}_s\}_{s\in\mathcal{M}}: \tilde{m}_s\neq M_s\ \forall s\in\mathcal{H}, \{\tilde{m}_l\}_{l=1}^s\in\mathcal{B}(P_s),\right.\right.\right.$ $\left.\left.\left.s\in\mathcal{M}\right\}\right|\right].$ It is clear that the number of indices such that $\tilde{m}_s\neq M_s\ \forall s\in\mathcal{H}$ is given by:
\begin{equation*}
\prod_{s\in\mathcal{H}}(2^{n\hat{R}_s}-1)\leq 2^{n\sum_{s\in\mathcal{H}}\hat{R}_s}.
\end{equation*}
As all indices of the generated codewords are independently and uniformly distributed in each of the bins used in the encoders 1 and 2, the probability that each of the above indices $\{\tilde{m}_s\}_{s\in\mathcal{M}}$ belongs to the bins $\left\{\mathcal{B}(p_s)\right\}_{s\in\mathcal{M}}$ is given by $2^{-n \sum_{s=s_{\rm min}(\mathcal{H})}^{2K}R_s}$ for any sequence $\left\{p_s\right\}_{s\in\mathcal{M}}$. Then we can write:
\begin{align*}
\mathbb{E}&\left[\left|\mathcal{A}\left(\left\{m_{1,l},m_{2,l}\right\}_{l=1}^K\right)\right|\right]\nonumber\\
&\qquad\leq
\sum_{\mathcal{H}\in 2^{\mathcal{M}}}2^{n\left(\sum_{s\in\mathcal{H}}\hat{R}_s-\sum_{s=s_{\rm min}(\mathcal{H})}^{2K}R_s\right)}.
\end{align*}
Clearly, $\mathbb{E}\left[\left|\mathcal{A}\left(\left\{m_{1,l},m_{2,l}\right\}_{l=1}^K\right)\right|\right]\xrightarrow[n\rightarrow\infty]{}0$ if for each $\mathcal{H}\in 2^{\mathcal{M}}$:
\begin{equation}
\sum_{s\in\mathcal{H}}\hat{R}_s-\sum_{s=s_{\rm min}(\mathcal{H})}^{2K}R_s<0.
\label{eq:rate_decod_3}
\end{equation}
Consider this equation for every $\mathcal{H}\in 2^{\mathcal{M}}$. Clearly $s_{\rm min}(\mathcal{H})\in[1:2K]$ when $\mathcal{H}$ ranges over $2^{\mathcal{M}}$. Consider the sets $\mathcal{H}\in 2^{\mathcal{M}}$ such that $s_{\rm min}(\mathcal{H})=r$. It is clear that over these sets, the one which put the more stringent condition in (\ref{eq:rate_decod_3}) is $[r:2K]$. In this way, the $2^{2K}$ equations in (\ref{eq:rate_decod_3}) can be replaced by only $2K$ equations given by:
\begin{equation*}
\sum_{s=r}^{2K}(\hat{R}_s-R_s)<0,\ \ r\in[1:2K].
\end{equation*}
Using the relabelling in (\ref{eq:relabel}) it is easy to see that these equations can be put in the following manner in terms of $R_{i,l}$ and $\hat{R}_{i,l}$ with $l\in[1:K]$ and $i\in\{1,2\}$:
\begin{align*}
&\hat{R}_{1,l}+\hat{R}_{2,l}+\sum_{k=l+1}^{K}(\hat{R}_{1,k}+\hat{R}_{2,k})\nonumber\\
&\qquad<R_{1,l}+R_{2,l}+\sum_{k=l+1}^{K}(R_{1,k}+R_{2,k})
\end{align*}
\begin{equation*}
\hat{R}_{2,l}+\sum_{k=l+1}^{K}(\hat{R}_{1,k}+\hat{R}_{2,k})<R_{2,l}+\sum_{k=l+1}^{K}(R_{1,k}+R_{2,k}).
\end{equation*}
At this point we can use equations (\ref{eq:first_rate_cond}), (\ref{eq:first_rate_condb}), (\ref{eq:second_rate_cond}), (\ref{eq:third_rate_cond}), (\ref{eq:fourth_rate_cond}) jointly with the fact that the total rates at encoders 1 and 2 can be written as:
\begin{equation*}
R_1=\sum_{l=1}^KR_{1,l}\ ,\ \ R_2=\sum_{l=1}^KR_{2,l},
\end{equation*}
in a Fourier-Motzkin elimination procedure to obtain:
 \begin{align*}
&R_1> I(X_1;W_{1,K+1}|X_2),\\
&R_2> I(X_2;V_{2,K}|W_{2,K})+I(X_2;W_{2,K}|X_1),\\
&R_1+R_2> I(X_1X_2;W_{1,K+1}).
 \end{align*}
 
Now we are set to prove analyze the average level or relevance. Let us denote with $\mathcal{C}$ the random realization of one codebook and be $\mathcal{C}=c$ one of its realizations. The average level of relevance over all random codebooks can be written as:
\begin{align*}
\mathbb{E}_{\mathcal{C}}&\left[\frac{1}{n}I\left(Y^n;M_{W_{1,K+1}}\big|\mathcal{C}=c\right)\right]\nonumber\\
&\qquad\qquad\qquad=\frac{1}{n}I\left(Y^n;M_{W_{1,K+1}}\big|\mathcal{C}\right)\\
&\qquad\qquad\qquad=\frac{1}{n}I\left(Y^n;M_{W_{1,K+1}}\mathcal{C}\right)\\
&\qquad\qquad\qquad\geq \frac{1}{n}I\left(Y^n;M_{W_{1,K+1}}\right),
\end{align*}
using the independence of the random generated codes with $Y^n$. The following decomposition can be obtained introducing the indices recovered at encoder 3, which will denote as $\hat{M}_{W_{1,K+1}}$:
\begin{align*}
\frac{1}{n}I&\left(Y^n;M_{W_{1,K+1}}\right)\nonumber\\
&=\frac{1}{n}H\left(Y^n\right)-\frac{1}{n}H\left(Y^n|M_{W_{1,K+1}}\hat{M}_{W_{1,K+1}}\right)\nonumber\\
&\qquad-\frac{1}{n}I\left(Y^n;\hat{M}_{W_{1,K+1}}|M_{W_{1,K+1}}\right)\\
&\geq H(Y)-\frac{1}{n}H\left(Y^n|M_{W_{1,K+1}}\hat{M}_{W_{1,K+1}}\right)\nonumber\\
&\qquad-\frac{1}{n}H\left(\hat{M}_{W_{1,K+1}}|M_{W_{1,K+1}}\right).
\end{align*}
Last term in the above expression can be negligible when $n\rightarrow\infty$ by a simple application of Fano inequality. To bound the other conditional entropy term we consider the following random variable:
\begin{equation*}
 \hat{Y}^n = \left\{
  \begin{array}{l l}
    Y^n & \quad \text{if } \;M_{W_{1,K+1}}=\hat{M}_{W_{1,K+1}}\\
    &\qquad\quad\wedge\;
\left(Y^n,W_{1,K+1}^n \right)\in\mathcal{T}^n_{\left[YW_{1,K+1}\right]\epsilon_{K+1}}\\
     \emptyset & \quad \text{else}
   \end{array} \right.
\end{equation*}
where, $W^n_{1,K+1}\triangleq  W^n_{1,K+1}\left(\hat{M}_{W_{1,K+1}}\right)$. This auxiliary variable allows us to bound the conditional entropy as follows:
\begin{align*}
H&\left(Y^n|M_{W_{1,K+1}}\hat{M}_{W_{1,K+1}}\right)\nonumber\\
&\;= H\left(Y^n,\hat{Y}^n|M_{W_{1,K+1}}\hat{M}_{W_{1,K+1}}\right),\\
&\;\leq  H\left(\hat{Y}^n|M_{W_{1,K+1}}\hat{M}_{W_{1,K+1}}\right)+H\left(Y^n|\hat{Y}^n\right),\\
&\;\overset{(a)}{\leq}\log\left\{\left| \mathcal{T}^n_{\left[Y|W_{1,K+1}\right]}\right|+1\right\}+\delta+H\left(Y^n|\hat{Y}^n\right),\\
&\;\overset{(b)}{\leq} n\left(H\left(Y|W_{1,K+1}\right)+\epsilon\right)+\delta+H\left(Y^n|\hat{Y}^n\right),\\
&\;\overset{(c)}{\leq} n\left[H\left(Y|W_{1,K+1}\right)+\epsilon\right]+\delta+1\nonumber\\
&\qquad+n\mbox{Pr}\left(Y^n\neq\hat{Y}^n\right)\log|\mathcal{Y}|,
\end{align*}	
where $(a)$ follows from the fact that the uniform distribution maximize entropy,  $(b)$ stems from standard properties of a conditional typical set and $(c)$ is consequence of Fano inequality. We define the error probability $P_e^n=\mbox{Pr}\left\{Y^n\neq\hat{Y}^n\right\}$. Then, the relevance condition can be bounded as
\begin{align*}
\frac{1}{n}I&\left(Y^n;M_{W_{1,K+1}}\right)\nonumber\\
&\geq H(Y)-H\left(Y|W_{1,K+1}\right)-P_e^n\log|\mathcal{Y}|-\kappa_n \\
&\geq I(Y;W_{1,K+1})-P_e^n\log|\mathcal{Y}|-\kappa_n,
\end{align*}
where $\kappa_n$ goes to zero with $n$ large enough. In this way,
\begin{equation*}
\mathbb{E}_{\mathcal{C}}\left[\frac{1}{n}I\left(Y^n;M_{W_{1,K+1}}\big|\mathcal{C}=c\right)\right]\geq 
 I(Y;W_{1,K+1})-\epsilon_n,
\end{equation*}
with $\epsilon_n\rightarrow 0$ when $n\rightarrow\infty$. This show that  every relevance level $\mu\leq  I(Y;W_{1,K+1})$ is achievable in an average sense over all random codebook. For that reason, there must exists at least one good codebook. 

\subsection{Achievability in Theorem~\ref{theo:p1}}

The coding scheme is basically the same as the previous one. In this case encoder 1 and 2 operate sequentially in the same manner as above until the last round. As there is no encoder 3, only the first part of error probability analysis given is relevant for this case. The calculation of the relevance levels at encoder 1 and 2 follows also the same lines and for that reason is also omitted. We should mention that, at a given round, the bins generated, for example, at encoder 1 needs to contain only the index of latest generated description. It is not needed to generate larger bins in order to contain also all previous generated descriptions at encoder 1 and 2. In this way, instead of (\ref{eq:first_rate_cond}), only the following are to be satisfied in the bins generation:
\begin{equation*}
R_{1,l}<\hat{R}_{1,l}\ ,\ \ R_{2,l}<\hat{R}_{2,l},\ \ \ l\in[1:K],
\end{equation*}
which simplifies the analysis and the needed Fourier-Motzkin elimination procedure. The reason for this difference is given by the absence of the decoder in node 3. 

\section{Corner points for $\mathcal{R}_{\mbox{\tiny CDIB}}^{\mbox{\tiny outer}}(1)$}
\label{sec:extreme}
 Let any fix distribution of $U_1$ and $U_2$ according to the corresponding  Markov chains. This  distribution induces $4$ different corner points in $\mathcal{R}_{\mbox{\tiny CDIB}}^{\mbox{\tiny outer}}(1)$, namely:
\begin{align*}
Q_1&=\left[
I(X_1;U_1|X_2), I(U_1U_2;X_2), I(Y;U_1U_2)\right],\\
Q_2&=\left[I(X_1;U_1), I(U_2;X_2|U_1), I(Y;U_1U_2)\right],\\
Q_3&=\left[I(U_1;X_1), 0, I(Y;U_1)-I(U_2;X_2|U_1Y)\right],\\
Q_4&=\left[I(U_1;X_1|X_2), 0, I(X_1;U_1|X_2)-I(U_1U_2;X_1X_2|Y)\right]
\end{align*}
The involved directions are given by the vectors $(0,1)$ and $(1,0)$ and do not enter in the analysis. The inclusion of  $Q_1$ and $Q_2$ in $\mathcal{R}_{\mbox{\tiny CDIB}}^{\mbox{\tiny inner}}(1)$ is easily proved by simply choosing $V_1=U_1$ and $V_2=U_2$. For $Q_3$ simply choose $V_1=U_1$ and $V_2=v_2$ with $v_2\in\mathcal{V}_2$.  The analysis of $Q_4$ is slightly more sophisticated. We need to use time sharing.  We define a random variable $Z\sim\mbox{Bern}\left(\lambda\right)$, with $\lambda\in(0,1)$, and independent of everything else. We select
\begin{align*}
V_1&=U_1\mathds{1}\left\{Z=1\right\}+v_1\mathds{1}\left\{Z=0\right\},\quad v_1\in\mathcal{V}_1,\\
V_2&=v_2,\quad v_2\in\mathcal{V}_2.
\end{align*}
Thanks to an appropriate choice of the time-sharing parameter $\lambda$, we will show that the point $Q_4$ is in $\mathcal{R}_{\mbox{\tiny CDIB}}^{\mbox{\tiny inner}}(1)$. That choice is given by
 \begin{equation*}
 \lambda\triangleq \frac{I(X_1;U_1|X_2)}{I(X_1;U_1)}=1-\frac{I(X_2;U_1)}{I(X_1;U_1)}.
 \end{equation*}
It is easy to see that the following conditions are met:
\begin{IEEEeqnarray*}{rCl}
R_1&\geq& \lambda I(U_1;X_1|X_2),\\
R_2 &\geq&  0,\\
R_1+R_2 &\geq& \lambda I(U_1;X_1),\\
\mu&\leq& \lambda I(Y;U_1).
\end{IEEEeqnarray*}  
With this specific choice it is easy to show that we meet the rate conditions in  $\mathcal{R}_{\mbox{\tiny CDIB}}^{\mbox{\tiny inner}}(1)$ . It remains to analyze the relevance condition $\mu_{Q_4}\leq\lambda I(Y;U_1)$. To this end, let us consider: $A\triangleq  \lambda I(Y;U_1)-I(X_1;U_1|X_2)+I(U_1U_2;X_1X_2|Y)$.
We can easily check that:
\begin{align*}
A=&\lambda I(Y;U_1)-\lambda I(X_1;U_1)+I(U_1U_2;X_1X_2|Y) \\
=&-\lambda I(X_1;U_1|Y)+I(U_1U_2;X_1X_2|Y) \\
=&(1-\lambda)I(X_1;U_1|Y)+I(X_1X_2;U_2|U_1Y).
\end{align*}
We have clearly that $A\geq 0$ which implies the relevance condition. Then, $Q_4\in\mathcal{R}_{\mbox{\tiny CDIB}}^{\mbox{\tiny inner}}(1)$. For every choice of the distributions of $U_1$ and $U_2$ (with the appropriate Markov chains), the extreme points of the outer bound are contained in $\mathcal{R}_{\mbox{\tiny CDIB}}^{\mbox{\tiny inner}}(1)$, which implies that $\mathcal{R}_{\mbox{\tiny CDIB}}^{\mbox{\tiny outer}}(1)\subseteq\mathcal{R}_{\mbox{\tiny CDIB}}^{\mbox{\tiny inner}}(1)$ from which the desired conclusion is obtained.

\section{$\mathcal{R}_{\mbox{\tiny CDIB}}^{\mbox{\tiny inner}}(1)=\tilde{\mathcal{R}}_{\mbox{\tiny CDIB}}(1)$ when $X_1\mkv X_2\mkv Y$.}
\label{sec:equivalence}

As $\mathcal{R}_{\mbox{\tiny CDIB}}^{\mbox{\tiny inner}}(1)\supseteq\tilde{\mathcal{R}}_{\mbox{\tiny CDIB}}(1)$ is trivial, we consider only $\mathcal{R}_{\mbox{\tiny CDIB}}^{\mbox{\tiny inner}}(1)\subseteq\tilde{\mathcal{R}}_{\mbox{\tiny CDIB}}(1)$. Consider $(R_1,R_2,\mu)\in\mathcal{R}_{\mbox{\tiny CDIB}}^{\mbox{\tiny inner}}(1)$. Then $\exists\ V_1,V_2$ auxiliary random variables such that
\begin{IEEEeqnarray*}{rCl}
R_1&\geq& I(X_1;V_1|X_2),\\
R_2&\geq& I(X_2;V_2|V_1),\\
 R_1+R_2&\geq& I(X_1X_2;V_1V_2),\\
 \mu&\leq &I(Y;V_1V_2),
\end{IEEEeqnarray*}
and such that (\ref{eq:third_mkv_cond}) is satisfied. For the given $V_1,V_2$, the above region presents two extreme points:
\begin{align*}
Q_1&=\left[I(X_1;V_1), I(X_2;V_2|X_1), I(Y;V_1V_2)\right],\\
Q_2&=\left[I(X_1;V_1|X_2), I(X_2;V_1V_2), I(Y;V_1V_2)\right].
\end{align*}
If we choose $\tilde{V}_1=V_1$ and $\tilde{V}_2=V_2$ we see that $Q_1\in\tilde{\mathcal{R}}_{\mbox{\tiny CDIB}}(1)$. Let us analyze $Q_2$. Consider the random variables $\tilde{V}_1=(V_1',Z)$ and $\tilde{V}_2=(V_2',Z)$ where
\begin{IEEEeqnarray*}{rCl}
V_1'&=& V_1\mathds{1}\left\{Z=1\right\}+v_1\mathds{1}\left\{Z=0\right\}, v_1\in\mathcal{V}_1\\
V_2'&=& V_2\mathds{1}\left\{Z=1\right\}+W\mathds{1}\left\{Z=0\right\},
\end{IEEEeqnarray*}
where $Z\sim\mbox{Bern}\left(\lambda\right)$ is independent of everything else with $\lambda\in(0,1)$ and $W$ is random variable that satisfies $W\mkv X_2\mkv X_1Y$. From these definitions we see that the following are satisfied:
\begin{equation}
\label{eq:markov_tilde}
\tilde{V}_1\mkv X_1\mkv (X_2,Y)\ ,\ \tilde{V}_2\mkv \tilde{V}_1X_2\mkv (X_1,Y).
\end{equation}
Consider $\lambda=\frac{I(X_1;V_1|X_2)}{I(X_1;V_1)}=1-\frac{I(X_2;V1)}{I(X_1;V_1)}$. The following relations are easy to obtain:
\begin{align*}
I(X_1;\tilde{V}_1)&= I(X_1;V_1|X_2),\\
I(X_2;\tilde{V}_2|\tilde{V}_1)&= \lambda I(X_1;V_1|V_2)+(1-\lambda)I(X_2;W),\\
I(Y;\tilde{V}_1\tilde{V}_2)&= \lambda I(Y;V_1V_2)+(1-\lambda)I(Y;W).
\end{align*}
From these equations, and in order to show that $Q_2\in\tilde{\mathcal{R}}_{\mbox{\tiny CDIB}}(1)$, we can obtain the following conditions on random variable $W$:
\begin{align}
I(X_2;W)&\leq I(X_2;V_1V_2)+I(X_1;V_1|X_2),\label{eq:W_conditions1}\\ I(Y;V_1V_2)&\leq I(Y;W).\label{eq:W_conditions2}
\end{align}
Consider the distribution $p_{V_1V_2|X_2}$ given by:
\begin{equation*}
p_{V_1V_2|X_2}(v_1,v_2|x_2)=\sum_{x_1}p(x_1|x_2)p(v_1|x_1)p(v_2|v_1x_2).
\end{equation*}
We choose random variable $W$ such that $p_{W|X_2}\sim p_{V_1V_2|X_2}$. With this choice we obtain $I(X_2;W)=I(V_1V_2;X_2)$ which clearly satisfies condition \eqref{eq:W_conditions1}. Up to this point we have not used the condition $X_1\mkv X_2\mkv Y$. Using this condition we can obtain $p_{WY}\sim p_{V_1V_2Y}$, which implies that $I(Y;W)=I(Y;V_1V_2)$, satisfying condition in \eqref{eq:W_conditions2}. So, we were able to find $(\tilde{V}_1,\tilde{V}_2)$ that satisfies \eqref{eq:markov_tilde} and 
\begin{IEEEeqnarray*}{rcl}
I(X_1;\tilde{V}_1)=I(X_1;V_1|X_2),\\
I(X_2;\tilde{V}_2|\tilde{V_1})\leq I(X_2;V_1V_2),\\
I(Y;\tilde{V}_1\tilde{V}_2)=I(Y;W).
\end{IEEEeqnarray*}
This shows definitely show that  $Q_2\in\tilde{\mathcal{R}}_{\mbox{\tiny CDIB}}(1)$. As for any pair $(V_1,V_2)$ we have that $(Q_1,Q_2)\in\tilde{\mathcal{R}}_{\mbox{\tiny CDIB}}(1)$, then $\mathcal{R}_{\mbox{\tiny CDIB}}^{\mbox{\tiny inner}}(1)\subseteq\tilde{\mathcal{R}}_{\mbox{\tiny CDIB}}(1)$ and $\mathcal{R}_{\mbox{\tiny CDIB}}^{\mbox{\tiny inner}}(1)=\tilde{\mathcal{R}}_{\mbox{\tiny CDIB}}(1)$. 

\section{Proof of theorem \ref{theo:mu_D}}
\label{app:mu_D}

As the proof relies heavily on convex analysis notions, we begin recalling basic facts of convex analysis that will be used during the proof. These results are presented without proofs which can be consulted in several well-known references on convex analysis as \cite{rockafellar_convex_1970}. The works by Witsenhausen and Wyner \cite{witsenhausen75}, \cite{Wit_1980} provide a good summary of convex analysis for information-theoretic problems.  Consider a compact and connected set $\mathcal{A}\in\mathbb{R}^n$. We define $\mathcal{C}\triangleq  \mbox{co}\left(\mathcal{A}\right)$ to be the \emph{convex hull} of $\mathcal{A}$. Let $m\leq n$ be the dimension of $\mathcal{C}$ (that is, the dimension of its \emph{affine hull}). We say that $x\in \mathcal{C}$ is an \emph{extreme point} of $\mathcal{C}$ if there not exist $\lambda\in(0,1)$ and $x_1,x_2\in \mathcal{C}$  such that $x=\lambda x_1+(1-\lambda)x_2$. 
We say that $f:\mathbb{R}^n\rightarrow \mathbb{R}$ is \emph{convex} if its \emph{effective domain} (the set where $f(x)<\infty$) is convex and:
\begin{equation*}
f(\lambda x_1+(1-\lambda)x_2)\leq \lambda f(x_1)+(1-\lambda)f(x_2),
\end{equation*}
with $\lambda\in[0,1]$ and $x_1,x_2\in\mathbb{R}^n$. When the inequality is strict for every $\lambda\in(0,1),\ x_1,x_2\in\mathbb{R}^n$ we say that $f(x)$ is \emph{strictly convex}. When $-f(x)$ is convex (strictly convex), we say that $f(x)$ is \emph{concave (strictly concave)}. Some useful results are presented without proof:
\begin{enumerate}[(i)]
\item $\mathcal{C}$ is compact;
\item Every extreme point of $\mathcal{C}$ belongs to $\mathcal{A}$ and it is on the boundary of $\mathcal{C}$;
\item \emph{Fenchel-Eggleston's theorem \cite{Eggleston}:} If $\mathcal{A}$ has $m$ or less connected components, every point of $\mathcal{C}$ is the convex combination of no more that $m$ points $\mathcal{A}$;
\item \emph{Dubin's theorem \cite{Dubin}:} Every point of the intersection of $\mathcal{C}$ with $k$ hyperplanes is the convex combination of no more that $k+1$ extreme points of $\mathcal{C}$;
\item \emph{Krein-Milman's theorem \cite{Krein}:} $\mathcal{C}$ is the convex hull of its extreme points;
\item \emph{Supporting hyperplanes \cite{rockafellar_convex_1970}:} On every point of the boundary (relative boundary if $m<n$) of $\mathcal{C}$ there exists a \emph{supporting hyperplane} of dimension $m-1$ such that $\mathcal{C}$ is contained in one of the half-spaces determined by that hyperplane. Indeed, $\mathcal{C}$ is the intersection of all half-spaces that contain $\mathcal{C}$;
\item Consider the functions $f:\mathbb{R}^{n-1}\rightarrow \mathbb{R}$ and $g:\mathbb{R}^{n-1}\rightarrow \mathbb{R}$ defined as
 \begin{IEEEeqnarray*}{rCl}
  f(y)&\triangleq  &\inf{\left\{x: (x,y)\in \mathcal{C}\right\}},\\
g(y)&\triangleq  &\sup{\left\{x: (x,y)\in \mathcal{C}\right\}}.
\end{IEEEeqnarray*}
Then $f(y)$ is convex and $g(y)$ is concave. Moreover, the points in the graphs of $f(y)$ and $g(y)$ are  extreme points\footnote{For the points $y\in\mathbb{R}^{n-1}$ where $f(y)$ and $g(y)$ are strictly convex and concave respectively it is immediate to show that $(y,f(y))$ and $(y,g(y))$ are extreme points of $\mathcal{C}$. If they are simply convex and concave, it means that they could be affine functions over some closed set of their effective domains. In such a case, that part of the graph of $f(y)$ and $g(y)$ constitutes a non-zero dimensional \emph{face} \cite{rockafellar_convex_1970} of the set $\mathcal{C}$ which can thought as the set of points of $\mathcal{C}$ where a certain linear functional achieves its maximum over $\mathcal{C}$. But any linear functional achieves its maximum over a compact and convex set $\mathcal{C}$  at an extreme point of $\mathcal{C}$.} of $\mathcal{C}$;
\item Be $\left\{f_\alpha(y)\right\}$ and $\left\{g_{\beta}(y)\right\}$ families of convex and concave functions respectively. Then $\sup_{\alpha}{f_\alpha(y)}$ and $\inf_{\beta}{g_\beta(y)}$ are convex and concave functions;
\item Let $f:\mathbb{R}^{n-1}\rightarrow \mathbb{R}$ be an arbitrary lower semi-continuous function that nowhere has the value $-\infty$. We define the \emph{convex envelope} $\mbox{cvx}(f)(y)$ of $f(y)$ as the point-wise supremum of all affine functions that are smaller than $f(y)$. Similarly, if $f:\mathbb{R}^{n-1}\rightarrow \mathbb{R}$ is an arbitrary upper semi-continuous function that nowhere has the value $+\infty$ we define the \emph{concave envelope} $\mbox{conc}(f)(y)$ of $f(y)$ as the point-wise infimum of all affine functions that are greater than $f(y)$.
\end{enumerate}

Let us consider the set $\mathcal{P}\left(\mathcal{U}\right)$ where $\mathcal{U}$ is an arbitrary finite alphabet (cardinality equal to 3 suffices). From Theorem~\ref{theo:p1}, it is clear that we can write:
\begin{align*}
\mathcal{R}_{\mbox{\tiny TW-CIB}}^{\mbox{\tiny D}}(1/2)&=\left\{(R,\mu): R\geq I(X_1;U|X_2),\ \mu\leq I(Y;UX_2),\right.\nonumber\\
&\hspace{-0.3cm}\left.U\sim p(u)\in\mathcal{P}(\mathcal{U}), \ U\mkv X_1\mkv X_2Y\right\}.
\end{align*}
The desired function $\mu_{\mbox{\tiny TW-CIB}}^{\mbox{\tiny D}}(R)$ can be obtained from
\begin{equation*}
\mu_{\mbox{\tiny TW-CIB}}^{\mbox{\tiny D}}(R)=\sup{\big\{\mu: (R,\mu)\in\mathcal{R}_{\mbox{\tiny TW-CIB}}^{\mbox{\tiny D}}(1/2)\big\}}.
\end{equation*}
We define the following functions:
\begin{eqnarray}
g(r)&\triangleq &h_2(r\ast q)-h_2(r)\nonumber\\
&= & h_2(q)-(1-q\ast r)h_2\left(\frac{qr}{1-q\ast r}\right)\nonumber\\
&&\quad-(q\ast r)h_2\left(\frac{(1-q)r}{q\ast r}\right)
\label{eq:g_fun}\\
f(r)&\triangleq &h_2(p\ast q)-(1-q\ast r)h_2\left(p\ast\frac{qr}{1-q\ast r}\right)\nonumber\\
&&\quad-(q\ast r)h_2\left(p\ast\frac{(1-q)r}{q\ast r}\right)\nonumber\\
&=& h_2(r\ast q)-(1-p\ast q)h_2\left(r\ast\frac{pq}{1-p\ast q}\right)\nonumber\\
&&\quad-(p\ast q) h_2\left(r\ast\frac{p(1-q)}{p\ast q}\right)
\label{eq:f_fun}
\end{eqnarray}
where $p,q\in(0,1/2)$ and $r\in[0,1]$. It can be easily shown that these functions are strictly convex, continuous and twice continuously differentiable as functions of $r$. In addition, they are symmetric with respect to $r=\frac{1}{2}$ and $0\leq f(r)\leq g(r)$ for all $r\in[0,1]$. In fact, it is not difficult to check that:
\begin{equation*}
g(r)=I(X_1;U|X_2),\qquad f(r)=I(Y;U|X_2),
\end{equation*}
when $(X_1,X_2,Y)\sim \mbox{Bern}(1/2)$, $U\mkv X_1\mkv X_2Y$ and $X_1=U\oplus V$ with $V\sim\mbox{Bern}(r)$, $U\sim\mbox{Bern}(1/2)$ and $U\bot V$. The following lemma can be easily proved:

\begin{lemma}[Alternative characterization of $\mathcal{R}_{\mbox{\tiny TW-CIB}}^{\mbox{\tiny D}}(1/2)$ for Binary sources]
Consider Binary sources  $(X_1, X_2,Y)\sim \mbox{Bern}(1/2)$ with $X_2\mkv X_1\mkv Y$ such that  $X_1=X_2\oplus Z$ with $Z\sim\mbox{Bern}(q)$, $q\in(0,1/2)$, $Z\bot X_2$ and $Y=X_1\oplus W$ with $W\sim\mbox{Bern}(p)$, $p\in(0,1/2)$, $W\bot (X_1,X_2)$ . Region $\mathcal{R}_{\mbox{\tiny TW-CIB}}^{\mbox{\tiny D}}(1/2)$ is equivalent to:
\begin{align*}
\mathcal{R}_{\mbox{\tiny TW-CIB}}^{\mbox{\tiny D}}&(1/2)=\left\{(R,\mu): R\geq\sum_{u\in\mathcal{U}}p(u)g(r(u)),\right.\nonumber\\
& \left.\mu\leq 1-h(p*q)+\sum_{u\in\mathcal{U}}p(u)f(r(u)),\right.\nonumber\\
& \left.\frac{1}{2}=\sum_{u\in\mathcal{U}}p(u)r(u),\ r(u)\in[0,1]\ \forall u\in\mathcal{U}\right\}
\end{align*}
\label{lemma:alter_charac_D}
\end{lemma}
\begin{IEEEproof}
Consider $(R,\mu)\in\mathcal{R}_{\mbox{\tiny TW-CIB}}^{\mbox{\tiny D}}(1/2)$. Then, it should exist $U\mkv X_1\mkv (X_2,Y)$ with $p(u)\in\mathcal{P}\left(\mathcal{U}\right)$ such that $R\geq I(X_1;U|X_2)$ and $\mu\leq I(Y;UX_2)$. In a first place, we consider $I(X_1;U|X_2)$:
\begin{align*}
&I(X_1;U|X_2)=H(X_1|X_2)-H(X_1|UX_2)\\
&=h(q)-H(X_1|UX_2)\\
&= h_2(q)-\hspace{-0.4cm}\sum_{(x_2,u)\in\mathcal{X}_2\times\mathcal{U}}\hspace{-0.2cm}p(x_2,u)H(X_1|U=u, X_2=x_2)
\end{align*} 
Using the fact that $U\mkv X_1\mkv (X_2,Y)$, it is not difficult to check that:
\begin{align*}
&p(X_1=1|U=u,X_2=0)=\frac{q r(u)}{1-q\ast r(u)},\\
&p(X_1=1|U=u,X_2=1)=\frac{(1-q)r(u)}{q\ast r(u)},\\
&p(X_2=0,U=u)=(1-q\ast r(u))p(u),\\ &p(X_2=1,U=u)=(q\ast r(u))p(u),
\end{align*}
where $r(u)\triangleq p(X_1=1|U=u)$. Using these equations, and from the fact that $X_1$ conditioned on $X_2$ and $U$ is a binary random variable we have that: 
\begin{equation*}
H(X_1|U=u, X_2=x_2)=h_2\Big(p(X_1=1|U=u,X_2=x_2)\Big),
\end{equation*}
from which $I(X_1;U|X_2)=\sum_{u\in\mathcal{U}}p(u)g(r(u))$ is easily obtained. For  $I(Y;UX_2)$ we have:
\begin{align*} 
I(Y;UX_2)&=I(Y;X_2)+I(Y;U|X_2)\\
&=1-h_2(p\ast q)+I(Y;U|X_2).
\end{align*}
The analysis of $I(Y;U|X_2)$ is similar to that of $I(X_1;U|X_2)$, obtaining:
\begin{equation*}
I(Y;U|X_2)=\sum_{u\in\mathcal{U}}p(u)f(r(u)).
\end{equation*}
The requirement that $\sum_{u\in\mathcal{U}}p(u)r(u)=\frac{1}{2}$ follows from the fact that $p(X_1=1)=\frac{1}{2}$.
\end{IEEEproof}
Consider the continuous mapping $L:[0,1]\rightarrow [0,1]\times[0,h_2(q)]\times[0,1-h_2(p)]$ given by $L(r)=(r,g(r),1-h_2(p*q)+f(r))$. Consider the image of this mapping to be $\mathcal{A}$. As $[0,1]$ is a compact and connected subset of $\mathbb{R}$ and $L(r)$ is continuous, $\mathcal{A}$ is compact and connected. Let us consider $\mathcal{C}=\mbox{co}(\mathcal{A})$. This set, thanks to Fenchel-Eggleston theorem, we have:
\begin{align*}
\mathcal{C}=&\left\{(r,\xi,\eta): \left\{\lambda_i,r_i\right\}_{i=1}^3\in[0,1],\ \sum_{i=1}^3\lambda_i=1,\right.\nonumber\\
&\qquad\left. r=\sum_{i=1}^3\lambda_ir_i ,\ \xi=\sum_{i=1}^3\lambda_ig(r_i),\ \right.\nonumber\\
&\qquad\left. \eta=1-h_2(p\ast q)+\sum_{i=1}^3\lambda_if(r_i)\right\}.
\end{align*}
We also define the convex set
\begin{equation*} 
\mathcal{C}_{1/2}=\mathcal{C}\cap\left\{(r,\xi,\eta):r=\frac{1}{2}\right\}\Big|_{(\xi,\eta)},
\end{equation*}
that is the projection of $\mathcal{C}\cap\left\{(r,\xi,\eta):r=\frac{1}{2}\right\}$ onto the plane $(\xi,\eta)$. Define the concave function $\tilde{\mu}(R)$ as:
\begin{equation*} 
\tilde{\mu}(R)=\sup{\left\{\eta:\left(\frac{1}{2},R,\eta\right)\in \mathcal{C}\right\}}=\sup{\left\{\eta:(R,\eta)\in \mathcal{C}_{1/2}\right\}}.
\end{equation*}
As $\mathcal{C}$ is compact, $\mathcal{C}_{1/2}$ is also compact. Moreover, it is easy to see that it is not empty for $R\in[0,h(q)]$. This means that:
\begin{equation*} 
\tilde{\mu}(R)=\max{\left\{\eta:(R,\eta)\in \mathcal{C}_{1/2}\right\}}.
\end{equation*}
As the graph of $\tilde{\mu}(R)$ is the upper boundary of the convex set $\mathcal{C}_{1/2}$, by (vii), each point $(R,\tilde{\mu}(R))$ is a extreme point of $\mathcal{C}_{1/2}$ or a convex combination of extreme points of $\mathcal{C}_{1/2}$. From Dubin's theorem, as $\mathcal{C}_{1/2}$ is the intersection of $\mathcal{C}$ with one hyperplane, every extreme point of $\mathcal{C}_{1/2}$ is a convex combination of no more that 2 extreme points of $\mathcal{C}$ which also belong to $\mathcal{A}$. This means that there exist $\lambda^{*}\in[0,1]$ and $r_1^{*},r_2^{*}\in[0,1]$ such that:
\begin{equation*} 
\tilde{\mu}(R)=1-h_2(p\ast q)+\lambda^{*}f(r_1^{*})+(1-\lambda)f(r_2^*)
\end{equation*}
with $R=\lambda^{*}g(r_1^*)+(1-\lambda^{*})g(r_2^*)$. Notice that is not necessarily true that $\frac{1}{2}=\lambda^{*}r_1^{*}+(1-\lambda^{*})r_{2}^{*}$. However, using the symmetry of functions $f(r)$ and $g(r)$ it is not difficult to show that:
\begin{align*}
\tilde{\mu}(R)=\max& \Big\{1-h_2(p\ast q)+\lambda f(r_1)+(1-\lambda)f(r_2):\nonumber\\
&\lambda g(r_1)+(1-\lambda)g(r_2)=R,\ (\lambda,r_1,r_2)\in[0,1],\nonumber\\
& \frac{1}{2}=\lambda r_1+(1-\lambda)r_{2}\Big\} ,
\end{align*}
obtaining an alternative characterization for $\tilde{\mu}(R)$, from which it is easy to show that is an upper semi-continuous function. From the Lemma~\ref{lemma:alter_charac_D} and the definition of $\mathcal{C}$ it is clear that we can write:
\begin{equation*} 
\mathcal{R}_{\mbox{\tiny TW-CIB}}^{\mbox{\tiny D}}(1/2)=\left\{(\xi,\eta):\exists (\xi',\eta')\in \mathcal{C}_{1/2},\ \xi\geq \xi',\ \eta\leq\eta'\right\}.
\end{equation*}
This clearly implies that $\tilde{\mu}(R)\leq \mu_{\mbox{\tiny TW-CIB}}^{\mbox{\tiny D}}(R)$. It is easy to show that if $R\mapsto \tilde{\mu}(R)$ is not decreasing  then $\tilde{\mu}(R)\geq \mu_{\mbox{\tiny TW-CIB}}^{\mbox{\tiny D}}(R)$, which implies that  $\tilde{\mu}(R)=\mu_{\mbox{\tiny TW-CIB}}^{\mbox{\tiny D}}(R)$. The following lemma establish the non-decreasing property of $\tilde{\mu}(R)$.
\begin{lemma}
Consider random binary sources $(X_1,X_2, Y)\sim \mbox{Bern}(1/2)$ with $X_2\mkv X_1\mkv Y$ such that  $X_1=X_2\oplus Z$ with $Z\sim\mbox{Bern}(q)$, $q\in(0,1/2)$, $Z\bot X_2$ and $Y=X_1\oplus W$ with $W\sim\mbox{Bern}(p)$, $p\in(0,1/2)$, $W\bot (X_1,X_2)$. Then, for all $R\in[0,h_2(q)]$, 
\begin{equation*}
1-h_2(p\ast q)+\frac{h_2(p\ast q)-h_2(p)}{h_2(q)}R\leq \tilde{\mu}(R)\leq 1-h_2(p\ast q)+R
\end{equation*}
and $\tilde{\mu}(R)$ is not decreasing in $R$.
\label{lemma:properties_mu_D}
\end{lemma}
\begin{IEEEproof}
From the assumptions, Lemma~\ref{lemma:alter_charac_D}, definitions of $\mathcal{C}$ and $\tilde{\mu}(R)$, we have
\begin{align*} 
\tilde{\mu}(R)=\max&\left\{I(Y;UX_2): I(X_1;U|X_2)=R,\right.\nonumber\\  
&\left.U\sim p(u)\in\mathcal{P}(\mathcal{U}), \ U\mkv X_1\mkv (X_2,Y) \right\}.
\end{align*}
From \emph{data processing} inequality it is easy to show that for all variables $U$ such that $U\mkv X_1\mkv (X_2,Y)$, $I(Y;UX_2)\leq I(Y;X_1X_2)\leq 1-h_2(p)$ and $I(X_1;U|X_2)\leq H(X_1|X_2)=h_2(q)$. This implies that $\tilde{\mu}(R)\leq 1-h_2(p)$ for all $R\in[0,h_2(q)]$. Consider $U=X_1$. In this case $R=h_2(q)$ and $I(Y;UX2)=1-h_2(p)$, allowing us to conclude that $\tilde{\mu}(h_2(q))=1-h_2(p)$. When $U$ is constant, we obtain $I(X_1;U|X_2)=0$ and $I(Y;UX_2)=1-h(p\ast q)$. In fact, it is not hard to check that $\tilde{\mu}(0)=1-h(p\ast q)$. As $\tilde{\mu}(R)$ is concave, the lower bound on $\tilde{\mu}(R)$ follows immediately. The proof of the upper bound is straightforward and for that reason is omitted. To prove the non-decreasing property consider any $R\in[0,h_2(q)]$ and $R_1\leq R$. Then, exists $\lambda\in[0,1]$ such that $R=\lambda R_1+(1-\lambda)h_2(q)$. As $R\mapsto \tilde{\mu}(R)$ is concave, we have:
\begin{equation*} 
\tilde{\mu}(R)\geq \lambda \tilde{\mu}(R_1)+(1-\lambda)\tilde{\mu}(h_2(q))\geq \tilde{\mu}(R_1),
\end{equation*}
from which the result follows.
\end{IEEEproof}

From the previous results we can conclude that:
\begin{align*}
\mu_{\mbox{\tiny TW-CIB}}^{\mbox{\tiny D}}&(R)=\max_{(\lambda,r_1,r_2)\in[0,1]}1-h_2(p\ast q)+\lambda f(r_1)\\
&+(1-\lambda)f(r_2)\qquad\mbox{s.t.}\ \ \lambda g(r_1)+(1-\lambda)g(r_2)=R.\nonumber
\end{align*}
This problem can be solved numerically to obtain, for each $R$, the exact value of $\mu_{\mbox{\tiny TW-CIB}}^{\mbox{\tiny D}}(R)$. However, more can be said of $\mu_{\mbox{\tiny TW-CIB}}^{\mbox{\tiny D}}(R)$. As the graph of $\mu_{\mbox{\tiny TW-CIB}}^{\mbox{\tiny D}}(R)$ is an upper boundary of $C_{1/2}$, which is convex and compact, on each point of this boundary exists a supporting hyperplane. Consider point $(R_0,\mu_{\mbox{\tiny TW-CIB}}^{\mbox{\tiny D}}(R_0))$. The supporting hyperplane for this point is defined by the pair $(\alpha,\psi(\alpha))$, such that $\mu_{\mbox{\tiny TW-CIB}}^{\mbox{\tiny D}}(R_0)=\alpha R_0+\psi(\alpha)$ and 
$\mu_{\mbox{\tiny TW-CIB}}^{\mbox{\tiny D}}(R)\leq\alpha R+\psi(\alpha)$ for other $R\in[0,h_2(q)]$. This implies that:
\begin{align}
\psi(\alpha)&=\max{\left\{\mu_{\mbox{\tiny TW-CIB}}^{\mbox{\tiny D}}(R)-\alpha R: R\in[0,h_2(q)]\right\}}\nonumber\\
&=\max{\left\{\eta-\alpha \xi: (\xi,\eta)\in \mathcal{C}_{1/2}\right\}},\nonumber\\
&=\max{\left\{\eta-\alpha \xi: (1/2,\xi,\eta)\in \mathcal{C}\right\}}.
\label{eq:psi}
\end{align}
From (viii) above is immediate to see that $\psi(\alpha)$ is a convex function of $\alpha$. From its concavity and upper semi-continuity we know that $\mu_{\mbox{\tiny TW-CIB}}^{\mbox{\tiny D}}(R)$ can be expressed alternatively as the point-wise infimum of affine functions that are greater that $\mu_{\mbox{\tiny TW-CIB}}^{\mbox{\tiny D}}(R)$. In fact, it is not difficult to show that:
\begin{equation}
\mu_{\mbox{\tiny TW-CIB}}^{\mbox{\tiny D}}(R)=\min{\left\{\psi(\alpha)+\alpha R: \alpha\in \mathbb{R}\right\}}.
\label{eq:alter_mu_D}
\end{equation}
From the results of Lemma~\ref{lemma:properties_mu_D} is not difficult to see that in (\ref{eq:alter_mu_D}), it suffices to restrict $\alpha$ to the interval $[0,1]$. Consider now a fixed value of $\alpha\in[0,1]$ and define $\tilde{\nu}(r,\alpha)=1-h_2(p\ast q)+\nu(r,\alpha)$ where $\nu(r,\alpha)=f(r)-\alpha g(r)$ with $r\in[0,1]$. Define $A^{\alpha}$ to be the graph of $\tilde{\nu}(r,\alpha)$ and $C^{\alpha}=\mbox{co}(A^{\alpha})$. It is not hard to see that:
\begin{equation*}
C^{\alpha}=\left\{(r,\eta-\alpha\xi): (r,\xi,\eta)\in \mathcal{C}\right\},
\end{equation*}
and that the upper-boundary of $C^{\alpha}$ (which is compact) is the graph of the concave envelope of $\nu(r,\alpha)$. In fact, if we define $\psi(r,\alpha)$ as:
\begin{eqnarray*}
\psi(r,\alpha)&=&\max{\left\{\omega:(r,\omega)\in C^{\alpha}\right\}}\nonumber\\
&=&\max{\left\{\eta-\alpha\xi:(r,\xi,\eta)\in \mathcal{C}\right\}},
\end{eqnarray*}
we have that $\mbox{conc}(\tilde{\nu})(r,\alpha)=\psi(r,\alpha)$. It is clear that $\psi(1/2,\alpha)$ is equal to $\psi(\alpha)$ defined in (\ref{eq:psi}). That is:
\begin{align*}
\psi(\alpha)&=\mbox{conc}(\tilde{\nu})(r,\alpha)\Big|_{r=1/2}\\
&=1-h_2(p\ast q)+\mbox{conc}(\nu)(r,\alpha)\Big|_{r=1/2}.
\end{align*}
Note that $\nu(r,\alpha)$ is symmetric with respect to $r=1/2$ for every $\alpha$ and that $\nu(1/2,\alpha)=0$.  This symmetry implies that:
\begin{align*}
\mbox{conc}(\nu)(r,\alpha)\Big|_{r=1/2}&=\max_{r\in[0,1/2]}{\nu(\alpha,r)}\\
&=\max_{r\in[0,1/2]}{\left\{f(r)-\alpha g(r)\right\}}.
\end{align*}
and using (\ref{eq:alter_mu_D}) we have
\begin{align}
\mu_{\mbox{\tiny TW-CIB}}^{\mbox{\tiny D}}(R)=&1-h_2(p\ast q)\nonumber\\
&+\min_{\alpha\in[0,1]}\max_{r\in[0,1/2]}{\left\{f(r)+\alpha(R-g(r))\right\}}.
\label{eq:dual_mu_D}
\end{align}
It can also be shown that $\nu(r,\alpha)$ is a continuous and twice continuously differentiable as function of $r$ and $\alpha$. Moreover, when $\alpha=1$ we clearly have that $\nu(r,1)=f(r)-g(r)\leq 0$ for all $r\in[0,1/2]$ and when $\alpha=0$,  $\nu(r,0)=f(r)\geq 0$ for all $r\in[0,1/2]$. In both extreme cases the equality is obtained only when $r=1/2$. 
It is not difficult to obtain the second partial derivative of $\nu(r,\alpha)$ with respect to $r$:
\begin{align*}
&\frac{\partial^2 \nu(r,\alpha)}{\partial r^2}=-\frac{(1-\alpha)(1-2q)^2}{(q\ast r)(1-q\ast r)}-\frac{\alpha}{r(1-r)}\nonumber\\
&\qquad\quad+\frac{(1-p\ast q)(1-2\gamma)^2}{(\gamma\ast r)(1-\gamma\ast r)}+\frac{(p\ast q)(1-2\delta)^2}{(\delta\ast r)(1-\delta\ast r)},
\end{align*}
where
\begin{equation*}
\gamma\triangleq \frac{pq}{1-p\ast q},\ \ \delta\triangleq \frac{p(1-q)}{p\ast q}.
\end{equation*}
As $f(r)$ is strictly convex it is clear that $\frac{\partial^2 \nu(r,0)}{\partial r^2}> 0$ for all $r\in[0,1/2]$. On the other hand, $p,q\in(0,1/2)$ implies that $(1-2\gamma)^2<1$, $(1-2\delta)^2<1$ and $(\gamma\ast r)(1-\gamma\ast r)\geq r(1-r)$, $(\delta\ast r)(1-\delta\ast r)\geq r(1-r)$ for all $r\in[0,1/2]$ which leads us to:
\begin{equation*}
\frac{(1-p\ast q)(1-2\gamma)^2}{(\gamma\ast r)(1-\gamma\ast r)}+\frac{(p\ast q)(1-2\delta)^2}{(\delta\ast r)(1-\delta\ast r)}< \frac{1}{r(1-r)},
\end{equation*}
\begin{figure}[t]
	\centering{\includegraphics[width=0.48\textwidth]{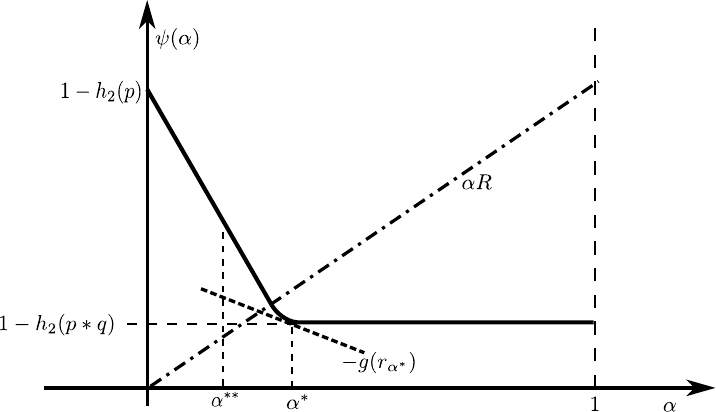}}
	\caption{Graph of $\psi(\alpha)$.}
\label{fig:psi}
\end{figure}
for all $r\in[0,1/2]$, implying that $\frac{\partial^2 \nu(r,1)}{\partial r^2}< 0$ for all $r\in[0,1/2]$, and thus  $r\mapsto \nu(r,1)=f(r)-g(r)$ is an strictly concave function. In fact, as $\alpha\mapsto \frac{\partial^2 \nu(r,\alpha)}{\partial r^2}$ is a continuous function, this strict concavity should hold for $\alpha$ in an open interval around $1$, where as a consequence, $\nu(r,\alpha)\leq 0$ for any $r\in[0,1/2]$. Consider $\alpha^{*}\in(0,1)$ to be the minimal value such that for all $\alpha\in[\alpha^*,1]$:
 \begin{equation*}
 \nu(r,\alpha)\leq 0\ , \ \forall r\in[0,1/2].
 \end{equation*}
Again as $r\mapsto \nu(r,0)$ is strictly convex and $\alpha\mapsto \displaystyle \frac{\partial^2 \nu(r,\alpha)}{\partial r^2}$ is continuous, it must exists a maximal value $\alpha^{**}<\alpha^{*}$ such that for all $\alpha\in[0,\alpha^{**}]$
\begin{equation*}
\frac{\partial^2 \nu(r,\alpha)}{\partial r^2}\geq 0\ , \  \forall r\in[0,1/2],
 \end{equation*}
 which implies the convexity of $\nu(\alpha,r)$ for all $r\in[0,1/2]$ for $\alpha\in[0,\alpha^{**}]$. For every $\alpha<\alpha^{*}$ we must have that there exists $r\in(0,1/2)$ such that $\nu(r,\alpha)>0$. This, jointly with the continuity of $(r,\alpha)\mapsto \nu(r,\alpha)$, implies that for $\alpha^{*}$ there must exist $r_{\alpha^{*}}\in(0,1/2)$ with  $\nu(r_{\alpha^{*}},\alpha^{*})=0$. Similarly, it can be argued that $\frac{\partial \nu(r,\alpha^{*})}{\partial r}\Big|_{r_{\alpha^{*}}}=0$. This means that for $\alpha\in[\alpha^{*},1]$,  $\max_{r\in[0,1/2]}{\nu(\alpha,r)}=0$ and $\psi(\alpha)=1-h_2(p\ast q)$. When $\alpha^{**}<\alpha<\alpha^{*}$,  $\max_{r\in[0,1/2]}{\nu(\alpha,r)}>0$ and $\psi(\alpha)>1-h_2(p\ast q)$. Consider $r_{\alpha}\in(0,1/2)$ to be the point at which the maximum is achieved. At this point $\frac{\partial \nu(r,\alpha)}{\partial r}\Big|_{r_\alpha}= 0$. We have, by the implicit function theorem, that:
\begin{align*}
\alpha&=\frac{f'(r_{\alpha})}{g'(r_\alpha)},\\
\psi(\alpha)&=1-h_2(p\ast q)+f(r_\alpha)-\alpha g(r_{\alpha}),\\ \psi'(\alpha)&=-g(r_\alpha)<0.
 \end{align*}
At point $\alpha^{*}$,  the derivative of $\psi(\alpha)$ could not exist, but the limit from the left exists and satisfies:
\begin{equation*}
\lim_{\alpha\uparrow\alpha^{*}}\psi'(\alpha)=-g(r_{\alpha^{*}})<0.
 \end{equation*}
Finally, when $\alpha\in[0,\alpha^{**}]$, as $\nu(r,\alpha)$ is convex, its maximum value has to be achieved at a boundary point of $[0,1/2]$. It is clear that this point should be $r=0$. In this manner  $\max_{r\in[0,1/2]}{\nu(\alpha,r)}=f(0)-\alpha g(0)=h(p\ast q)-h_2(p)-\alpha h_2(q)$ and $\psi(\alpha)=1-h_2(p)-\alpha h_2(q)$, which is an affine function in $\alpha$ with slope $h_2(q)$. With these results, we see that $\psi(\alpha)$ must have the shape shown in Fig.~\ref{fig:psi}. From (\ref{eq:alter_mu_D}) and (\ref{eq:dual_mu_D}), the obtained properties of $\psi(\alpha)$ and the fact that beyond $R>h_2(q)$, $\mu_{\mbox{\tiny TW-CIB}}^{\mbox{\tiny D}}(R)$ takes the value of $1-h_2(p)$, it is easy show that:
\begin{align*}
&\mu_{\mbox{\tiny TW-CIB}}^{\mbox{\tiny D}}(R)=\label{eq:final_mu_D}\\
&\left\{\begin{array}{ll}
1-h_2(p\ast q)+\alpha^{*}R & 0\leq R\leq g(r_{\alpha^{*}}),\nonumber\\
1-h_2(p\ast q)+f\left(g^{-1}(R)\right) & g(r_{\alpha^{*}})<R\leq h_2(q),\\
1-h_2(p) & R>h_2(q).
\end{array}\right.\nonumber
\end{align*}
Let us define $R_{c} \triangleq g(r_{\alpha^{*}})$. As $\mu_{\mbox{\tiny TW-CIB}}^{\mbox{\tiny D}}(R)$ is concave it is not difficult to see that $R_c$ and $\alpha^{*}$ should satisfy:
\begin{equation*}
\frac{f'\left(g^{-1}(R_c)\right)}{g'\left(g^{-1}(R_c)\right)}=\frac{f\left(g^{-1}(R_c)\right)}{R_c},\ \ \alpha^{*}=\frac{f\left(g^{-1}(R_c)\right)}{R_c}.
\end{equation*}

From the final expression in (\ref{eq:final_mu_D}), it is pretty clear how should be the scheme to be used to achieve $\mu_{\mbox{\tiny TW-CIB}}^{\mbox{\tiny D}}(R)$. When $R>R_c$, auxiliar random variable $U$ should be chosen such that: $U=X_1\oplus V,$ where $V\sim\mbox{Bern}\left(g^{-1}(R)\right)$. When $R\leq R_c$ a time-sharing scheme should be used. It is not difficult to show that $U$ should be chosen as: $U\triangleq \mathds{1}\left\{T=0\right\}+V_c\mathds{1}\left\{T=1\right\},$ where $V_c\sim\mbox{Bern}\left(g^{-1}(R_c)\right)$ and $T\sim \mbox{Bern}\left(\frac{R}{R_c}\right)$. When $R>h_2(q)$, $U\triangleq X_1$.
\section*{Acknowledgement}
The authors wish to thank the Associate Editor and the anonymous reviewers for the detailed suggestions and comments which significantly improved the manuscript.

\bibliographystyle{IEEEtran}
\bibliography{matias.bib}

\begin{IEEEbiographynophoto}{Mat\'{i}as Vera}
	(S'16) received the B.Sc. and M.Sc. degrees in electrical engineering from the University of Buenos Aires, Buenos Aires, Argentina, in 2014. He is currently a PhD student at the Department of Electronics, School of Engineering, Universidad de Buenos Aires, where he also works as a Teaching Assistant. His current research interests include machine learning, information theory and speaker recognition.
\end{IEEEbiographynophoto}
\begin{IEEEbiographynophoto}
	{Leonardo Rey Vega}
	(M'11) received the M.Sc (with honors) and PhD (summa cum laude) degrees in Electrical 
	Engineering from the University of Buenos Aires (Argentina) in 2004 and 2010, respectively. In 2007 and 2008 he was invited at the INRS-EMT in Montreal, Canada and in 2012 he was a visitor at the Department of Telecommunications at SUPELEC, France. He is currently an Associate Professor at the University of Buenos Aires and member of the National Scientific and Technical Research Council in Argentina. Dr. Rey Vega's research interests include  statistical signal processing, information theory, representation learning and wireless sensor networks.
\end{IEEEbiographynophoto}
\begin{IEEEbiographynophoto}{Pablo Piantanida} (SM'16)
	received both B.Sc. in Electrical Engineering and the  M.Sc (with honors)  from the University of Buenos Aires (Argentina) in 2003, and the Ph.D. from Universit\'e Paris-Sud (Orsay, France) in 2007. Since October 2007 he has joined the Laboratoire des Signaux et Syst\`emes (L2S), at CentraleSup\'{e}lec together with CNRS (UMR 8506) and Universit\'e Paris-Sud, as an Associate Professor of Network Information Theory. He is currently associated with Montreal Institute for Learning Algorithms (MILA) at Universit\'{e} de Montr\'{e}al. He is an IEEE Senior Member, and General Co-Chair of the 2019 IEEE International Symposium on Information Theory (ISIT). His research interests lie broadly in information theory and its interactions with other fields, including multi-terminal information theory, Shannon theory, machine learning and representation learning, statistical inference, cooperative communications, communication mechanisms for security and privacy.  
\end{IEEEbiographynophoto}
\clearpage
\end{document}

%% file: packages.tex
\usepackage[latin1]{inputenc}
\usepackage[T1]{fontenc}
\usepackage[english]{babel}
\usepackage{amssymb,amsmath,amsfonts}
\usepackage{times}
\usepackage{graphicx}
\usepackage{color}
\usepackage{verbatim}
\usepackage{cite}

\usepackage{subfig}		

\usepackage{stmaryrd}

\usepackage{cancel}
\usepackage[normalem]{ulem}

\usepackage{verbatim}

\newtheorem{definition}	{Definition}
\newtheorem{theorem}	{Theorem}

\newtheorem{lemma}		{Lemma}
\newtheorem{corollary}	{Corollary}
\newtheorem{remark}		{Remark}

%% file: commandes.tex
\newcommand{\mkv}{-\!\!\!\!\minuso\!\!\!\!-}













%% file: it_journal.bbl
\begin{thebibliography}{10}
\providecommand{\url}[1]{#1}
\csname url@samestyle\endcsname
\providecommand{\newblock}{\relax}
\providecommand{\bibinfo}[2]{#2}
\providecommand{\BIBentrySTDinterwordspacing}{\spaceskip=0pt\relax}
\providecommand{\BIBentryALTinterwordstretchfactor}{4}
\providecommand{\BIBentryALTinterwordspacing}{\spaceskip=\fontdimen2\font plus
\BIBentryALTinterwordstretchfactor\fontdimen3\font minus
  \fontdimen4\font\relax}
\providecommand{\BIBforeignlanguage}[2]{{%
\expandafter\ifx\csname l@#1\endcsname\relax
\typeout{** WARNING: IEEEtran.bst: No hyphenation pattern has been}%
\typeout{** loaded for the language `#1'. Using the pattern for}%
\typeout{** the default language instead.}%
\else
\language=\csname l@#1\endcsname
\fi
#2}}
\providecommand{\BIBdecl}{\relax}
\BIBdecl

\bibitem{Chandrasekaran2013Computational}
V.~Chandrasekaran and M.~I. Jordan, ``Computational and statistical tradeoffs
  via convex relaxation,'' \emph{Proc. of the Nat. Academy of Sciences}, vol.
  110, no.~13, pp. E1181--E1190, 2013.

\bibitem{Shannon1993CodingTheoremsForADiscreteSourceWithAFidelityCriterion}
C.~E. Shannon, ``Coding theorems for a discrete source with a fidelity
  criterion,'' in \emph{Claude Elwood Shannon: collected papers}, N.~J.~A.
  Sloane and A.~D. Wyner, Eds.\hskip 1em plus 0.5em minus 0.4em\relax IEEE
  Press, 1993, pp. 325--350.

\bibitem{tishby99}
N.~Tishby, F.~C. Pereira, and W.~Bialek, ``The information bottleneck method,''
  in \emph{Proceedings of the Annual Allerton Conference on Communication,
  Control and Computing}, 1999, pp. 368--377.

\bibitem{1057738}
R.~Dobrushin and B.~Tsybakov, ``Information transmission with additional
  noise,'' \emph{IRE Transactions on Information Theory}, vol.~8, no.~5, pp.
  293--304, September 1962.

\bibitem{Courtade_2014}
T.~Courtade and T.~Weissman, ``Multiterminal source coding under logarithmic
  loss,'' \emph{Information Theory, IEEE Trans. on}, vol.~60, no.~1, pp.
  740--761, 2014.

\bibitem{witsenhausen75}
H.~Witsenhausen and A.~Wyner, ``{A conditional entropy bound for a pair of
  discrete random variables},'' \emph{Information Theory, IEEE Transactions
  on}, vol.~21, no.~5, pp. 493--501, 1975.

\bibitem{GeorgISIT2015B}
G.~Pichler, P.~Piantanida, and G.~Matz, ``Distributed information-theoretic
  biclustering of two memoryless sources,'' in \emph{Proceedings of the 53-rd
  Annual Allerton Conference on Communication, Control and Computing}, 2015.

\bibitem{1057194}
R.~Ahlswede and I.~Csiszar, ``Hypothesis testing with communication
  constraints,'' \emph{Information Theory, IEEE Transactions on}, vol.~32,
  no.~4, pp. 533--542, Jul 1986.

\bibitem{Gil2015}
\BIBentryALTinterwordspacing
G.~Katz, P.~Piantanida, and M.~Debbah, ``Distributed binary detection with
  lossy data compression,'' \emph{Information Theory, IEEE Transactions on},
  October 2015, (submitted). [Online]. Available:
  \url{http://arxiv.org/abs/1601.01152}
\BIBentrySTDinterwordspacing

\bibitem{1055800}
J.~Korner and K.~Marton, ``Images of a set via two channels and their role in
  multi-user communication,'' \emph{Information Theory, IEEE Transactions on},
  vol.~23, no.~6, pp. 751--761, Nov 1977.

\bibitem{ahlswede1976}
\BIBentryALTinterwordspacing
R.~Ahlswede and P.~Gacs, ``Spreading of sets in product spaces and
  hypercontraction of the markov operator,'' \emph{Ann. Probab.}, vol.~4,
  no.~6, pp. 925--939, 12 1976. [Online]. Available:
  \url{http://dx.doi.org/10.1214/aop/1176995937}
\BIBentrySTDinterwordspacing

\bibitem{Erkip1998efficiency}
E.~Erkip and T.~M. Cover, ``The efficiency of investment information,''
  vol.~44, no.~3, pp. 1026--1040, May 1998.

\bibitem{PhD-Tung}
S.~Y. Tung, ``Multiterminal source coding,'' Ph.D. Dissertation, Electrical
  Engineering, Cornell University, Ithaca, NY, May 1978.

\bibitem{berger_ceo_1996}
T.~Berger, Z.~Zhang, and H.~Viswanathan, ``The {CEO} problem,''
  \emph{Information Theory, IEEE Trans. on}, vol.~42, no.~3, pp. 887--902,
  1996.

\bibitem{prabhakaran_rate_2004}
V.~Prabhakaran, D.~Tse, and K.~Ramachandran, ``Rate region of the quadratic
  gaussian {CEO} problem,'' in \emph{{IEEE} International Symp. on Information
  Theory, {ISIT} 2004}, 2004, p. 119.

\bibitem{oohama_rate-distortion_2005}
Y.~Oohama, ``Rate-distortion theory for {Gaussian} multiterminal source coding
  systems with several side informations at the decoder,'' \emph{IEEE
  Transactions on Information Theory}, vol.~51, no.~7, pp. 2577--2593, Jul.
  2005.

\bibitem{DBLP:journals/corr/PichlerPM16}
\BIBentryALTinterwordspacing
G.~Pichler, P.~Piantanida, and G.~Matz, ``Distributed information-theoretic
  biclustering,'' \emph{CoRR}, vol. abs/1602.04605, 2016. [Online]. Available:
  \url{http://arxiv.org/abs/1602.04605}
\BIBentrySTDinterwordspacing

\bibitem{kaspi_two-way_1985}
A.~Kaspi, ``Two-way source coding with a fidelity criterion,'' \emph{{IEEE}
  Transactions on Information Theory}, vol.~31, no.~6, pp. 735 -- 740, Nov.
  1985.

\bibitem{ma_ishwar2011}
N.~Ma and P.~Ishwar, ``Some results on distributed source coding for
  interactive function computation.'' \emph{IEEE Transactions on Information
  Theory}, vol.~57, no.~9, pp. 6180--6195, 2011.

\bibitem{Permuter_2010}
H.~H. Permuter, Y.~Steinberg, and T.~Weissman, ``Two-way source coding with a
  helper,'' \emph{IEEE Transactions on Information Theory}, vol.~56, no.~6, pp.
  2905--2919, June 2010.

\bibitem{Chia_2012}
Y.~K. Chia, H.~H. Permuter, and T.~Weissman, ``Cascade, triangular, and two-way
  source coding with degraded side information at the second user,'' \emph{IEEE
  Transactions on Information Theory}, vol.~58, no.~1, pp. 189--206, Jan 2012.

\bibitem{ours_2015}
L.~R. Vega, P.~Piantanida, and A.~O. Hero, ``The three-terminal interactive
  lossy source coding problem,'' \emph{IEEE Transactions on Information
  Theory}, vol.~63, no.~1, pp. 532--562, Jan 2017.

\bibitem{our_isit15}
M.~Vera, L.~R. Vega, and P.~Piantanida, ``The two-way cooperative information
  bottleneck,'' in \emph{{IEEE} International Symp. on Information Theory,
  {ISIT} 2015}, 2015, pp. 2131--2135.

\bibitem{prabhakaran04}
V.~Prabhakaran, K.~Ramchandran, and D.~Tse, ``On the role of interaction
  between sensors in the {CEO} problem,'' in \emph{Proceedings of the Annual
  Allerton Conference on Communication, Control and Computing}, 2004.

\bibitem{ELGamal-Kim-book}
A.~E. Gamal and Y.-H. Kim, \emph{Network Information Theory}.\hskip 1em plus
  0.5em minus 0.4em\relax New York, NY, USA: Cambridge University Press, 2012.

\bibitem{rockafellar_convex_1970}
R.~T. Rockafellar, \emph{Convex Analysis}.\hskip 1em plus 0.5em minus
  0.4em\relax Princeton University Press, Jun. 1970.

\bibitem{rioul_information_2011}
O.~Rioul, ``Information theoretic proofs of entropy power inequalities,''
  \emph{IEEE Transactions on Information Theory}, vol.~57, no.~1, pp. 33--55,
  Jan. 2011.

\bibitem{wyner_ziv76}
A.~D. Wyner and J.~Ziv, ``The rate-distortion function for source coding with
  side information at the decoder,'' \emph{IEEE Trans. Inform. Theory},
  vol.~22, pp. 1--10, 1976.

\bibitem{Wagner_2008b}
A.~B. Wagner, S.~Tavildar, and P.~Viswanath, ``Rate region of the quadratic
  gaussian two-encoder source-coding problem,'' \emph{IEEE Trans. on
  Information Theory}, vol.~54, no.~5, pp. 1938--1961, May 2008.

\bibitem{wyner_ziv73}
A.~D. Wyner and J.~Ziv, ``A theorem on the entropy of certain binary sequences
  and applications: Part {I},'' \emph{{IEEE} Transactions on Information
  Theory}, vol.~19, no.~6, pp. 769--772, 1973.

\bibitem{Csiszar81}
I.~Csisz{\'a}r and J.~K{\"o}rner, \emph{Information Theory: Coding Theorems for
  Discrete Memoryless Systems}.\hskip 1em plus 0.5em minus 0.4em\relax New
  York: Academic, 1981.

\bibitem{GML_ours_2014}
P.~Piantanida, L.~Vega, and A.~O. Hero, ``A proof of the generalized {Markov}
  lemma with countable infinite sources,'' in \emph{Information Theory (ISIT),
  2014 IEEE International Symposium on}, June 2014, pp. 591--595.

\bibitem{Wit_1980}
H.~Witsenhausen, ``Some aspects of convexity useful in information theory,''
  \emph{IEEE Transactions on Information Theory}, vol.~26, no.~3, pp. 265--271,
  May 1980.

\bibitem{Eggleston}
H.~Eggleston, \emph{Convexity}.\hskip 1em plus 0.5em minus 0.4em\relax New
  York:Cambridge University, 1963.

\bibitem{Dubin}
L.~Dubin, ``On extreme points of convex sets,'' \emph{Journal of Mathematical
  Analysis and Applications}, vol.~5, pp. 237--244, 1962.

\bibitem{Krein}
M.~Krein and D.~Milman, ``On extreme points of regular convex sets,''
  \emph{Studia Mathematica}, vol.~9, pp. 133--138, 1940.

\end{thebibliography}
